\documentclass{emulateapj}
\newcommand {\bdot}{\hbox{\Huge .}}
\newcommand {\mdot}{m\kern-0.1667em\bdot}
\newcommand {\jdot}{j\kern-0.1667em\bdot}
\newcommand {\zdot}{z\kern-0.1667em\bdot}

\shorttitle{The Reddest DR3 SDSS/XMM-Newton Quasars}
\shortauthors{M. Young et al.}

\begin{document}

\title{The Reddest DR3 SDSS/XMM-Newton Quasars
}

\author{
M. Young\altaffilmark{1,2}, M. Elvis\altaffilmark{1}, 
G. Risaliti\altaffilmark{1,3}
} 
\email{myoung@cfa.harvard.edu}

\altaffiltext{1}{Harvard-Smithsonian Center for Astrophysics, 60 Garden St. 
Cambridge, MA 02138 USA}
\altaffiltext{2}{Boston University, Astronomy Department, 725 Commonwealth Ave., 
Boston, MA 02215}
\altaffiltext{3}{INAF - Osservatorio di Arcetri, Largo E. Fermi 5,
Firenze, Italy}
\begin{abstract}

We have cross-correlated the SDSS DR3 \citet{Schnei05} quasar catalog with 
the XMM-Newton archive.  Color and redshift selections ($g - r$ $\geq$ 0.5 and 0.9 $<$ z $<$ 2.1) 
result in a sample of 17 red, moderate redshift quasars.  The redshift selection 
minimizes possible contamination due to host galaxy emission and Ly$\alpha$ forest absorption.  
Both optical and X-ray information are required to distinguish between 
the two likely remaining causes of the red colors: 1) dust-reddening and 2) an 
intrinsically red continuum.  We find that 7 of 17 quasars can be classified as probable 
`intrinsically red' objects.  These 7 quasars have unusually broad MgII emission lines 
($<$FWHM$>$=10,500 km s$^{-1}$), moderately flat, but unabsorbed X-ray spectra 
($<\Gamma>$=1.66$\pm$0.08), and low accretion rates ($\dot{M}/\dot{M_{Edd}} \sim$ 0.01).  
We suggest low accretion rates as a a possible physical explanation for quasars with 
intrinsically red optical continua.  We find that 8 of 17 quasars can be classified as 
dust-reddened.  Three of these have upper-limits on the absorption column from X-ray spectral fits 
of N$_H$ = 3-13 x 10$^{22}$ cm$^2$, while the other five quasars must be absorbed by at least 
N$_H$ = 10$^{23}$ cm$^2$ in order to be consistent with a comparably selected 
$\alpha_{ox}-l_{uv}$ distribution.  Two objects in the sample are unclassified.  
\end{abstract}

\keywords{ accretion disks --- galaxies: active --- quasars: general}

\section{INTRODUCTION}

Until recently, quasars have typically been selected as point-like objects 
with colors bluer than stars (e.g. U $ - $ B $<$ -0.44, Schmidt \& Green 1983)\nocite{SG83}.  
This limited the possible spectral energy diagrams (SEDs) that known quasars could have.  
A composite SED shows that quasars' blue colors are due to the ‘Big Blue Bump’ (BBB) 
continuum feature that dominates the optical and UV spectrum \citep{MS82, Elvis94}.  
However, surveys using different selection mechanisms 
discovered that quasars can have red colors as well, indicating a diminished 
or missing BBB.  While in many cases, this reddening is due to dust obscuration, 
some cases may be due to changes in the accretion disk thought 
to power the optical/UV continuum.  We identify such a population in this paper.

Red quasars were first discovered in the \citet{Web95} sample of radio-loud 
quasars.  \citet{Web95} suggested dust-reddening as the cause of the red $B_J-K$ colors 
and estimated that red colors could cause as much as 80\% of the quasar population 
to go undetected in existing optical surveys.  Some 200 members of a new population 
of red, radio-quiet quasars were discovered in the near-infrared (1-2 $\mu$m) with the 
Two Micron All Sky Survey (2MASS) by \citet{Cutri02}.  The space density of the 
red 2MASS quasars was found to be at least equal to that of optical and UV-selected 
quasars.  \cite{Cutri02} also suggested dust-reddening to explain the red colors of 
the 2MASS red quasars because the near-IR color distribution is consistent with a 
reddening of $A_V$ = 1-5 magnitudes.  In addition, $\sim$10\% of the predominantly 
radio-quiet sample is highly polarized ($P > 3\%$) \citep{Smith02}.  Supporting 
this conclusion, Chandra and XMM-Newton observations of red 2MASS AGN find spectra 
that are either unusually hard or absorbed by N$_H$­ = 10$^{21}$-10$^{23}$ cm$^{-2}$ 
\citep{Wilkes02, Wilkes05, Urr05}. 

Unlike UV-excess based surveys, the Sloan Digital Sky Survey (SDSS) allows detection 
of red quasars in the optical band because the algorithm for selecting candidates 
for spectroscopy uses a four-dimensional multicolor selection criterion \citep{Rich03} 
similar to that of the 2dF QSO Redshift Survey \citep{Croom04}.  
This allows the SDSS to select all objects lying outside of the stellar locus, 
including atypically red quasars.  \citet{Rich03} explored the color 
distribution of SDSS quasars and found that while a population of intrinsically red 
quasars exists, the majority of the red tail of the color distribution is explained 
by dust reddening.  In a later study \citep{Hopkins04}, the curvatures of the SDSS 
spectra were found to be best-fit by a dust extinction curve similar to that of the 
Small Magellanic Cloud (SMC) \citep{Prevot84}.  However, only mildly dust-reddened 
quasars can be detected (E(B-V) $<$ 0.5) in the SDSS before dust extinction causes them 
to drop below the detection threshold of the survey \citep{Rich03}.  

While dust-reddening is a common explanation for red colors in quasars, other possible 
explanations include: (1) host galaxy contamination, (2) absorption by the Ly$\alpha$ forest, 
(3) optical synchrotron emission superimposed on a normal, blue spectrum to create a red 
spectrum \citep{SR96, Francis00, Whit01}, and (4) an intrinsically red continuum. 
The redshift selection in this paper minimizes contamination via options (1) and (2).  

X-ray observations used in conjunction with optical information can constrain the amount 
of intrinsic absorption in a source, thereby allowing an investigation of dust-reddened 
versus intrinsically red optical continua.  
Previous studies have found intrinsically red quasar candidates in addition to dust-reddened 
quasars.  For example, \citet{Hall06} selected a sample of 12 red quasars from the SDSS, 
but Chandra X-ray observations show evidence for NH absorption (N$_H­ \sim~10^{22}$ cm$^{-2}$) 
for only 4 of their 12 quasars.  The other seven quasars show no evidence for even moderate 
intrinsic absorption and three of these show evidence for an intrinsically red optical continuum.  

In a similar vein, \citet{Risaliti03} observed 16 optically-selected, X-ray weak quasars 
with Chandra.  While the sample was not selected to be red, the average color of the sample 
is redder than that of the parent Hamburg Quasar Survey ($\Delta(B-R) \sim$ 1 vs. 
$\Delta(B-R) \sim$ 0.5).  The X-ray weak quasars have a flat average photon index,  
$<\Gamma> = 1.5$ (where $\Gamma$ = -$\alpha$ + 1 for F$_{\nu} \propto \nu^{\alpha}$) and 
12 of the 16 quasars have steeper optical-to-X-ray indices than normal. 
(The optical-to-X-ray continuum is characterized by $\alpha_{ox}$ = log(F$_{2keV}$/F$_{uv}$) 
/ log($\nu_{2keV}$/$\nu_{uv})$, where F$_{2keV}$ and F$_{uv}$ are the intrinsic flux densities 
at 2 keV and 2500 $\mbox{\AA}$, respectively \citep{Tan79}.)  While X-ray absorption
remains a possibility, \citet{Risaliti03} argue that these quasars are intrinsically underluminous 
in the X-rays.  The evidence, however, is not conclusive due to the low signal-to-noise for 
many of the Chandra spectra.  

Larger surveys including high quality optical and X-ray spectra are now possible with the advent 
of the SDSS and the XMM-Newton archive of X-ray observations, both of which cover large areas 
of the sky.  SDSS data can constrain dust-reddening in the optical, while XMM-Newton observations 
constrain absorption in the X-rays.  In this paper we investigate the optical and X-ray properties 
of 17 red SDSS quasars.  \S2 introduces the data sources and sample selection and \S3 covers the 
optical and X-ray reduction and data analysis.  In \S4, we classify the quasars as dust-reddened, 
intrinsically red or unclassified, and in \S5 we discuss low accretion rates as a possible 
explanation for the steep optical/UV spectra of intrinsically red quasars.  We assume throughout 
the paper that H$_0$ = 70 km s$^{-1}$ Mpc$^{-1}$, $\Omega_M = 0.3$, and $\Omega_\Lambda = 0.7$.

\section{SAMPLE SELECTION}

\subsection{\it Data Sources: SDSS \& XMM Newton}

The third version of the SDSS quasar catalog \citep{Schnei05} is based on Data 
Release 3 (DR3), which covers an area of 5,282 deg$^2$ \citep{Aba05}.  The quasar 
catalog contains 46,240 objects from the SDSS that have: (1) 15.0 $< i <$ 22.2; (2) 
M$_i <$ -22.0; (3) at least one emission line with FWHM greater than 1000 km s$^{-1}$, 
or (4) unambiguous broad absorption line characteristics.  SDSS photometry in the 
$u$, $g$, $r$, $i$ and $z$ bands covers 3,250 - 10,000 $\mbox{\AA}$.  SDSS spectroscopy covers 
3,800 - 9,200 $\mbox{\AA}$ (so the $u$ band and part of the $z$ band lie outside the spectral 
range), with a spectral resolving power of $\sim$ 2,000.  

The SDSS DR3 quasar catalog overlaps with 1\% of the area of archival XMM-Newton observations.  
XMM-Newton imaging is carried out by the three European Photon Imaging Cameras (EPIC): 
MOS-1, MOS-2, and pn \citep{Turner01, Str01}.  The XMM-Newton telescope 
is good for retrieving serendipitous X-ray spectra of SDSS quasars due to a large 
collecting area (922 cm$^2$ for MOS and 1,227 cm$^2$ for PN at 1 keV), large field of view 
(33' x 33' for MOS and 27.5' x 27.5' for PN), and good spectral resolution 
(E/$\Delta$E $\sim$ 20 - 50 for both MOS and PN)
\footnote{http://imagine.gsfc.nasa.gov/docs/sats\_n\_data/missions/xmm.html}.  
The XMM-Newton spectral range covers the 0.5 - 12 keV band.  

The SDSS and XMM-Newton archives are well-matched in sensitivity.  A source with a 2-10 keV flux of 
10$^{-14}$ erg s$^{-1}$ cm$^{-2}$ and a standard radio quiet spectrum ($\Gamma$ $\sim$ 2.0) produces $\sim100$ 
counts in $\sim10$ ksec, sufficient for a crude spectral fit.  
For a typical value of $\alpha_{ox} = -1.5$ \citep{Vign03}, a 10$^{-14}$ ergs cm$^{-2}$ s$^{-1}$ quasar 
would have $r$ = 19.37 while an X-ray quiet quasar with $\alpha_{ox} = -2.0$ would have $r$ = 15.63.  Both 
of these magnitudes lie comfortably in the SDSS range.  

\subsection{\it The SDSS-XMM Red Quasar Sample}

We define our red sample of \citet{Schnei05} quasars lying in XMM-Newton fields by selecting the 
most extreme red colors, $g - r \geq 0.5$.  By further restricting our search to moderate redshifts (0.9 $<$ 
z $<$ 2.1), we minimize host galaxy contribution at low redshifts (z $<$ 0.5) and Ly$\alpha$ forest absorption 
at high redshifts ($z > 2.5$).  This selection results in a sample of 17 quasars.  

\begin{figure}
\centering
\plotone{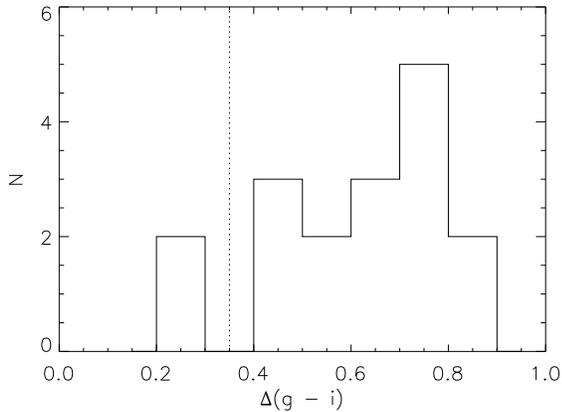}
\caption{The $\Delta(g-i)$ color distribution of the red sample.  The majority of sources in this 
sample meet the \citet{Rich03} requirement that $\Delta(g-i ) > 0.35$ for dust-reddened quasars.}
\label{fig:color dist}
\end{figure}

\citet{Rich03}, hereafter R03, defines dust-reddened quasars via their relative colors.  Relative 
colors compare a quasar's measured colors with the median colors in its redshift bin, where redshift bin 
sizes are 0.1 in redshift, so that $\Delta(g - i) = (g - i)~-~<(g - i)>_z$.  The use of relative colors 
corrects for the effect of typical emission lines on the photometry in a particular band.  The relative 
colors of the SDSS quasars match a Gaussian distribution on the blue side but require the addition of a 
tail on the red side (Fig. 3 in R03).  R03 identifies quasars belonging to the red tail as 
$\Delta(g - i) > 0.2$, since the $\Delta(g - i)$ color correlates best with the photometric spectral index.  
The $\Delta(g - i)$ color has been corrected for an SMC-like extinction curve with E(B - V) = 0.04, a value 
typical of high Galactic latitudes, for which E($g$ - $i$) = 0.07.  In practice, this yields a minimum of 
$\Delta(g - i) \sim 0.35$ for the quasars with redshift z$\sim$1.5.  The ($g$ - $i$) and ($g$ - $r$) colors 
behave similarly with redshift (Fig. 2 of R03), and the maximum median color is 0.2 for 1 $<$ z $<$ 2.  
In Figure~\ref{fig:color dist}, we show a histogram of the relative ($g$ - $i$) 
colors for our sample.  Only two have $\Delta(g - i) < 0.35$.  Thus, the $g$ - $r$ $\geq$ 0.5 color selection 
mimics $\Delta(g - i) > 0.35$, though we do not use relative colors.  

Our red quasar sample is unbiased in X-ray loudness since we do not require X-ray detections for selection.  
Nevertheless, the X-ray detection rates are high: 94\% for 2$\sigma$ detections (16/17) and 76\% for $> 
3\sigma$ detections (13/17), although most of the quasars have a relatively low X-ray signal-to-noise: 
10/17 quasars have S/N $< 10$.  

Table~\ref{table:sample} lists the sample, giving the source ID number used within this paper, the full SDSS name, 
radio classification, X-ray S/N (which is averaged if multiple observations are available), redshift, 
Galactic N$_H$, and the absolute $i$-band magnitude M$_i$.  A source is radio loud if 
F$_{5GHz}$ / F$_{4400}$ = R$_L >$ 10 \citep{Kell89}.  To calculate 
$R_L$, a power-law is interpolated between the optical magnitudes to get F$_\lambda$(4400), and the 1.4~
GHz radio flux is obtained from FIRST survey detections \citep{White97}, which are extrapolated to 5~
GHz using a radio power-law $\alpha_R$ = -0.8.  All the quasars lie in the area covered by the FIRST survey, 
so if there is no detection, we use the 5$\sigma$ upper-limit on the 1.4GHz radio flux to extrapolate to 5~GHz.

For five of 17 sources, there are two X-ray observations.  In these cases, both observations were processed 
and used in our analysis.  
\begin{table*}
\begin{center}
\caption{SDSS/XMM-Newton Red Quasar Sample}
\label{table:sample}
\begin{tabular}{llccccc}
\tableline
\tableline
Source & SDSS name & R$_L$\tablenotemark{a} & (S/N)$_x$\tablenotemark{b}  &  z  &  $N_{H,gal}$           & $M_i$\\
  ID   &           &                        &                             &     &  (10$^{20}$ cm$^{-2}$) &      \\ 
\tableline
1 & SDSS J094440.43+041055.5  & RQ                & 0.17  & 1.98  & 3.63 & -25.3\\
2 & SDSS J023217.68-073119.9  & RQ                & 3.23  & 1.16  & 3.05 & -23.6\\
3 & SDSS J165245.77+394722.3  & RQ                & 1.89  & 1.19  & 1.64 & -24.2\\
4 & SDSS J072843.03+370835.0  & RL (R$_L$ = 34.8) & 2.87  & 1.40  & 6.00 & -24.0\\
5 & SDSS J100201.50+020329.4  & RQ                & 2.85  & 2.01  & 2.59 & -25.0\\
6 & SDSS J221723.29-082139.4  & RQ                & 2.93  & 1.25  & 5.30 & -24.2\\
7 & SDSS J153322.80+324351.4  & RQ                & 4.04  & 1.90  & 2.05 & -24.8\\
8 & SDSS J122637.01+013016.1  & RQ                & 4.69  & 1.55  & 1.84 & -25.2\\
9 & SDSS J143513.89+484149.3  & RQ                & 4.46  & 1.89  & 2.07 & -24.7\\
10 & SDSS J221719.55-081234.4 & RQ                & 5.88  & 1.46  & 5.31 & -24.1\\
11 & SDSS J111452.84+531531.7 & RQ                & 6.33  & 1.21  & 0.923 & -24.1\\
12 & SDSS J113342.73+490025.8 & RQ                & 8.65  & 1.30  & 1.59 & -24.6\\
13 & SDSS J125456.82+564941.4 & (R$_L$ $<$18.3)    & 14.43 & 1.27  & 1.26 & -22.8\\
14 & SDSS J095918.70+020951.4 & RL (R$_L$ = 36.3)  & 21.87 & 1.16  & 2.64 & -23.4\\
15 & SDSS J095857.34+021314.4 & RQ                & 27.94 & 1.02  & 2.64 & -23.0\\
16 & SDSS J105316.76+573550.7 & RQ                & 35.21 & 1.21  & 0.556 & -23.9\\
17 & SDSS J091301.01+525928.9 & RQ (R$_L$ = 4.9)  & 43.48 & 1.38  & 1.60 & -26.9\\
\tableline
\end{tabular}
\end{center}
\tablenotetext{a}{The radio loud parameter R$_L$ is defined as F$_{5GHz}$/F$_{4400}$, where R$_L>$10 indicates 
radio-loudness \citep{Kell89}.}

\tablenotetext{b}{When more than one X-ray observation exists for a single source, the S/N listed is the rms of 
the observations.} 
\end{table*}    

\subsection{\it SDSS Data} 

SDSS photometry was extracted from the online database\footnote{http://www.sdss.org/dr3/access/index.html} 
via SQL-based queries in CasJobs.  The individual, calibrated SDSS spectra were also downloaded 
from the SDSS database.  Table~\ref{table:SDSS sample} gives the source ID, the abbreviated SDSS 
name, the $g - r$ color (which determined selection), the apparent $i$ band magnitude, and the
full-width half-maximum (FWHM [km s$^{-1}$]) of the broad MgII component.  The MgII information 
is taken from \citet{Shen07} for all but three objects.  \citet{Shen07} simultaneously fit the 
FeII emission, a continuum power-law, and broad and narrow Gaussians to the MgII line.  
The spectral signal-to-noise is too low for two objects (\#13 and \#15) to use this method.  
Another object (\#17) has good S/N, but line values are not given in \citet{Shen07}.  
We fit the MgII lines for these three sources by estimating the continuum values by eye and then 
simultaneously fitting broad and narrow Gaussians with IRAF\footnote{http://iraf.noao.edu/}.  
\begin{table}
\begin{center}
\caption{Red Quasar Sample in the SDSS}
\label{table:SDSS sample}
\begin{tabular}{llccccccccc}
\tableline
\tableline
ID & SDSS name & g - r   & i & MgII FWHM\\
   &           &         &   & \footnotesize{(km s$^{-1}$)}\\
\tableline
1  & SDSS J0944+0410                  & 0.50 & 18.22 &  8,800\\
2  & SDSS J0232-0731                  & 0.53 & 19.25 &  11,700\\
3  & SDSS J1652+3947                  & 0.54 & 18.65 &  8,800\\
4  & SDSS J0728+3708                  & 0.74 & 19.09 &  9,700\\
5  & SDSS J1002+0203                  & 0.55 & 18.58 &  10,800\\
6  & SDSS J2217-0821                  & 0.74 & 18.69 &  7,600\\
7  & SDSS J1533+3243                  & 0.97 & 18.69 &  6,700\\
8  & SDSS J1226+0130                  & 0.59 & 18.01 &  7,700\\
9  & SDSS J1435+4841                  & 0.50 & 18.78 &  7,000\\
10 & SDSS J2217-0812                  & 0.62 & 19.06 &  5,800\\
11 & SDSS J1114+5315                  & 0.76 & 18.77 &  6,400\\
12 & SDSS J1133+4900                  & 0.52 & 18.42 &  9,100\\
13 & SDSS J1254+5649\tablenotemark{a} & 0.57 & 20.12 &  8,200\\
14 & SDSS J0959+0209                  & 0.60 & 19.40 &  23,200\\
15 & SDSS J0958+0213\tablenotemark{a} & 0.59 & 19.66 &  10,500\\
16 & SDSS J1053+5735                  & 0.55 & 18.95 &  5,400\\
17 & SDSS J0913+5259\tablenotemark{a} & 0.64 & 16.20 &  6,000\\
\tableline
\end{tabular}
\end{center}
\tablecomments{FWHM measurements for the MgII broad emission lines are from \citet{Shen07}.}

\tablenotetext{a}{The line parameters for these sources are not included in \citet{Shen07}.  Instead, broad and narrow 
Gaussians were fit simultaneously to the line, with the continuum estimated by eye.  The FWHM for 
broad Gaussian component are listed.}
\end{table}  
 
There is one broad-absorption line quasar (BAL) in the sample, SDSS J0944+0410 (\#1), and two 
narrow absorption-line (NAL) quasars, SDSS J1226+0130 (\#8) and SDSS J1435+4851 (\#9).  Only 
five of the 17 quasars have the CIV line, which is the most reliable marker of BAL/NAL presence, 
visible in their spectra.  MgII, the BEL seen in all 17 quasars, rarely shows either BALs or 
NALs.  So the true number of BALs and NALs in the sample could be even greater.  

\subsection{\it X-ray Data Reduction} 

Using the XMM-Newton Science Analysis System, SAS v7.02\footnote{http://xmm.esac.esa.int/sas}, 
we filtered the XMM observations to 
remove time intervals of flaring high-energy background events using the standard cut-off 
of 0.35 cts s$^{-1}$ for the MOS cameras and 1.0 ct s$^{-1}$ for the PN camera.  The event rate 
plots were visually examined to determine that these cut-offs were appropriate.  We then 
extracted source and background regions for spectral analysis.  The SAS task \emph{eregionanalyse} 
was used to select the source radius with the optimal signal-to-noise, with typical radii 
ranging from 10'' to 60''.  Background regions were defined by eye, avoiding obvious X-ray 
sources and chip edges.  These regions were typically a circle of radius 2000-2500 pixels 
(100-125''), selected to lie on the same chip as the source and as close to the source as 
possible without overlapping the source extraction region.  

Where possible, observations were retrieved for all three XMM EPIC CCDs.  In seven observations, 
the source lies in a bad region in one of the cameras, either in a strip between two chips 
or, because the MOS and PN cameras have different shapes, the source may lie outside the 
field of view in one of the cameras.  In an eighth case, the source is within the field of 
view of all three cameras, but in the MOS-2 camera, most of the events 
are filtered out while removing the flaring high-energy background.  In all these cases 
(7 of 22 observations), we use the remaining images from the other cameras for analysis.  

Table~\ref{table:XMM sample} contains data for the XMM observations, ordered by increasing S/N.  
The table includes the source ID, XMM identification number, observation date, net exposure time, 
net counts, X-ray S/N and the cameras available for analysis.  

\begin{table*}
\begin{center}
\caption{XMM-Newton Observation Log}
\label{table:XMM sample}
\begin{tabular}{llcccccl}
\tableline
\tableline
Source & XMM ID & Date & Net Exp Time\tablenotemark{a}  &  Net Counts\tablenotemark{a}  &  S/N  &  Cameras\\
 ID    &        &      &   (ks)            &   (counts)       &       &  PN M1 M2\\
\tableline
1  & 0201290301  & 2005-6-09   & 56.7   & 0.03   & 0.2  & {\it--}  X  X\\
2  & 0200730401  & 2005-1-30   & 122.8  & 40.3   & 3.2  & X X X\\
3a & 0113060401  & 2002-07-14  & 4.55   & 7.3    & 1.2  & {\it--}  X  X\\
3b & 0113060201  & 2003-9-03   & 5.0    & 19.2   & 2.6  & {\it--}  X  X\\
4  & 0145450501  & 2003-12-04  & 9.2    & 25.1   & 2.9  &  X  X  X\\
5a & 0203360401  & 2005-12-10  & 29.6   & 39.0   & 2.8  &  X  X  X\\
5b & 0203360801  & 2005-12-07  & 89.4   & 11.8   & 2.9  &  X  X {\it--}\\
6  & 0009650201  & 2002-9-20   & 40.2   & 31.2   & 2.9  & {\it--}  X  X\\
7  & 0039140101  & 2003-8-29   & 25.3   & 70.8   & 4.0  &  X  X  X\\
8  & 0110990201  & 2002-8-11   & 32.3   & 70.3   & 4.7  &  X  X  X\\
9a & 0110930401  & 2004-2-04   & 21.1   & 77.5   & 4.2  &  X  X  X\\
9b & 0110930901  & 2004-2-04   & 17.7   & 98.3   & 4.7  &  X  X  X\\
10 & 0009650201  & 2002-9-20   & 40.2   & 84.5   & 5.9  & {\it--}  X  X\\
11 & 0143650901  & 2004-5-20   & 17.1   & 135.3  & 6.3  & X X X\\
12 & 0149900201  & 2004-12-15  & 36.0   & 196.3  & 8.7  & {\it--}  X  X\\
13 & 0081340201  & 2002-11-02  & 60.1   & 750.0  & 14.4 &  X  X  X\\
14 & 0203361801  & 2005-1-16   & 79.3   & 1486.8 & 21.9 &  X  X  X\\
15 & 0203361801  & 2005-1-16   & 79.    & 2696.2 & 27.9 &  X  X  X\\
16a & 0147511701  & 2004-2-01   & 282.4  & 4185.3 & 33.7 &  X  X  X\\
16b & 0147511801  & 2004-2-01   & 282.2  & 4775.3 & 36.8 &  X  X  X\\
17a & 0143150301  & 2004-9-06   & 20.8   & 3687.3 & 33.6 &  X  X  X\\
17b & 0143150601  & 2004-9-06   & 47.9   & 9130.1 & 53.4 &  X  X  X\\
\tableline
\end{tabular}
\end{center}
\tablenotetext{a}{Net exposure time and net counts include information from all available observations.  
Net counts are taken from 0.5-10 keV.}
\end{table*}  

\section{ANALYSIS}

\subsection{\it Optical Spectra and Continuum Dust Reddening Fits}

We wrote an IDL procedure to fit dust-reddening to the optical continuum.  The procedure 
calls a de-reddening routine, FM\_UNRED.PRO (Landsman 1999), which uses the Fitzpatrick 
(1999) parameterization to characterize the optical and UV extinction curve.  

Each source spectrum is first shifted to the rest-frame.  A composite quasar spectrum 
created from SDSS quasars \citep{VdB01} is then reddened incrementally with the 
SMC extinction curve \citep{Gordon03} and normalized to the red quasar spectrum 
by averaging the continuum between 3020 and 3066 $\mbox{\AA}$ \citep{VdB01} in order 
to calculate the $\chi^2$ value.  The best-fit parameters are those for which $\chi^2$ 
reached a minimum.  The total $\chi^2$ was allowed to vary by 2.7 to get the 90\% confidence 
interval errors. 

The SMC, rather than Galactic, extinction curve is used because of the
absence of the $\lambda$2175 feature in AGN spectra \citep{Pitman00, Kuhn01, Czerny04, Hopkins04}.  
The optical dust-reddening (hereafter denoted as E(B-V)$_{spec}$) 
is allowed to vary from E(B-V)$_{spec}$ = 0.0 to E(B-V)$_{spec}$ = 1.0.  Simulations have 
shown that quasars with E(B-V) $>$ 0.5 are not expected from the SDSS selection process 
\citep{Rich03}, so the fits to the spectral continuum cover the full E(B-V) parameter space.  

The Small Blue Bump, a blend of Balmer continuum emission and FeII lines, extends from 
$\sim$ 2000 - 4000~$\mbox{\AA}$ \citep{WNW85}.  Our color selection criteria 
($g - r \geq$ 0.5) bias our sample towards sources with atypically large FeII emission between 
$\sim$ 2200 - 2700~$\mbox{\AA}$.  Five of 17 sources have atypical FeII emission.  Excluding the region 
from 2200 - 3000~$\mbox{\AA}$ from the chi-square calculation for all the sources accounts for both 
FeII emission and the 2798~$\mbox{\AA}$ MgII line, so that the reddening fit to the continuum is not 
biased by unusually strong line emission.  The reddening fit was not sensitive to other 
broad emission lines, such as the 1549~$\mbox{\AA}$ CIV line, so we did not omit any additional lines 
from the fit.  The BAL quasar (SDSS J0944+0410, \#1) was not fit with reddening because numerous absorption 
lines and a strong Fe II bump left no reliable continuum points with which to calculate $\chi^2$.  

We also fit a power-law (F$_{\nu}$ $\propto \nu^{\alpha_{opt}}$) to the optical continuum.  
All major emission lines are removed before performing the fit, as is the region dominated 
by the blended FeII and Balmer emission.  A power-law will not be a good fit to spectral 
continua with obvious dust curvature, but for (E(B-V)$_{spec}$ $<$ 0.1), dust-reddening 
will look similar to a red power-law.  Section $\S$3.2 discusses this issue in more detail.
\begin{table}
\begin{center}
\caption{Optical Reddening Fits}
\label{table:E(B-V)}
\begin{tabular}{lcccc}
\tableline
\tableline
ID & E(B-V)$_{spec}$ & $N_{H,opt}$\tablenotemark{a}         & $N_{H,x}$\tablenotemark{b}           & $\alpha_{opt}$\tablenotemark{c}\\
   &                 & \footnotesize{(10$^{20}$ cm$^{-2}$)} & \footnotesize{(10$^{20}$ cm$^{-2}$)} & \\
\tableline
1 & $-^d$ & {\it-} & {\it-} & -1.49\\
2 & 0.03$\pm$0.02 & 1.7$\pm$1.2 & {\it-} & -0.92\\
3 & 0.08$\pm$0.02 & 4.6$\pm$1.2 & {\it-} & -1.05\\
4 & 0.25$\pm$0.02 & 14.5$\pm$1.2 & {\it-} & -2.92\\
5 & 0.052$\pm$0.005 & 3.0$\pm$0.3 & {\it-} & -1.36\\
6 & 0.17$\pm$0.02 & 9.9$\pm$1.2 & {\it-} & -1.95\\
7 & 0.12$\pm$0.02 & 7.0$\pm$1.2 & $\leq$696 & -1.85\\
8 & 0.15$\pm$0.01 & 8.7$\pm$0.6 & $\leq$1340 & -1.79\\
9 & 0.13$\pm$0.01 & 7.5$\pm$0.6 & $\leq$677 & -2.07\\
10 & 0.11$\pm$0.02 & 6.4$\pm$1.2 & $\leq$184 & -1.68\\
11 & 0.18$\pm$0.03 & 10.4$\pm$1.7 & $\leq$89.0 & -1.46\\
12 & 0.04$\pm$0.01 & 2.3$\pm$0.6 & $\leq$76.3 & -0.87\\
13 & 0.10$^{+0.11}_{-0.08}$ & 5.8$^{+6.4}_{-4.6}$ & $\leq$23.1 & -1.44\\
14 & 0.12$\pm$0.03 & 7.0$\pm$1.7 & $\leq$5.49 & -1.05\\
15 & 0.16$^{+0.09}_{-0.07}$ & 9.3$^{+5.2}_{-4.1}$ & $\leq$5.39 & -1.94\\
16 & 0.11$\pm$0.02 & 6.4$\pm$1.2 & $\leq$4.7 & -1.17\\
17 & 0.086$^{+0.004}_{-0.003}$ & 5.00$^{+0.23}_{-0.17}$ & $\leq$2.9 & -1.31\\
\tableline
\end{tabular}
\end{center}
\tablenotetext{a}{Gas column predicted from E(B-V)$_{spec}$ using the Galactic gas-to-dust ratio 
\citep{Bohlin78,Kent91}.}

\tablenotetext{b}{Gas column upper-limit measured via X-ray spectral fits (model D).}

\tablenotetext{c}{The optical slope as fit to the spectral continuum with emission lines removed.}

\tablenotetext{d}{The BAL quasar was not fit with E(B-V)$_{spec}$ because numerous absorption lines made it 
impractical to fit using chi-square minimization.}
\end{table}

\begin{figure}
\centering
\includegraphics[width=3.25in]{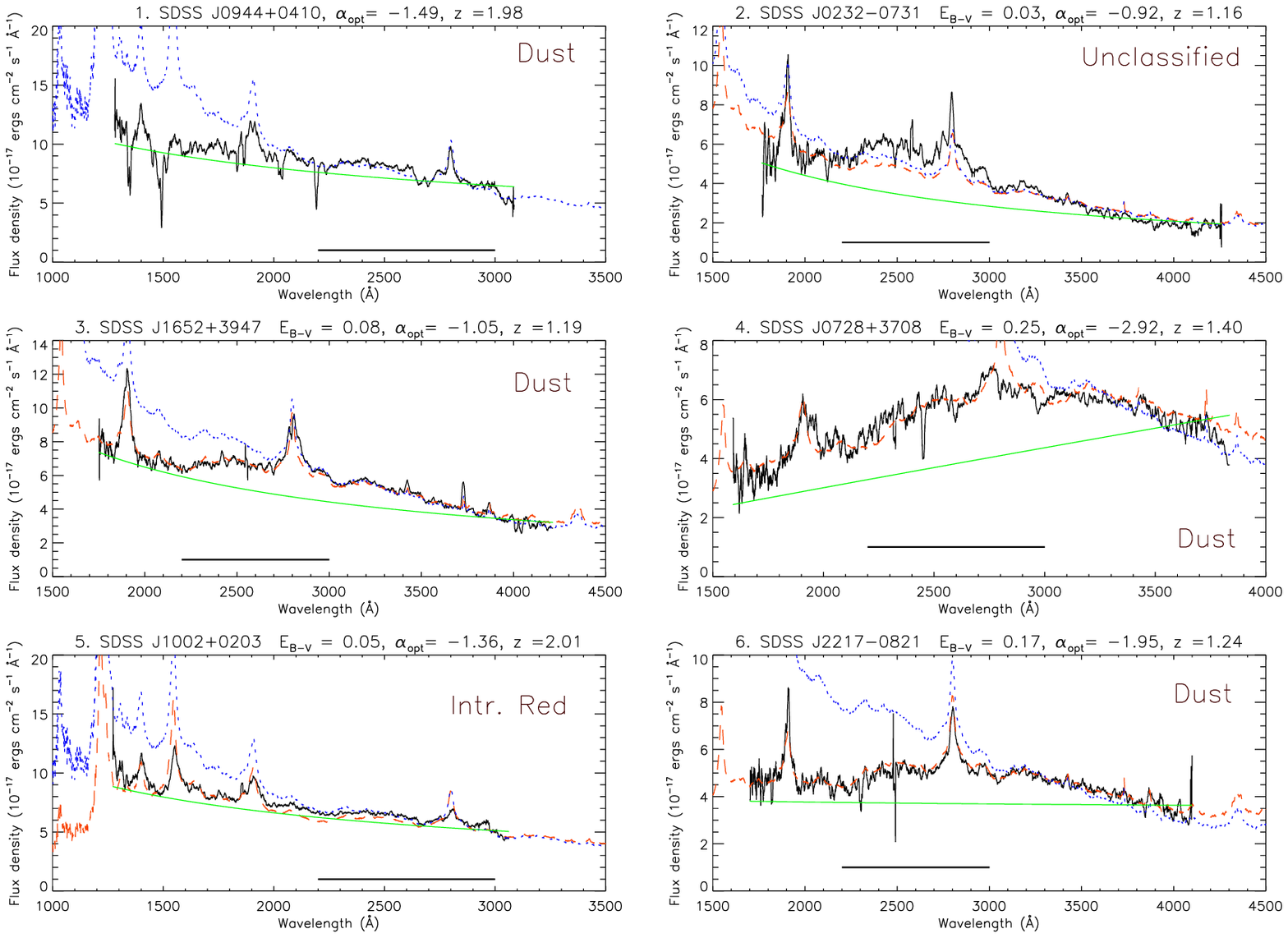}
\includegraphics[width=3.25in]{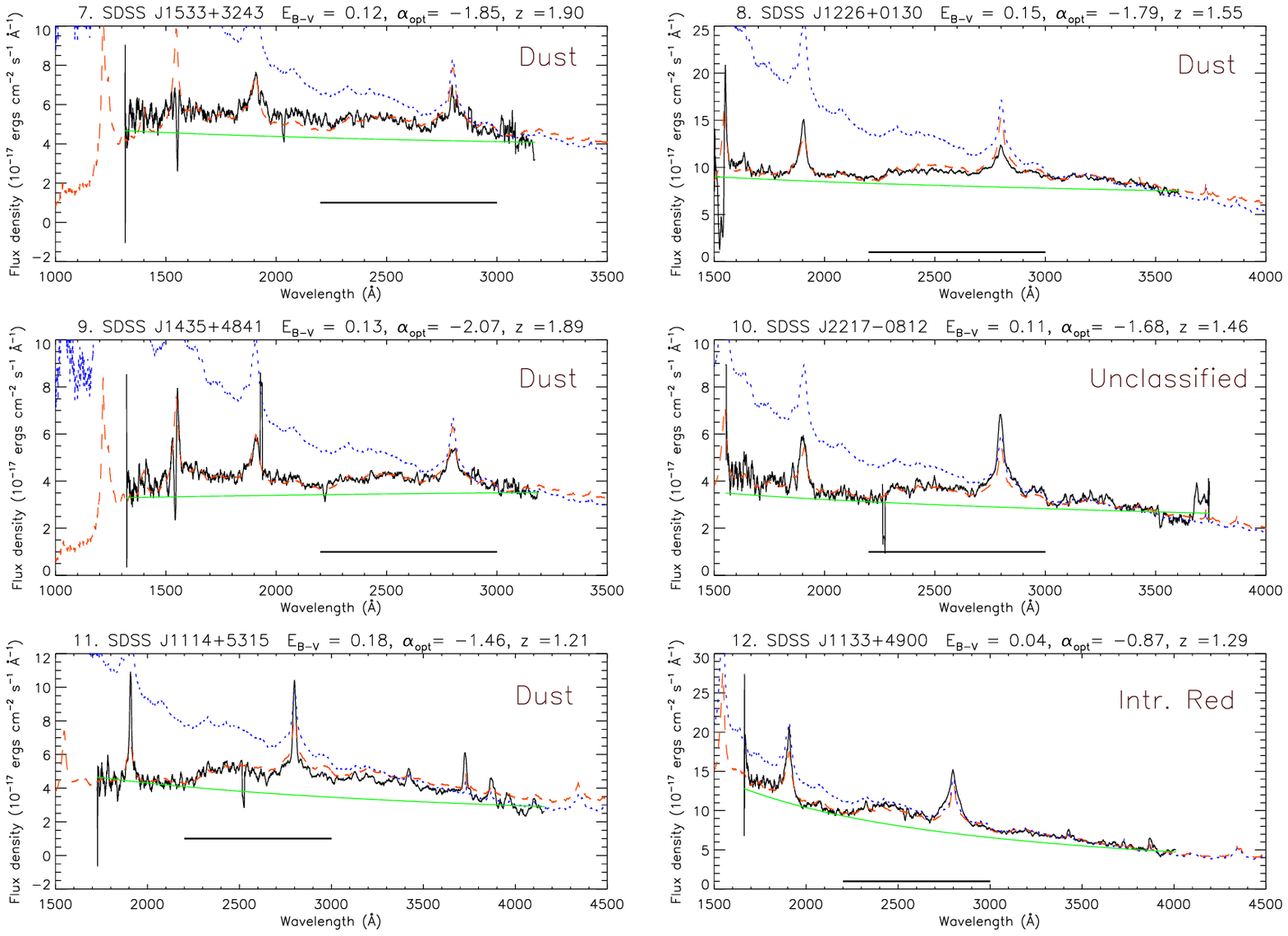}
\includegraphics[width=3.25in]{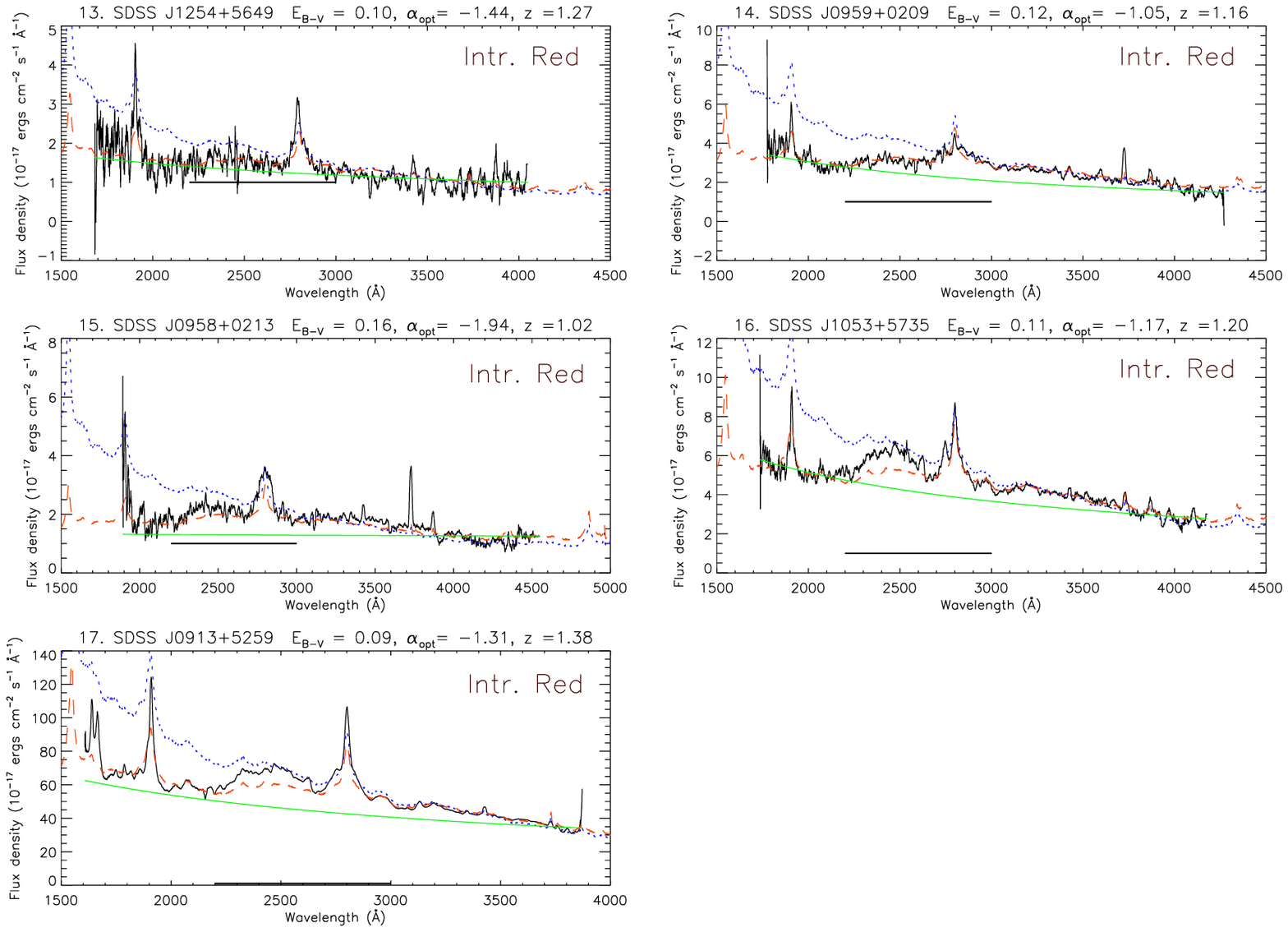}
\caption{The SDSS source spectra are plotted as solid black lines, F$_\lambda$ (10$^{-17}$ 
ergs cm$^{-2}$ s$^{-1}$ $\mbox{\AA}$) vs. $\lambda$ ($\mbox{\AA}$). The original \citet{VdB01} 
template (dotted, blue line) and the reddened template (dashed, red line) are overplotted.  
Note that the BAL quasar (SDSS J0944+0410, \#1) had no reliable continuum to fit reddening and so 
only the original template is plotted.  The solid line beneath the spectrum marks the FeII+MgII 
region ignored by the continuum-fitting program as described in the text.  A power-law fit to the
spectrum, omitting major emission lines, is also shown (solid, green line).}
\label{fig:SDSS spectra}
\end{figure}

Figure~\ref{fig:SDSS spectra} plots the SDSS source spectra in F$_\lambda$ (10$^{-17}$ ergs 
cm$^{-2}$ s$^{-1}$ $\mbox{\AA}^{-1}$) vs. $\lambda$ ($\mbox{\AA}$) space.  Overplotted are 
the original \citet{VdB01} template (dotted blue line), the reddened template (dashed red line), 
and the optical power-law (solid green line).  Table~\ref{table:E(B-V)} gives the reddening 
results: E(B-V)$_{spec}$, the expected intrinsic gas column density (N$_{H,opt}$) obtained by 
applying the Galactic gas-to-dust ratio \citep{Bohlin78, Kent91}, the intrinsic gas column 
density from the X-ray fits (N$_{H,x}$, see \S3.3), and the optical power-law index, $\alpha_{opt}$.  

Reddening fits to the optical spectra require E(B-V)$_{spec}$ in 14 out of 17 sources at$>3\sigma$.  
For these 14 sources, E(B-V)$_{spec}$ ranges from 0.04 to 0.25, with a weighted mean of 
$<E(B-V)_{spec}>$ = 0.1 corresponding to $<N_{H,opt}>$ = 5.8 x 10$^{20}$ cm$^{-2}$, using the Galactic 
gas-to-dust ratio.  

\subsection{\it Testing the Optical Continuum for Weak Curvature with Relative Colors}

We find that reddening is a good fit to most of the optical spectra.  However, for the  
redshifts of this sample (z $>$ 1), the optical continuum will not  have strong dust-induced 
curvature for mild reddening (E(B-V)$_{spec} \leq$ 0.1).  Under these conditions, a 
dust-reddened optical continuum will look similar to a red power-law.  Figure~\ref{fig:SDSS spectra} in this 
paper shows the similarity between mild dust-reddening (dashed red line) and a red power-law 
(solid green line).  The curvature in the relative broad-band photometry can help 
differentiate these two possibilities because the \emph{u} band (for which the FWHM ranges from 3251 
to 3851~$\mbox{\AA}$) is mostly outside the spectral range (3800 - 9200~$\mbox{\AA}$).  The \emph{u} band is a factor 
3.6 more absorbed by dust (for E(B-V) = 0.1) than the $r$ band, where the spectra are centered.

Relative colors give an idea of the shape of the continuum \citep{Hall06}.  A quasar 
with a typical $g - i$ color, for example, will have $\Delta(g-i) = 0$, while a quasar with 
red ($g - i$) color will have $\Delta(g-i) > 0$.  A dust-reddened quasar with typical 
emission lines will have $\Delta(u-r) > \Delta(g-i) > \Delta(r-z)$, because 
curvature increases on the blue side of the continuum.  A typical blue power-law 
continuum, on the other hand, will have $\Delta(u-r)$ $\sim$ $\Delta(g-i)$ $\sim$ 
$\Delta(r-z)$.  In the case of a 
power-law continuum that is redder than average, the relative colors will go as 
$\Delta(u-r) < \Delta(g-i) > \Delta(r-z)$.  Therefore, an object with $\Delta(u-r) - 
\Delta(g-i) < 0$ displays dust-reddened curvature, while an object with $\Delta(u-r) - 
\Delta(g-i) > 0$ displays a red power-law.  We use this criterion to classify quasars as 
intrinsically red, as shown in Figure~\ref{fig:NH/E(B-V)-color} (discussed in \S4.2).  

Since relative colors compare the observed quasar colors to the mean colors for quasars 
at that redshift, atypical emission lines will affect the relative color results.  
For the redshift range 1 $<$ z $<$ 2, the most significant lines are MgII and the FeII 
emission line blend.  To correct for atypical MgII lines, we calculate the equivalent width 
of MgII from the spectrum.  From this, we subtract the average MgII equivalent width 
obtained from the Vanden Berk composite spectrum (2001)\nocite{VdB01}.  The residual MgII line flux 
is calculated and added or subtracted from the band where MgII is found.  To correct for 
the FeII emission line blend, we redden the composite spectrum, shift it to 
the observed frame, and normalize to the source spectrum.  We then subtract this spectrum 
from the source for the range 2200-2700~$\mbox{\AA}$ and calculate the equivalent width of the 
residual FeII line.  We add or subtract the residual flux from the bands where FeII is 
found.  Since the \emph{u} band is outside of the spectral range, we cannot correct for 
atypical emission lines in this band.  The relative colors and the optical continuum 
shapes are listed in Table~\ref{table:relcolors}.  
\begin{table*}
\begin{center}
\caption{Relative Colors and the Optical Continuum Shape}
\label{table:relcolors}
\begin{tabular}{lcccc}
\tableline
\tableline
ID & $\Delta$(u-r) & $\Delta$(g-i) & $\Delta$(r-z) & Continuum Shape\\
\tableline
1 & 0.97 $\pm$0.06 & 0.68 $\pm$0.06 & 0.13 $\pm$0.04 & Dust\\
2 & 0.96 $\pm$0.07 & 0.25 $\pm$0.04 & 0.03 $\pm$0.07 & Dust\\
3 & 0.79 $\pm$0.04 & 0.41 $\pm$0.03 & 0.29 $\pm$0.04 & Dust\\
4 & 1.26 $\pm$0.08 & 0.81 $\pm$0.03 & 0.65 $\pm$0.04 & Dust\\
5 & 0.63 $\pm$0.04 & 0.71 $\pm$0.04 & 0.16 $\pm$0.04 & Red power-law ($\alpha_{opt}$ = -1.36)\\
6 & 1.06 $\pm$0.07 & 0.83 $\pm$0.03 & 0.52 $\pm$0.05 & Dust\\
7 & 1.10 $\pm$0.06 & 0.69 $\pm$0.02 & 0.23 $\pm$0.04 & Dust\\
8 & 1.13 $\pm$0.06 & 0.74 $\pm$0.04 & 0.58 $\pm$0.04 & Dust\\
9 & 0.99 $\pm$007 & 0.77 $\pm$0.03 & 0.56 $\pm$0.04 & Dust\\
10 & 0.77 $\pm$0.07 & 0.74 $\pm$0.03 & 0.41 $\pm$0.06 & Undefined\\
11 & 0.80 $\pm$0.07 & 0.74 $\pm$0.05 & 0.47 $\pm$0.04 & Undefined\\
12 & 0.14 $\pm$0.04 & 0.24 $\pm$0.03 & 0.06 $\pm$0.04 & Red power-law ($\alpha_{opt}$ = -0.87)\\
13 & -0.07 $\pm$0.07 & 0.54 $\pm$0.05 & 0.56 $\pm$0.1 & Red power-law ($\alpha_{opt}$ = -1.44)\\
14 & 0.60 $\pm$0.07 & 0.54 $\pm$0.03 & 0.38 $\pm$0.08 & Undefined\\
15 & 0.54 $\pm$0.08 & 0.63 $\pm$0.04 & 0.40 $\pm$0.09 & Red power-law ($\alpha_{opt}$ = -1.94)\\
16 & 0.33 $\pm$0.04 & 0.44 $\pm$0.03 & 0.26 $\pm$0.04 & Red power-law ($\alpha_{opt}$ = -1.17)\\
17 & 0.25 $\pm$0.02 & 0.40 $\pm$0.02 & 0.20 $\pm$0.03 & Red power-law ($\alpha_{opt}$ = -1.31)\\
\tableline
\end{tabular}
\end{center}
\tablecomments{The relation between the relative colors gives the optical continuum shape from the photometry.
$\Delta$(u - r) $>$ $\Delta$(g - i)$>$ $\Delta$(r - z) indicates dust, but
$\Delta$(u - r)$<$ $\Delta$(g - i)$>$ $\Delta$(r - z) indicates a red power-law.}
\end{table*}

\subsection{\it X-ray Spectral Fits}

We made fits to the extracted spectra using the \emph{Sherpa} 
package\footnote{http://cxc.harvard.edu/sherpa/threads /index.html} within 
CIAO\footnote{http://cxc.harvard.edu/ciao/}.  For each source, the available MOS+PN spectra were 
fit simultaneously by linking parameters.  We used a conservative spectral range of 0.5 - 10 keV 
for fitting.  The observations were fit according to their S/N, with more complicated models 
(models A-E) being applied as S/N increased.  Table~\ref{table:models} summarizes the model characteristics.  
All models applied in this paper include local absorption fixed to the Galactic hydrogen 
column density (N$_{H,gal}$) for a source's coordinates.  Values for N$_{H,gal}$ were taken from 
WebPIMMS\footnote{http://heasarc.gsfc.nasa.gov/Tools/w3pimms.html}.
\begin{table*}
\begin{center}
\caption{S/N-dependent Models Applied to XMM observations}
\label{table:models}
\begin{tabular}{cccccc}
\tableline
\tableline
     &  Model A  &  Model B  &  Model C  &  Model D  &  Model E\\
\tableline
S/N  &  S/N $<$ 4  & 2 $<$ S/N $<$ 4  & S/N $>$ 4  &  S/N $>$ 4  &  S/N $>$ 25\\
\tableline
Model & Cash fit & Cash fit  & $\chi^2$ fit & $\chi^2$ fit & $\chi^2$ fit\\
Description & Fixed power-law  & Free power-law & Free power-law & Free power-law & Free power-law\\
     & ($\Gamma$ = 1.9) &  &  &  & \\
     & Gal. abs. & Gal. abs. & Gal. abs. & Gal. abs. & Gal. abs.\\
     & (fixed) & (fixed) & (fixed) & (fixed) & (fixed)\\
     &  &  &  & Intrinsic & Intrinsic \\
     &  &  &  & abs. (free) & abs. (free) \\
     &  &  &  &            & Fe K$\alpha$ line \\
\tableline
\end{tabular}
\end{center}
\tablecomments{A total of 22 observations cover 17 sources.}
\end{table*}

For six observations with S/N $<$ 4, we fit Model A, where we freeze the photon index of the X-ray 
power-law to $\Gamma$ = 1.9.  
We leave normalization free to vary in order to get the flux or a 90\% upper-limit on the flux.  
For five observations with 2 $<$ S/N $<$ 4, we fit Model B, a single power-law with free $\Gamma$ and 
normalization.  Models A and B use the Cash statistic for fitting, which we discuss further below.  

For 14 observations with $S/N > 4$, we fit Model C, a single power-law, but using Chi-square 
statistics rather than Cash.  For the same observations, we also fit the spectra with Model D, 
a power-law with intrinsic absorption.  Finally, for five observations with $S/N > 25$, we fit 
model E, a power-law with intrinsic absorption, plus a Gaussian for the Fe K$\alpha$ line.  The 
Gaussian line energy is set to 6.4 keV shifted to the quasar's reference frame.  
For sources with $S/N > 4$ (where models C, D and E are applied), the Chi-square statistic is 
used and the data are binned by at least 15 counts/ bin.

For S/N $<$ 4, Chi-square is not an appropriate statistic because there are not enough counts 
per bin, so both models A and B are fit using the more time-consuming Cash statistic (1979) 
which gives more reliable results for low-count sources \citep{NS89}.  When using 
Cash statistics, the source and background counts are binned by one count/bin, to ensure that 
there are no empty bins.  The background is not subtracted and is instead fit simultaneously 
with the source.  The XMM background is fit with the three components described in the XMM 
Users’ Handbook\footnote{http://xmm.vilspa.esa.es/external/xmm\_user\_support/documentation}: 
a power-law fixed to $\Gamma = 1.4$ plus Galactic absorption (to account 
for the extragalactic X-ray spectrum), a broken power-law, with the break energy fixed to 
3.2 keV (to account for the quiescent soft proton spectrum), and lines at 1.5 keV for the 
MOS cameras and at 1.5 and 8 keV for the PN camera (to account for cosmic-ray interactions 
with the detector).  Since this background model has five free parameters for each camera, 
constraining all fit parameters results in large errors in the best-fit parameters.  
\begin{table*}
\begin{center}
\caption{Best-fit X-ray spectral fits}
\label{table:Best fits}
\begin{tabular}{llcccccc}
\tableline
\tableline
Source & Model & XMM ID & $\Gamma$ & F$_{0.5-2keV}$\tablenotemark{a} & F$_{2-10keV}$\tablenotemark{a} & $L_{2-10keV}$\tablenotemark{b} & $\chi^2/\nu$\tablenotemark{c}\\
 ID    &       &        &          &                                 &                                &                                &                               \\
\tableline
1  & A  & 0201290301 & 1.9\tablenotemark{d}  & $\leq$0.024 & $\leq$0.034 & $\leq$9.5 x 10$^{41}$ & 1791/1599\\
2  & A  & 0200730401 & 1.9  & 0.24 & 0.36 & 5.2 x 10$^{42}$ & 8904/5766\\
3a & A  & 0113060401 & 1.9  & $\leq$0.51 & $\leq$3.72 & $\leq$3.6 x 10$^{43}$ & 1225/1603\\
3b & A  & 0113060201 & 1.9  & 2.25 & 3.10 & 4.8 x 10$^{43}$ & 1018/1614\\
4  & A  & 0145450501 & 1.9  & 0.016 & 13.4 & 5.2 x 10$^{43}$ & 1608/1619\\
5a & A  & 0203360401 & 1.9  & 0.32 & 0.46 & 1.3 x 10$^{43}$ & 17119/5749\\
5b & A  & 0203360801 & 1.9  & 0.53 & 0.74 & 2.1 x 10$^{43}$ & 10447/4946\\
6  & A  & 0009650201 & 1.9  & 0.52 & 2.96 & 3.2 x 10$^{43}$ & 1831/1636\\
7  & C  & 0039140101 & 1.8$^{+0.5}_{-0.4}$ & 1.95 & 3.05 & 7.7 x 10$^{43}$ & 4.3/6\\
8  & C  & 0110990201 & 1.0$^{+0.4}_{-0.4}$ & 1.36 & 7.49 & 7.0 x 10$^{43}$ & 9.7/7\\
9a & C  & 0110930401 & 1.4$^{+0.4}_{-0.4}$ & 2.78 & 7.96 & 1.3 x 10$^{44}$ & 2.6/5\\
9b & C  & 0110930901 & 1.0$^{+0.4}_{-0.4}$ & 4.57 & 27.2 & 2.7 x 10$^{44}$ & 3.7/7\\
10 & C  & 0009650201 & 1.9$^{+0.4}_{-0.3}$ & 3.05 & 4.90 & 9.4 x 10$^{43}$ & 2.6/6\\
11 & C  & 0143650901 & 1.5$^{+0.3}_{-0.3}$ & 5.03 & 13.3 & 1.5 x 10$^{44}$ & 5.9/9\\
12 & C  & 0149900201 & 1.4$^{+0.5}_{-0.4}$ & 8.58 & 16.2 & 3.0 x 10$^{44}$ & 8.1/15\\
13 & C  & 0081340201 & 1.6$^{+0.1}_{-0.1}$ & 9.59 & 20.7 & 2.8 x 10$^{44}$ & 27/40\\
14 & C  & 0203361801 & 1.75$^{+0.09}_{-0.09}$ & 6.52 & 11.9 & 1.6 x 10$^{44}$ & 54/63\\
15 & C  & 0203361801 & 1.78$^{+0.07}_{-0.06}$ & 17.0 & 29.5 & 3.4 x 10$^{44}$ & 75/107\\
16a & C & 0147511701 & 1.77$^{+0.06}_{-0.05}$ & 12.1 & 20.1 & 2.9 x 10$^{44}$ & 100/165\\
16b & C & 0147511801 & 1.72$^{+0.06}_{-0.05}$ & 13.2 & 23.5 & 3.2 x 10$^{44}$ & 155/189\\
17a & C & 0143150301 & 1.68$^{+0.05}_{-0.06}$ & 60.6 & 120 & 1.8 x 10$^{45}$ & 94/142\\
17b & C & 0143150601 & 1.72$^{+0.04}_{-0.03}$ & 57.9 & 107 & 1.7 x 10$^{45}$ & 203/311\\
\tableline
\end{tabular}
\end{center}

\tablenotetext{a}{Fluxes are determined from the MOS-1 observations because it has the best (most well-known) 
calibration. Units are 10$^{-14}$ ergs cm$^{-2}$ s$^{-1}$.}

\tablenotetext{b}{ Luminosities are in rest-frame in units of ergs s$^{-1}$.}

\tablenotetext{c}{This column contains either the chi-square value and the degrees of freedom when chi-square statistics 
are applied (model C) or the likelihood of fit and the degrees of freedom when the Cash statistic is applied (model A).}

\tablenotetext{d}{$\Gamma$ is fixed to 1.9 for model A.}

\end{table*}

Table~\ref{table:Best fits} lists the source ID, the model used for each source, XMM observation ID, the 
best-fit spectral slopes, flux values from 0.5-2 keV and 2-10 keV, the rest-frame 2-10 keV luminosity, 
and the $\chi^2$ or Likelihood of Fit values and degrees of freedom for the best fit.  For 
sources with S/N$<$4, Table~\ref{table:Best fits} lists the best-fit parameters for Model A, where the spectral 
slope is fixed to 1.9, leaving the normalization free to vary.  (For sources with S/N $<$ 2, 
the best-fit upper-limits are listed.)  We list the best-fit spectral slopes from Model B in 
Table~\ref{table:abs fits} for reference, but due to the large spectral slope errors we do not use these values 
for further analysis.
\begin{table*}
\begin{center}
\caption{Alternative X-ray Spectral Fits: $\Gamma$ (S/N $<$ 4)and N$_{H,x}$ (S/N $>$ 4)}
\label{table:abs fits}
\begin{tabular}{llcccc}
\tableline
\tableline
Model & Source & XMM ID & $\Gamma$ & N$_{H,x}$                            & $\chi^2/\nu$\tablenotemark{a}\\
      &   ID   &        &          & \footnotesize{(10$^{22}$ cm$^{-2}$)} &                              \\
\tableline
B & 2   & 0200730401 & 1.8$^{+8.6}_{-4.0}$ & - & 8949/5767\\
B & 3a  & 0113060401 & 1.2$^{+2.1}_{-2.3}$ & - & 1015/1614\\
B & 4   & 0145450501 & 1.2$^{+2.6}_{-2.8}$ & - & 1418/1619\\
B & 5a  & 0203360401 & 2.0$^{+16.2}_{-3.9}$ & - & 17008/5750\\
B & 5b  & 0203360801 & 2.5$^{+18.5}_{-4.9}$ & - & 11003/4946\\
B & 6   & 0009650201 & 1.2$^{+2.7}_{-2.9}$ & - & 1822/1636\\
D & 7   & 0039140101 & 1.8$^{+0.5}_{-0.4}$ & $\leq$7.0 & 3.5/5\\
D & 8   & 0110990201 & 1.0$^{+0.4}_{-0.4}$ & $\leq$13.4 & 8.0/5\\
D & 9a  & 0110930401 & 1.4$^{+0.4}_{-0.4}$ & $\leq$6.8 & 2.6/4\\
D & 9b  & 0110930901 & 1.0$^{+0.4}_{-0.4}$ & $\leq$10.2 & 3.6/6\\
D & 10  & 0009650201 & 1.9$^{+0.4}_{-0.3}$ & $\leq$3.3 & 2.2/5\\
D & 11  & 0143650901 & 1.5$^{+0.3}_{-0.3}$ & $\leq$0.9 & 5.9/8\\
D & 12  & 0149900201 & 1.4$^{+0.5}_{-0.4}$ & $\leq$0.8 & 7.3/14\\
D & 13  & 0081340201 & 1.6$^{+0.1}_{-0.1}$ & $\leq$0.2 & 27/39\\
D & 14  & 0203361801 & 1.75$^{+0.09}_{-0.09}$ & $\leq$0.05 & 54/62\\
D & 15  & 0203361801 & 1.78$^{+0.07}_{-0.06}$ & $\leq$0.05 & 75/106\\
D & 16a & 0147511701 & 1.77$^{+0.06}_{-0.05}$ & $\leq$0.06 & 100/164\\
D & 16b & 0147511801 & 1.72$^{+0.06}_{-0.05}$ & $\leq$0.04 & 155/188\\
D & 17a & 0143150301 & 1.68$^{+0.05}_{-0.06}$ & $\leq$0.04 & 94/141\\
D & 17b & 0143150601 & 1.72$^{+0.04}_{-0.03}$ & $\leq$0.02 & 203/310\\
\tableline
\end{tabular}
\end{center}
\tablecomments{For sources with S/N$<$4, model parameters from the Model B (Cash power-law) fits are given. 
For sources with S/N$>$4, we list 90\% confidence upper-limits on Model D, power-law $+$ intrinsic absorption.}

\tablenotetext{a}{This column contains either the chi-square value and the degrees of freedom when chi-square statistics 
are applied (model C) or the likelihood of fit and the degrees of freedom when the Cash statistic is applied (model A).}
\end{table*}

For sources with S/N $> 4$, we determine the best-fit model via the 
F-test\footnote{http://cxc.harvard.edu/ciao/ahelp/ftest.html}, which measures the significance 
of the change in chi-square as components are added to a model.  In all cases, the more complex 
models did not have a significantly lower chi-square value, so for $S/N > 4$, Table~\ref{table:Best fits} lists the 
parameters for the power-law model (model C).  
\begin{table*}
\begin{center}
\caption{Alternative X-ray Spectral Fits - Fe K$\alpha$ line (S/N $>$ 25)}
\label{table:FeK fits}
\begin{tabular}{llcccc}
\tableline
\tableline
Source & Model & XMM ID & Fe K$\alpha$ FWHM\tablenotemark{a} & Fe K$\alpha$ line flux\tablenotemark{a}     & $\chi^2/\nu$\tablenotemark{b}\\
  ID   &       &        &   \footnotesize{(eV)}              & \footnotesize{(photons cm$^{-2}$ s$^{-1}$)} & \\
\tableline
15  & E & 0203361801 & $\leq$10 & $\leq$2.6e-6 & 72.3/102\\
16a & E & 0147511701 & $\leq$220 & $\leq$3.6e-7 & 100.2/160\\
16b & E & 0147511801 & $\leq$440 & $\leq$2.5e-6 & 145.8/184\\
17a & E & 0143150301 & $\leq$190 & $\leq$3.8e-4 & 92.2/137\\
17b & E & 0143150601 & $\leq$0 & $\leq$3.3e-6 & 194.5/306\\
\tableline
\end{tabular}
\end{center}
\tablenotetext{a}{Fe K$\alpha$ line parameter upper-limits are given at 90\% confidence.}
\tablenotetext{b}{The $\chi^2$ value and the degrees of freedom ($\nu$) for $\chi^2$ statistics.}
\end{table*}

\begin{figure*}
\centering
\includegraphics[width=1.4in]{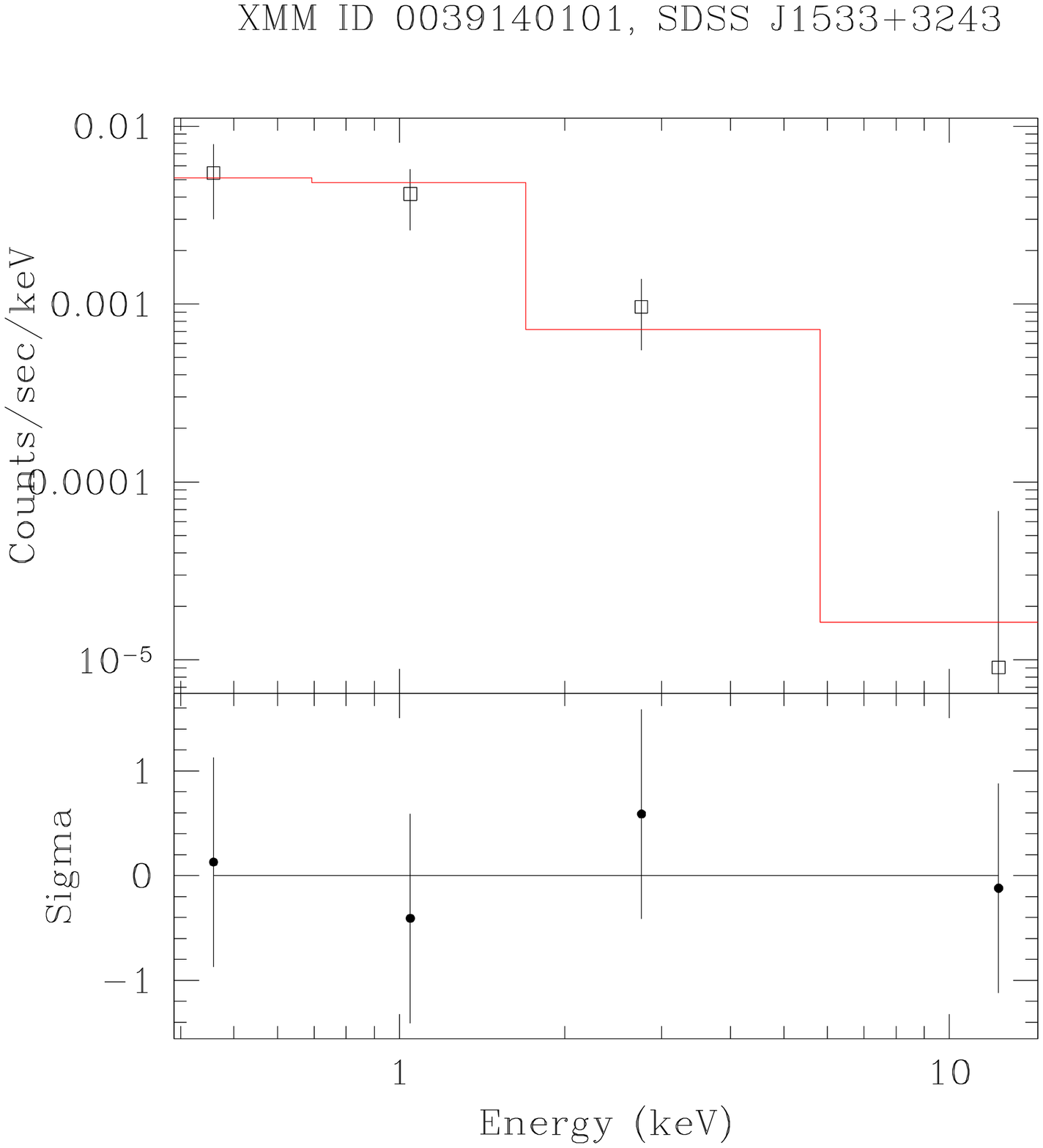}
\includegraphics[width=1.4in]{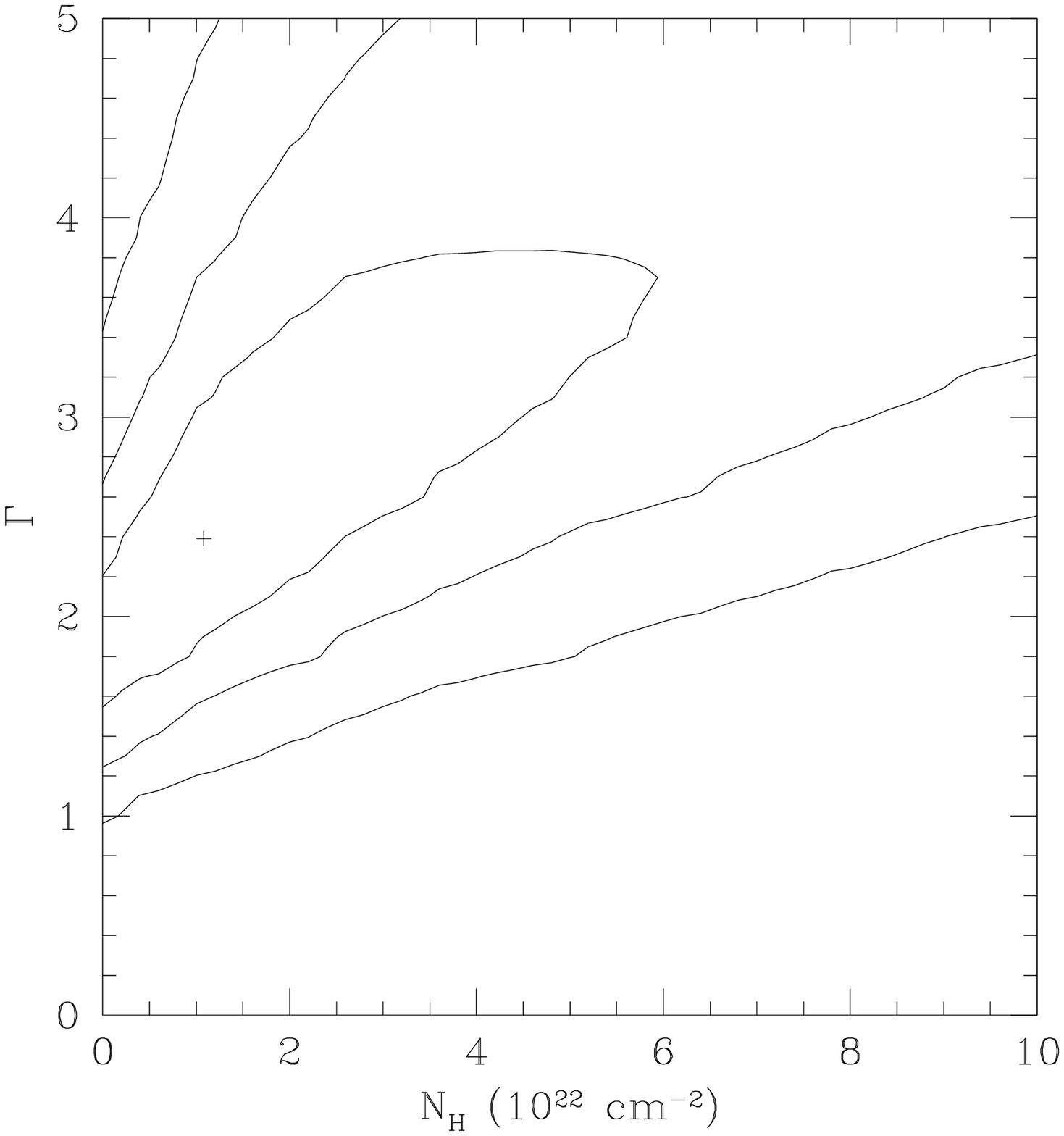}
\includegraphics[width=1.4in]{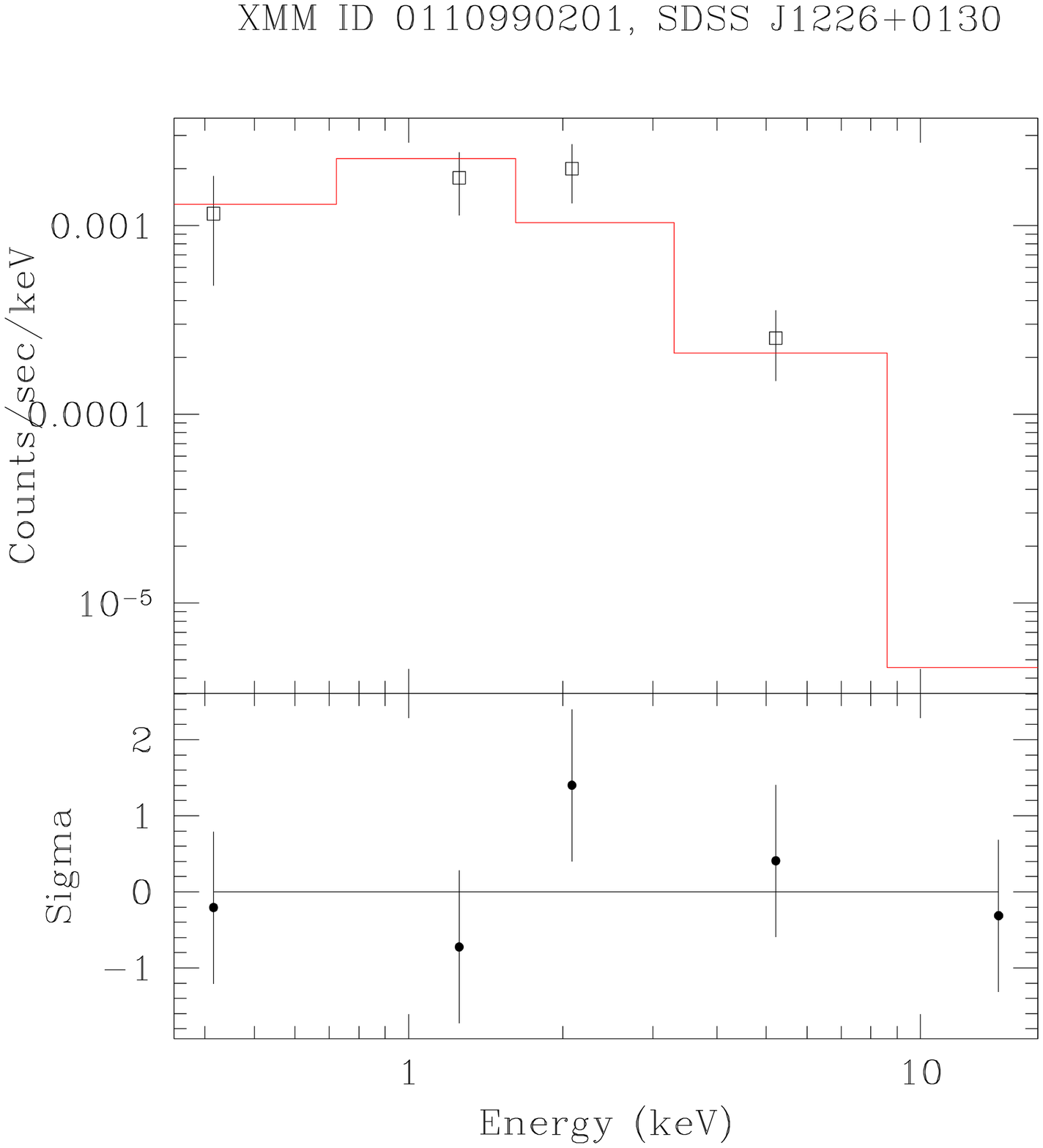}
\includegraphics[width=1.4in]{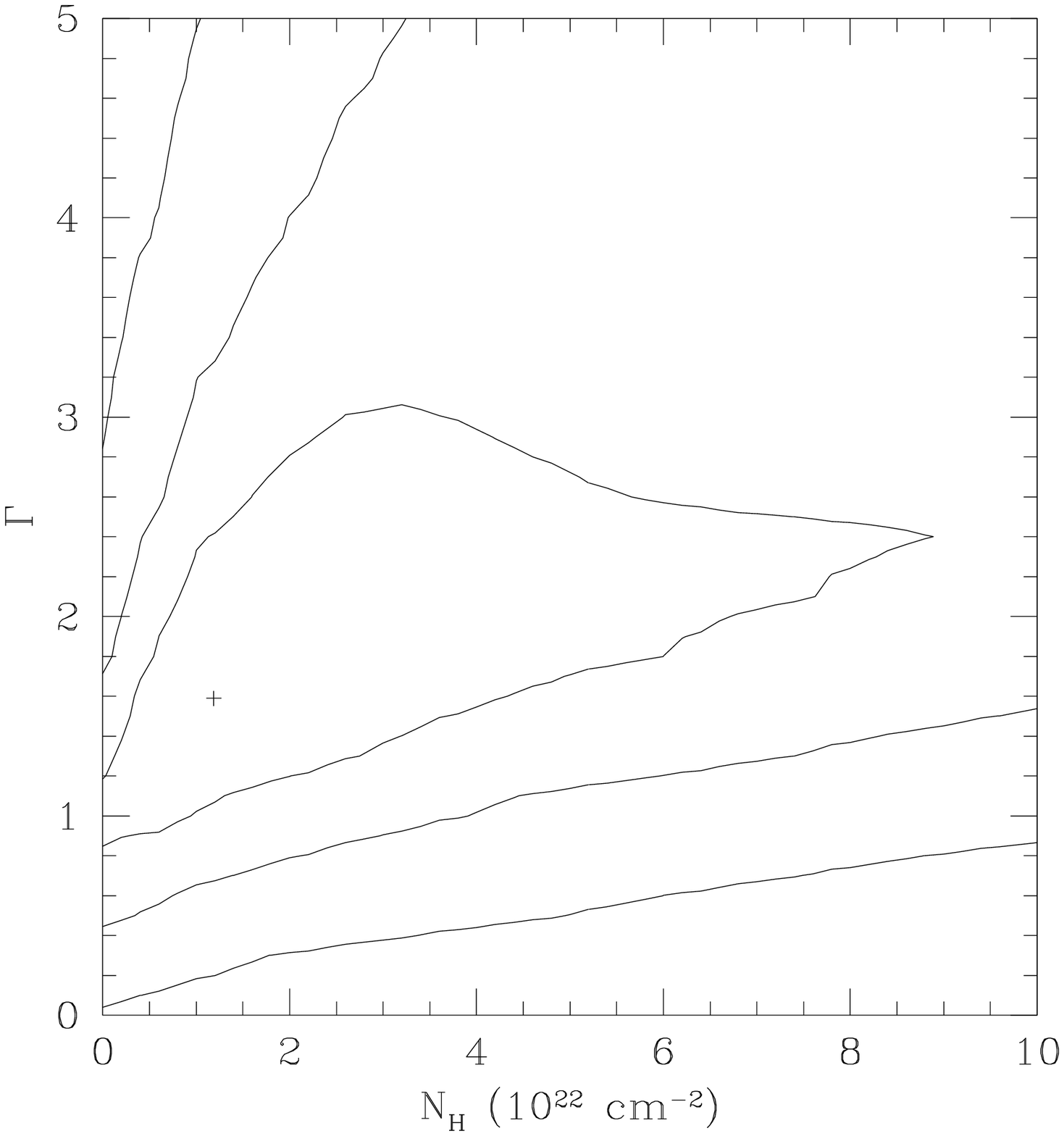}
\includegraphics[width=1.4in]{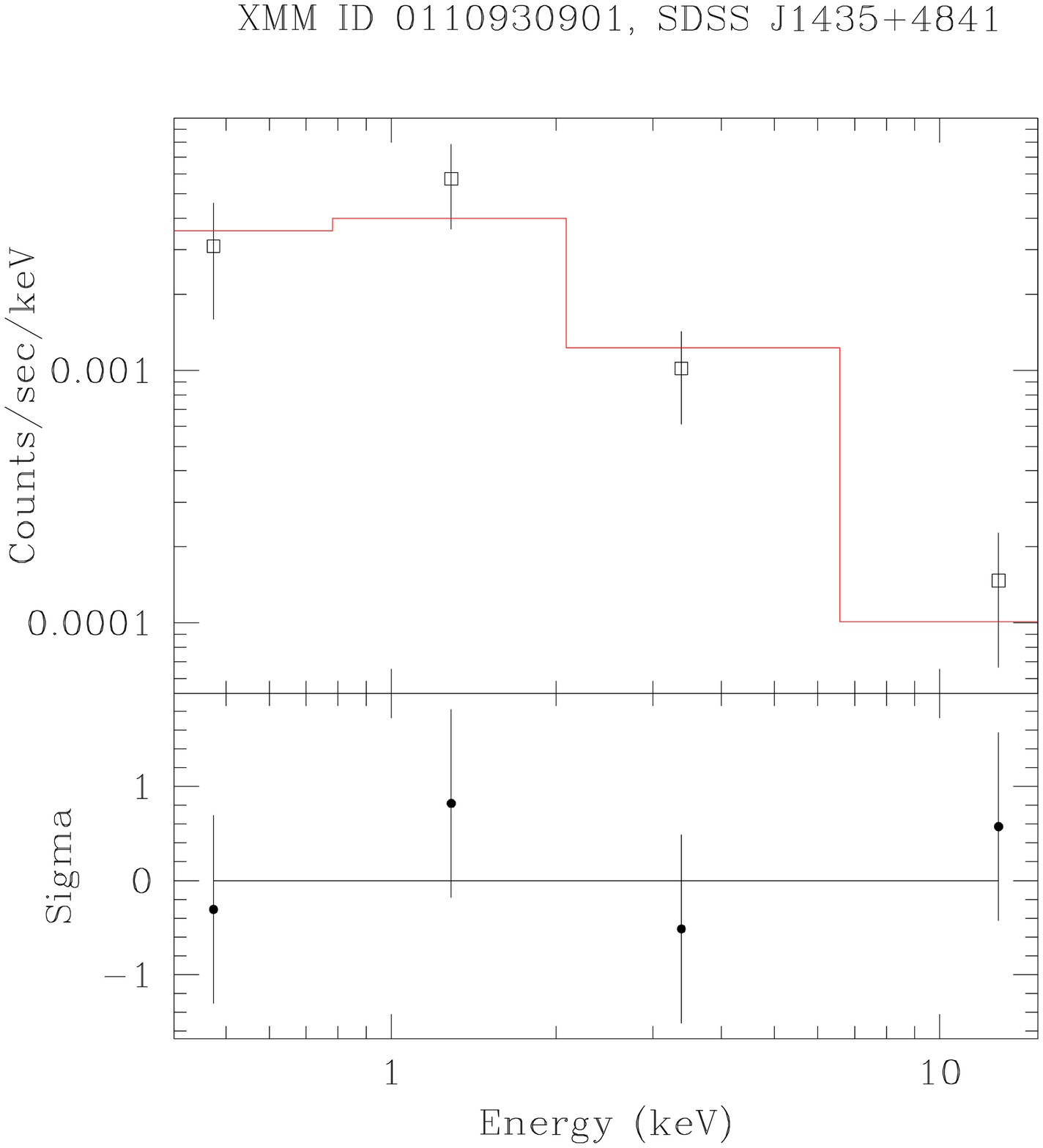}
\includegraphics[width=1.4in]{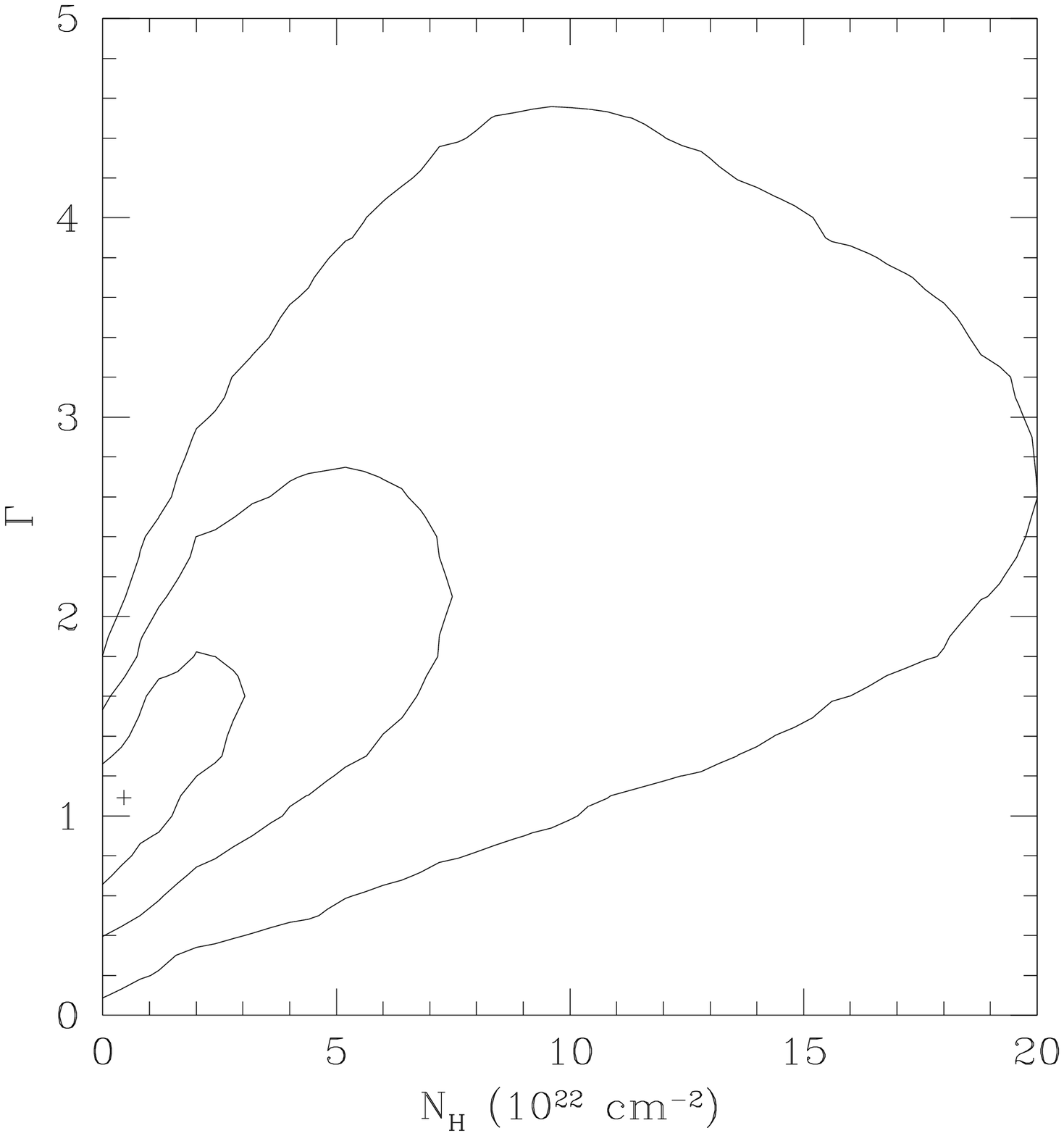}
\includegraphics[width=1.4in]{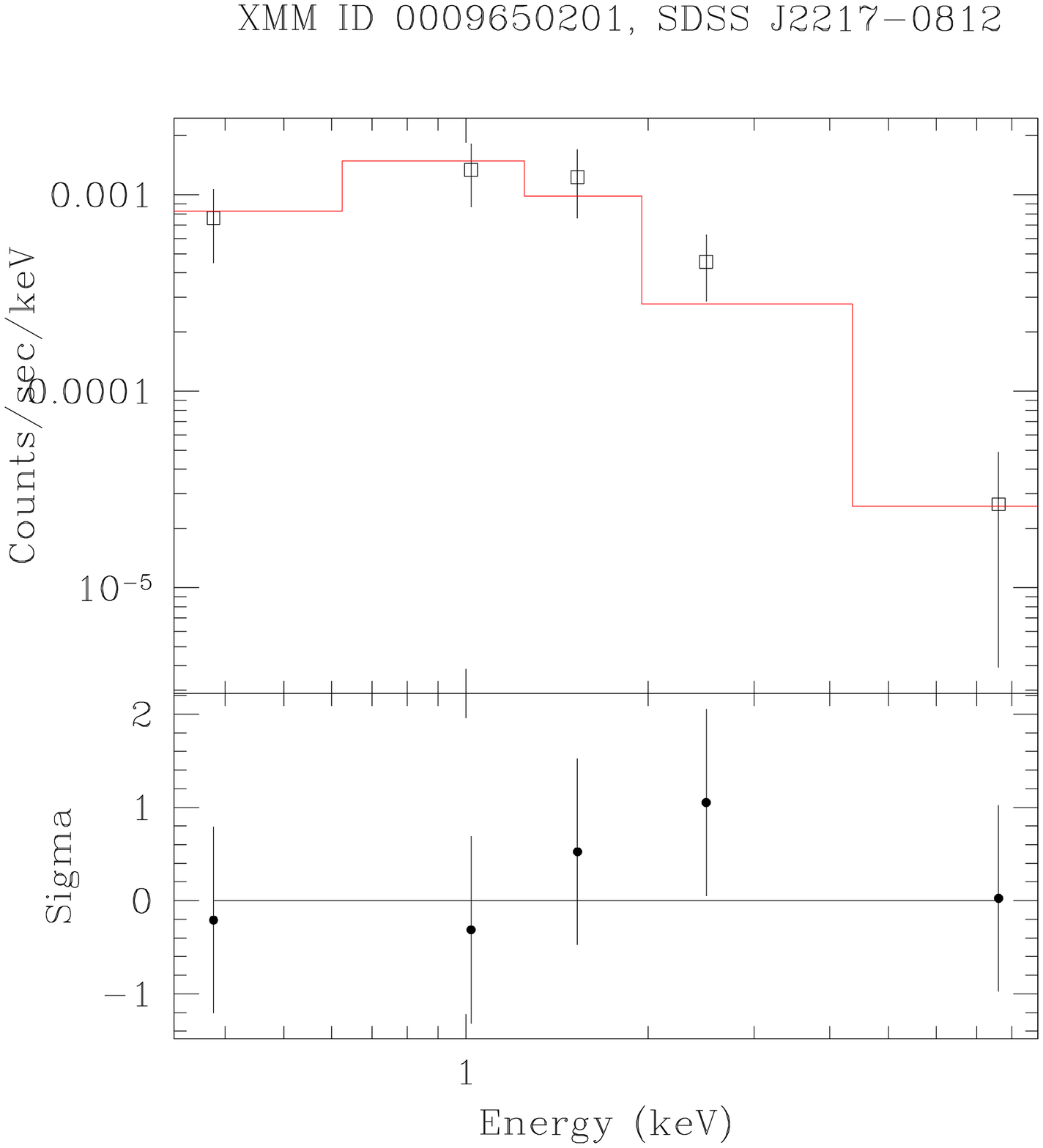}
\includegraphics[width=1.4in]{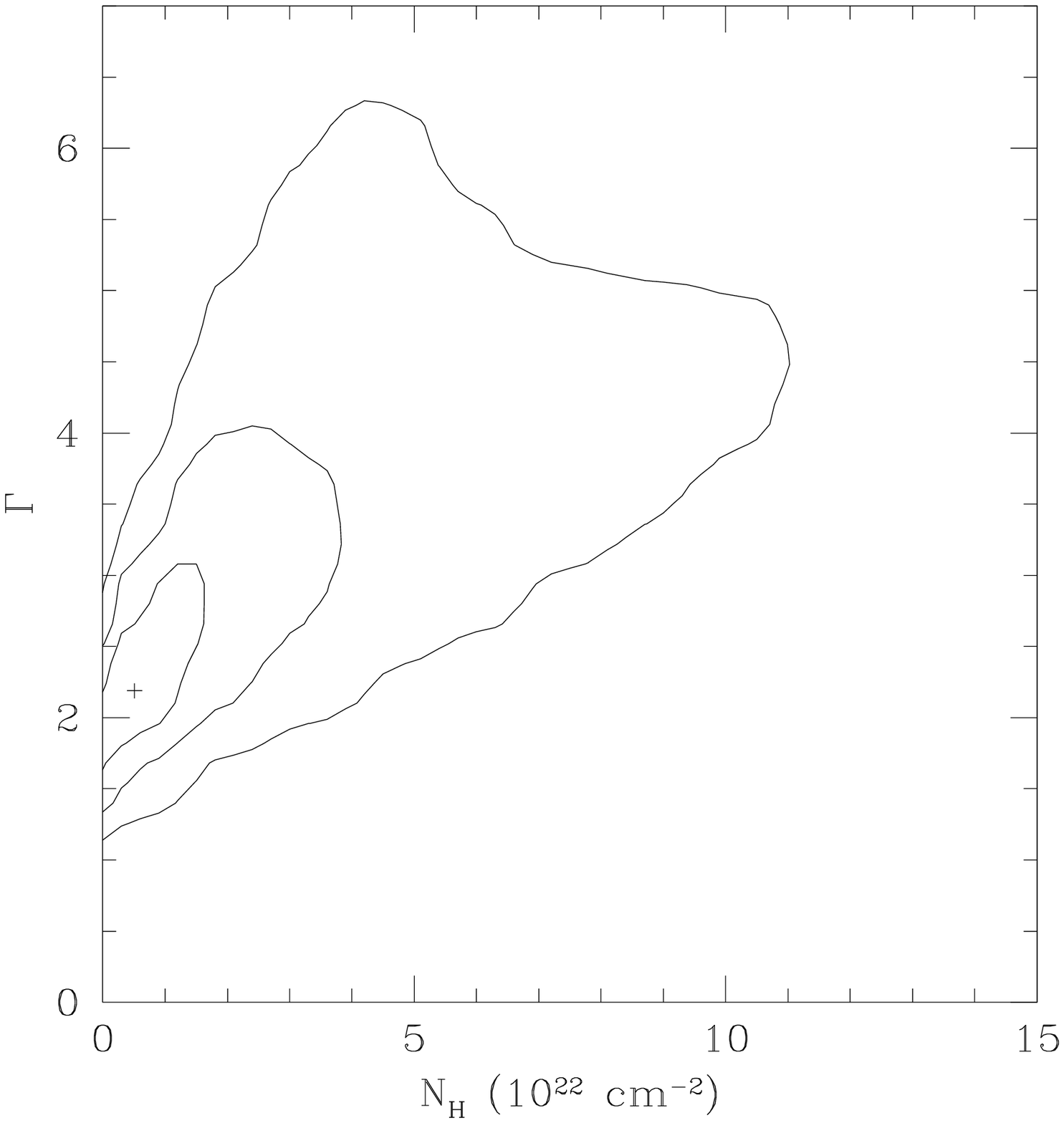}
\includegraphics[width=1.4in]{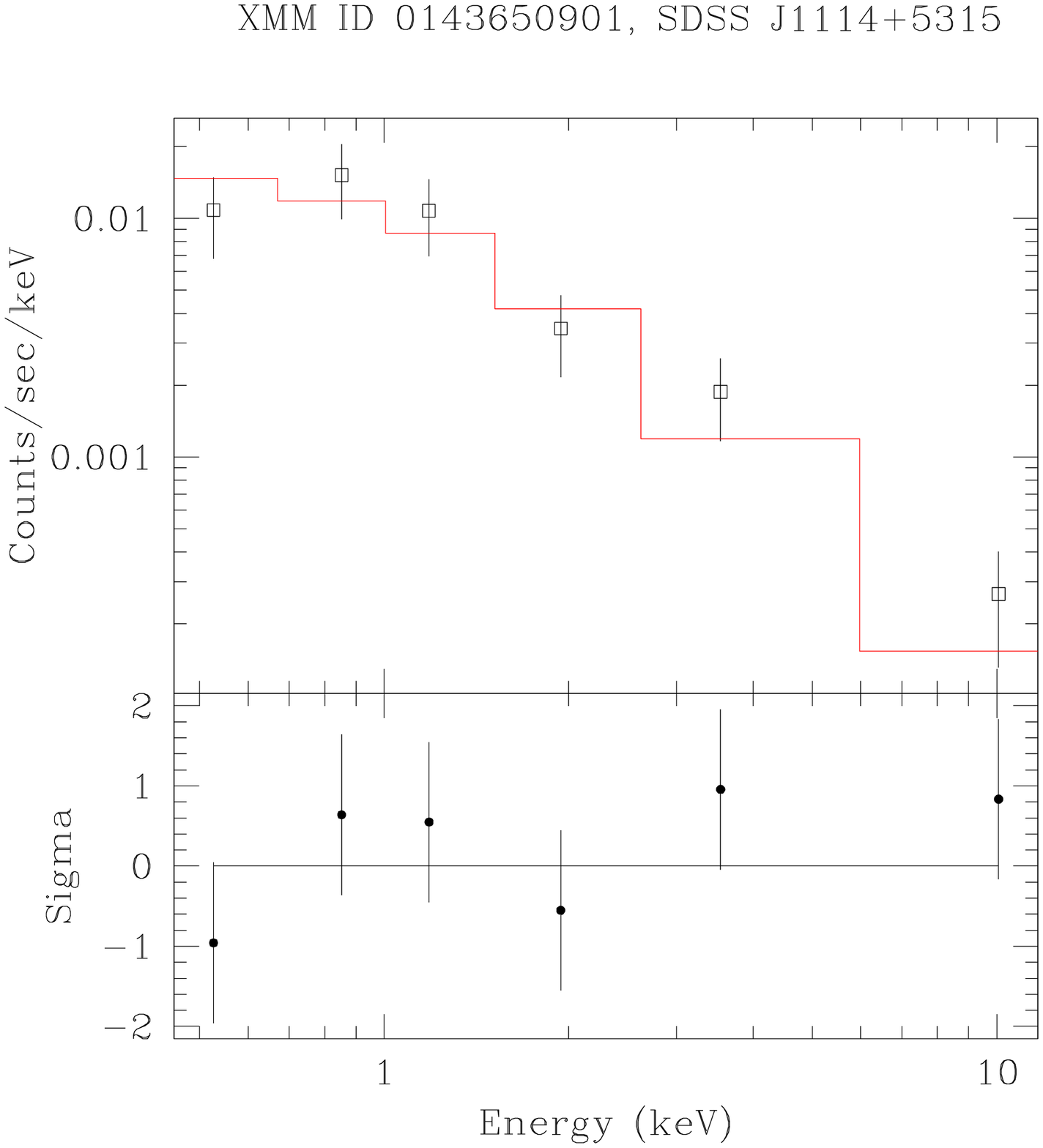}
\includegraphics[width=1.4in]{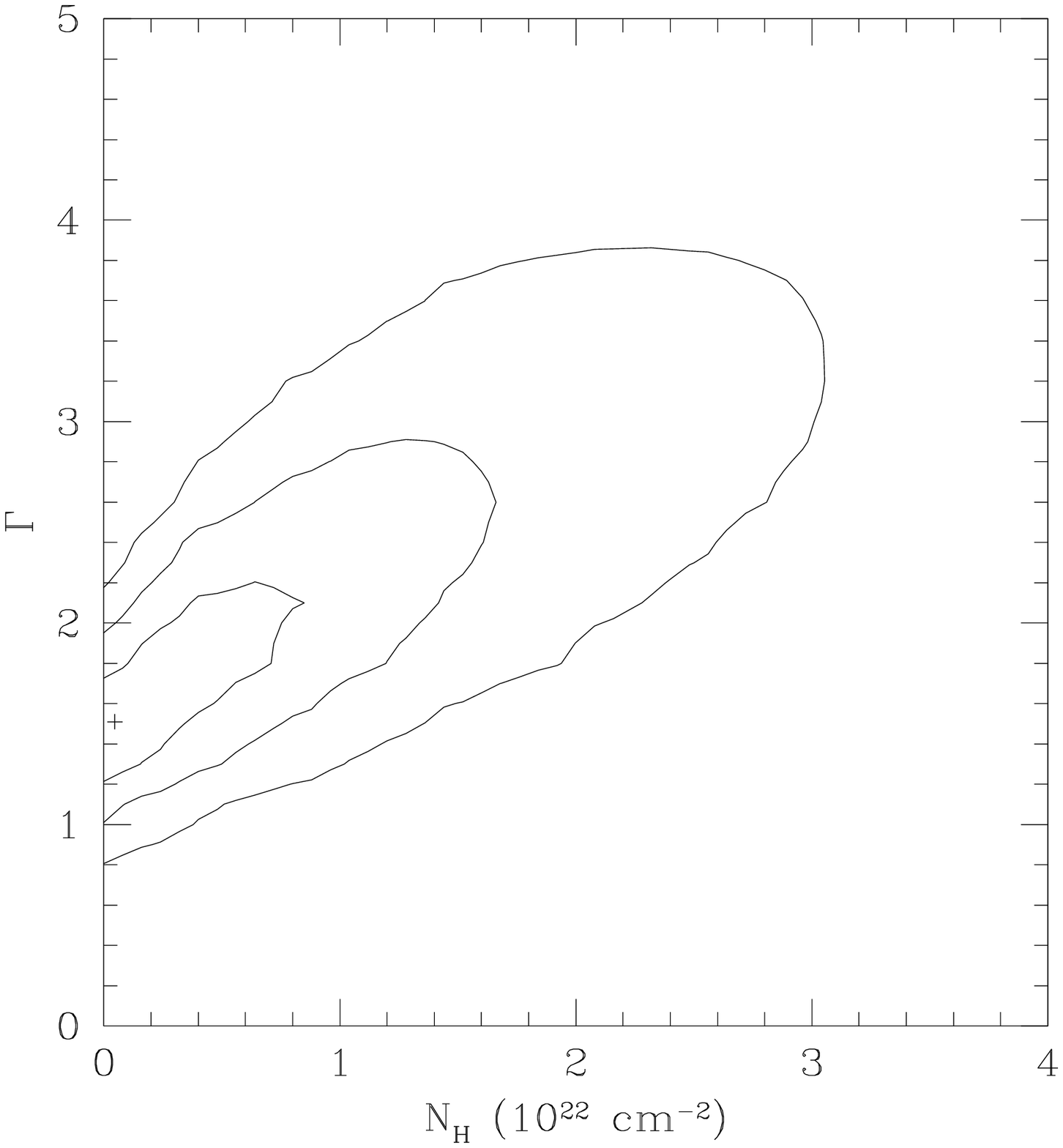}
\includegraphics[width=1.4in]{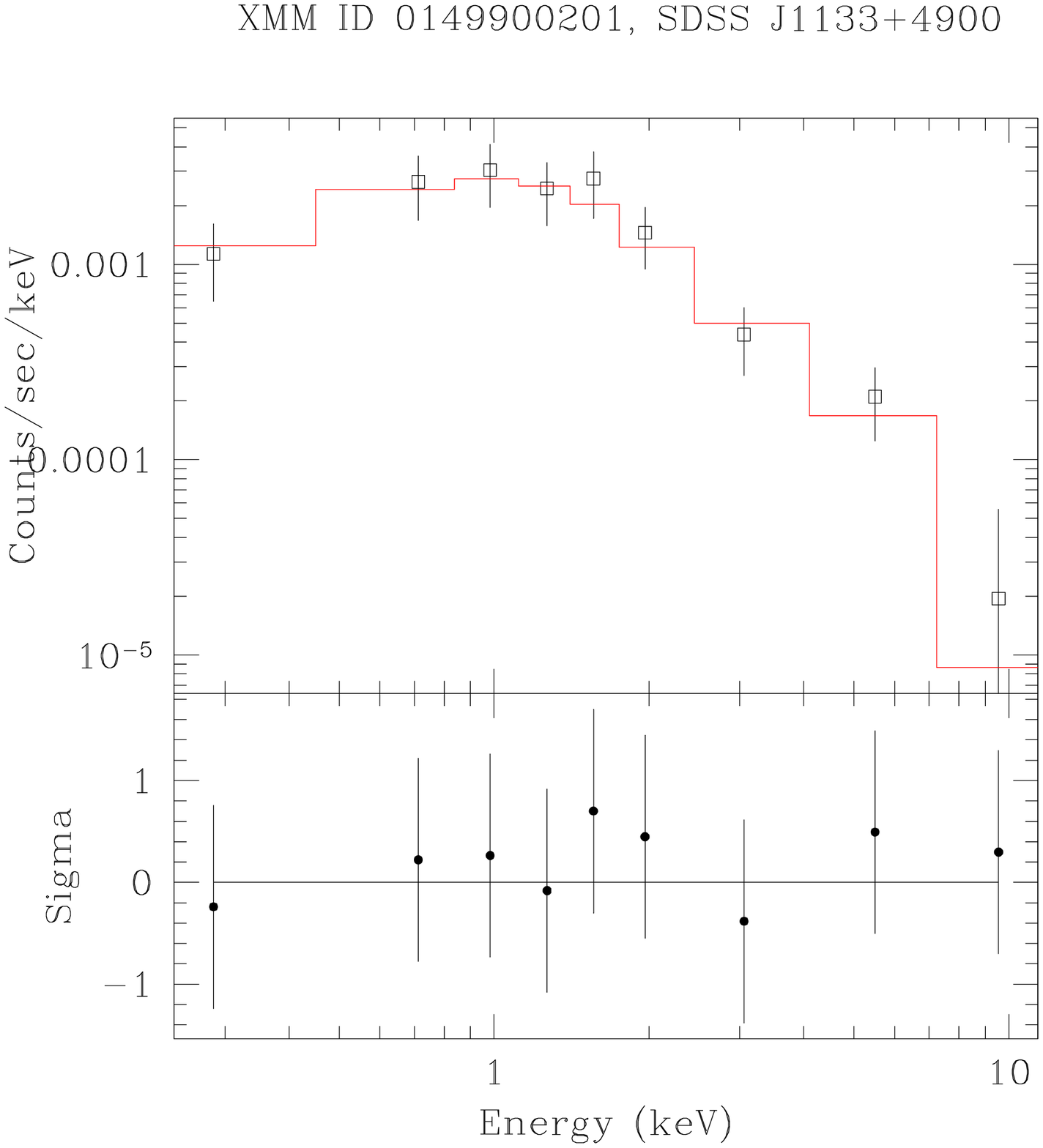}
\includegraphics[width=1.4in]{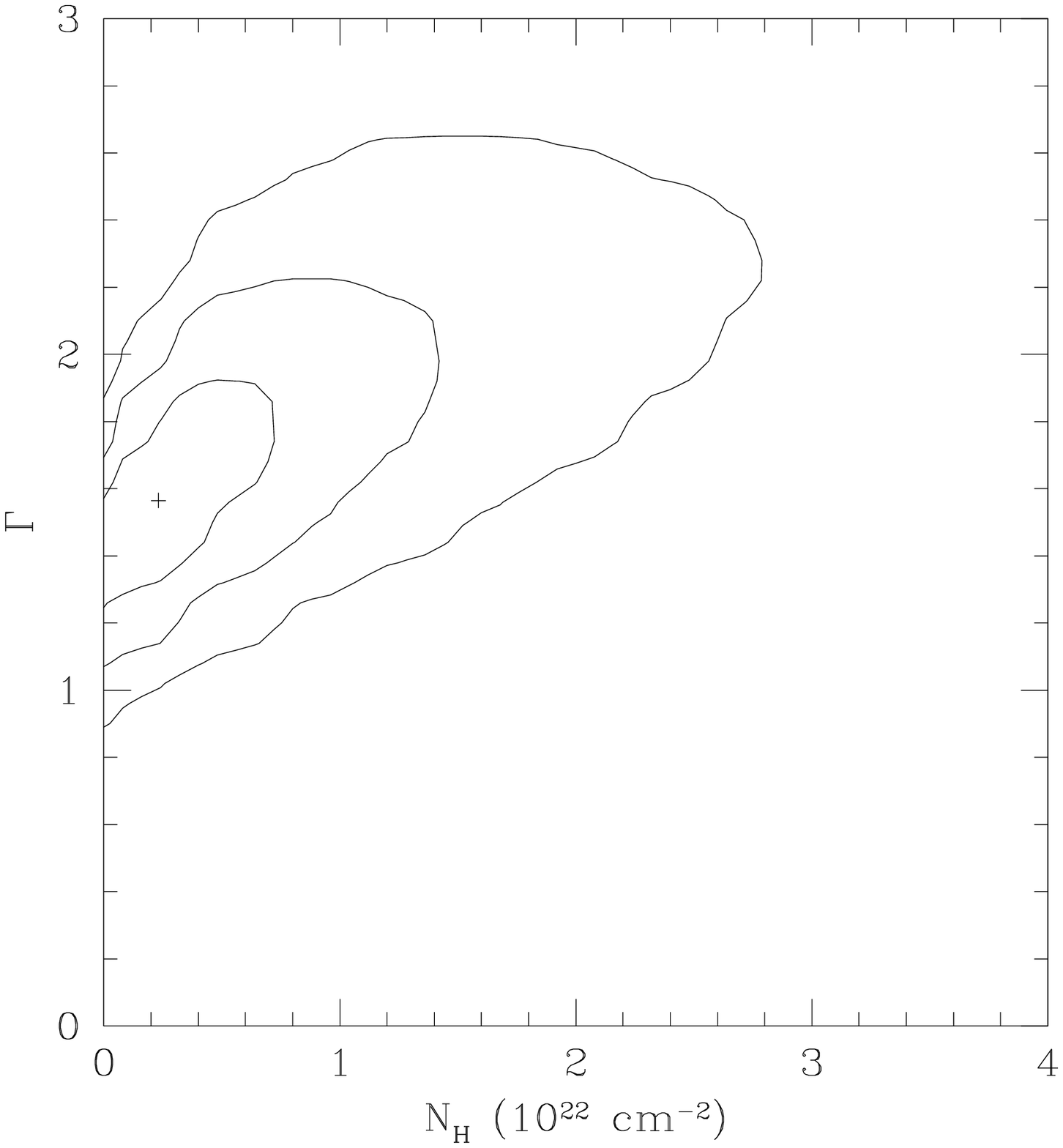}
\includegraphics[width=1.4in]{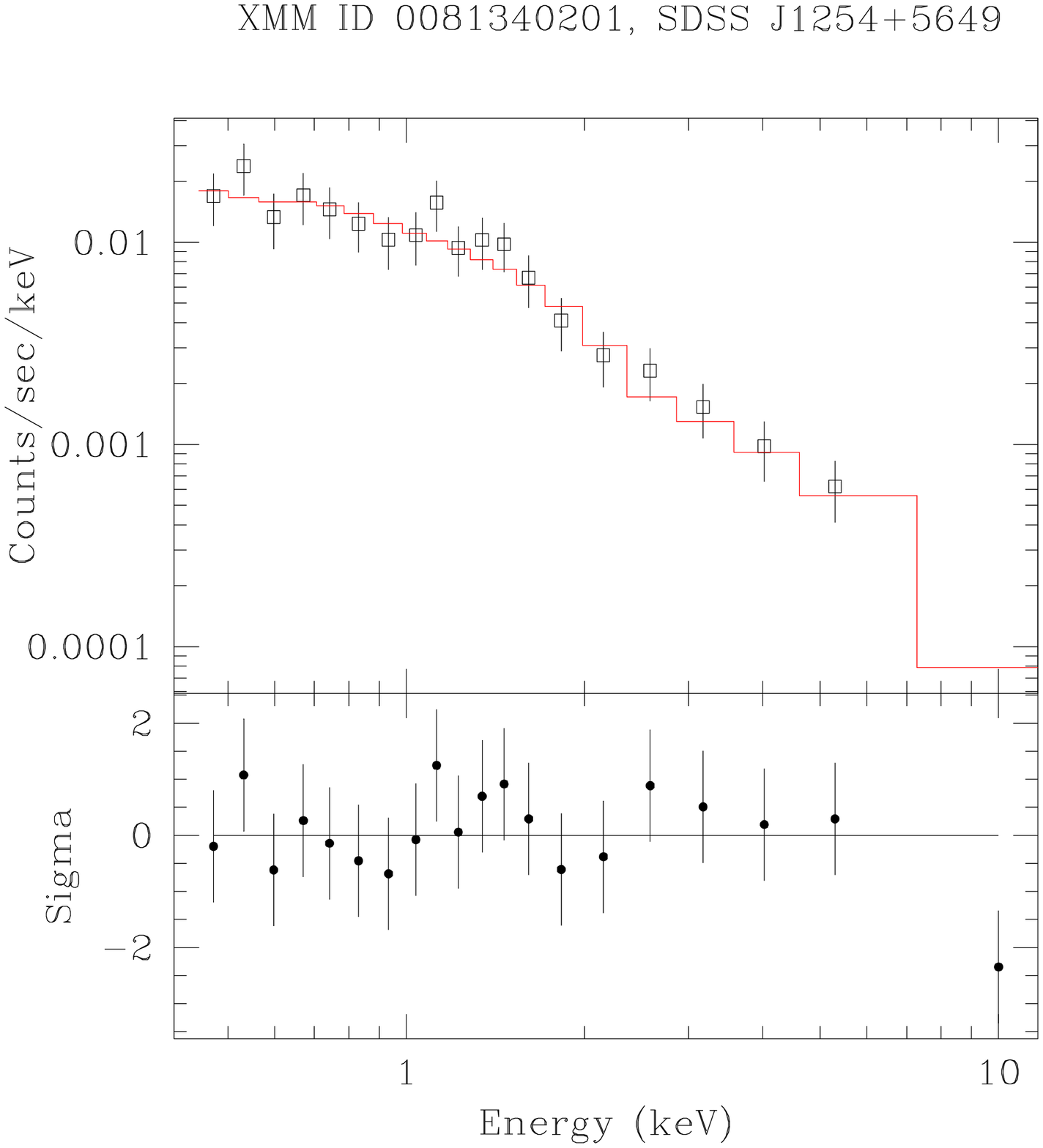}
\includegraphics[width=1.4in]{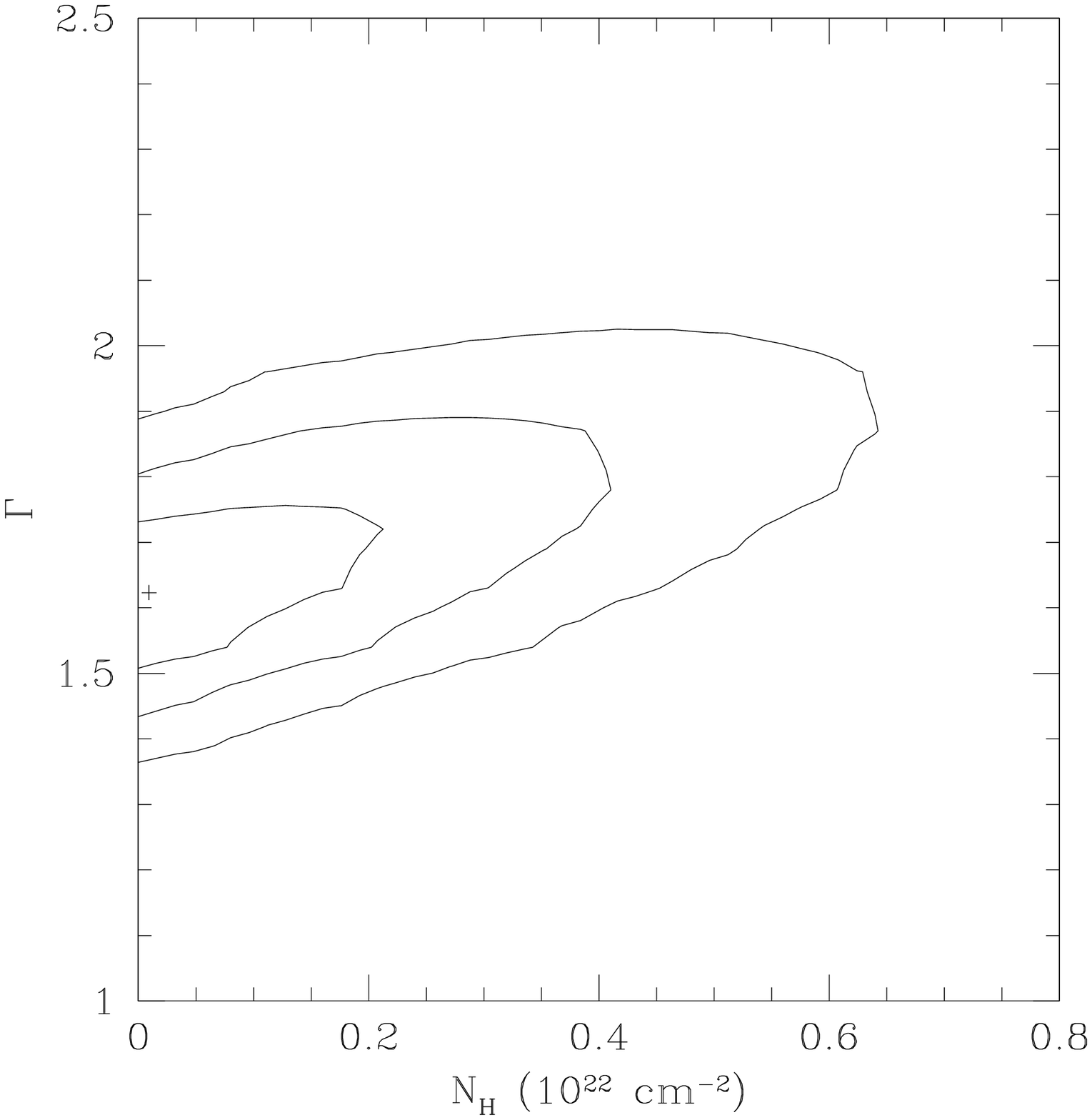}
\includegraphics[width=1.4in]{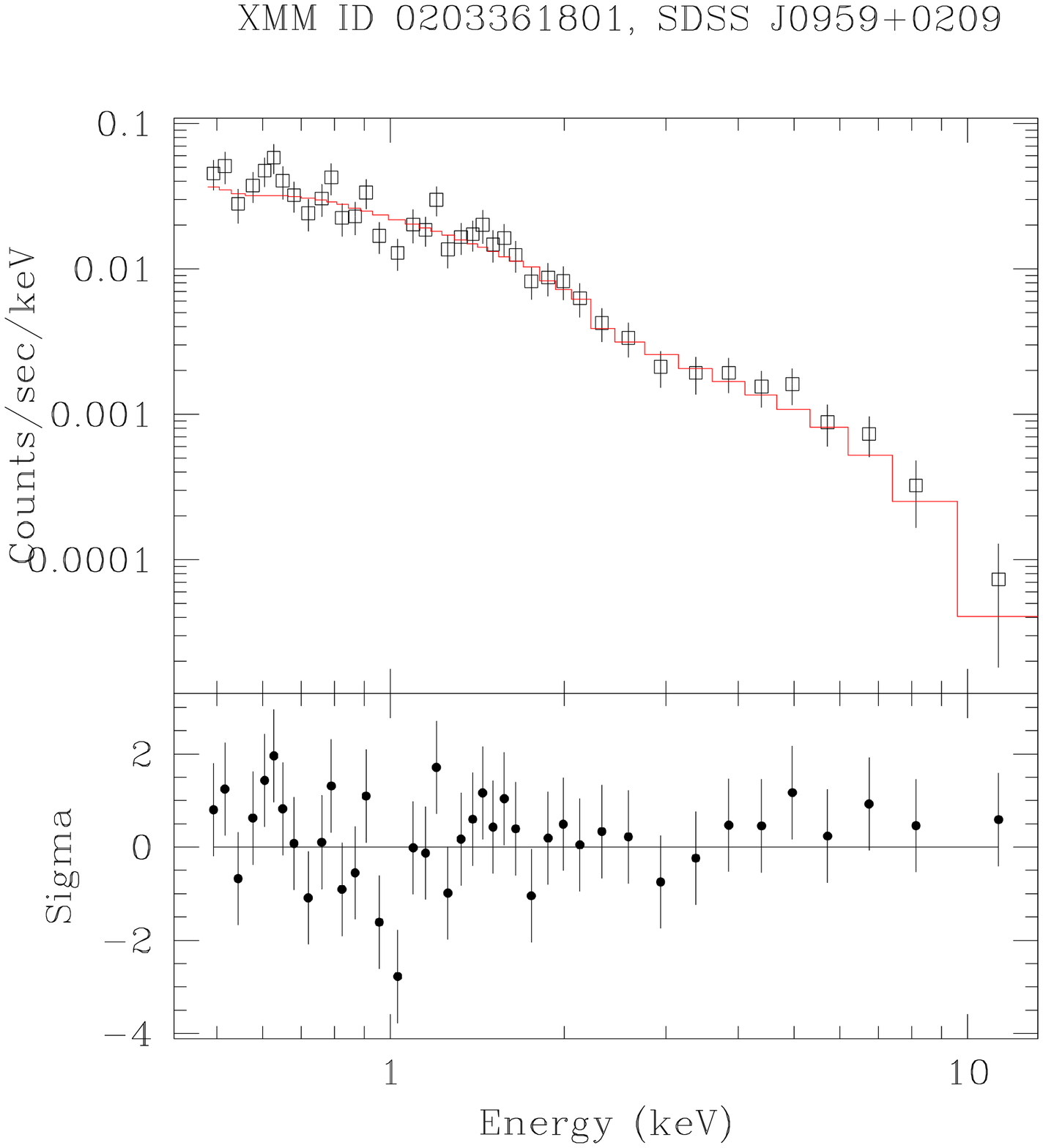}
\includegraphics[width=1.4in]{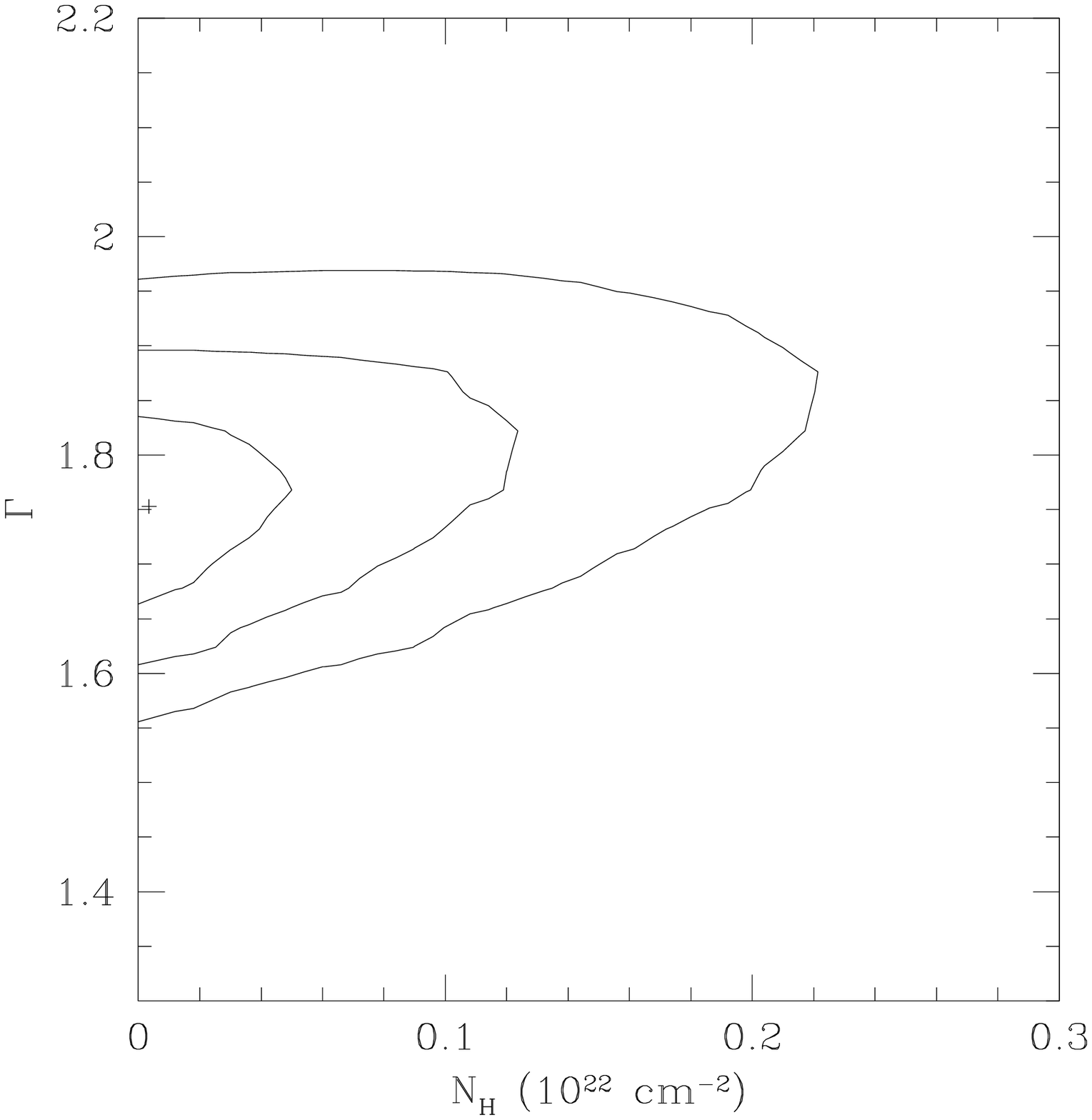}
\includegraphics[width=1.4in]{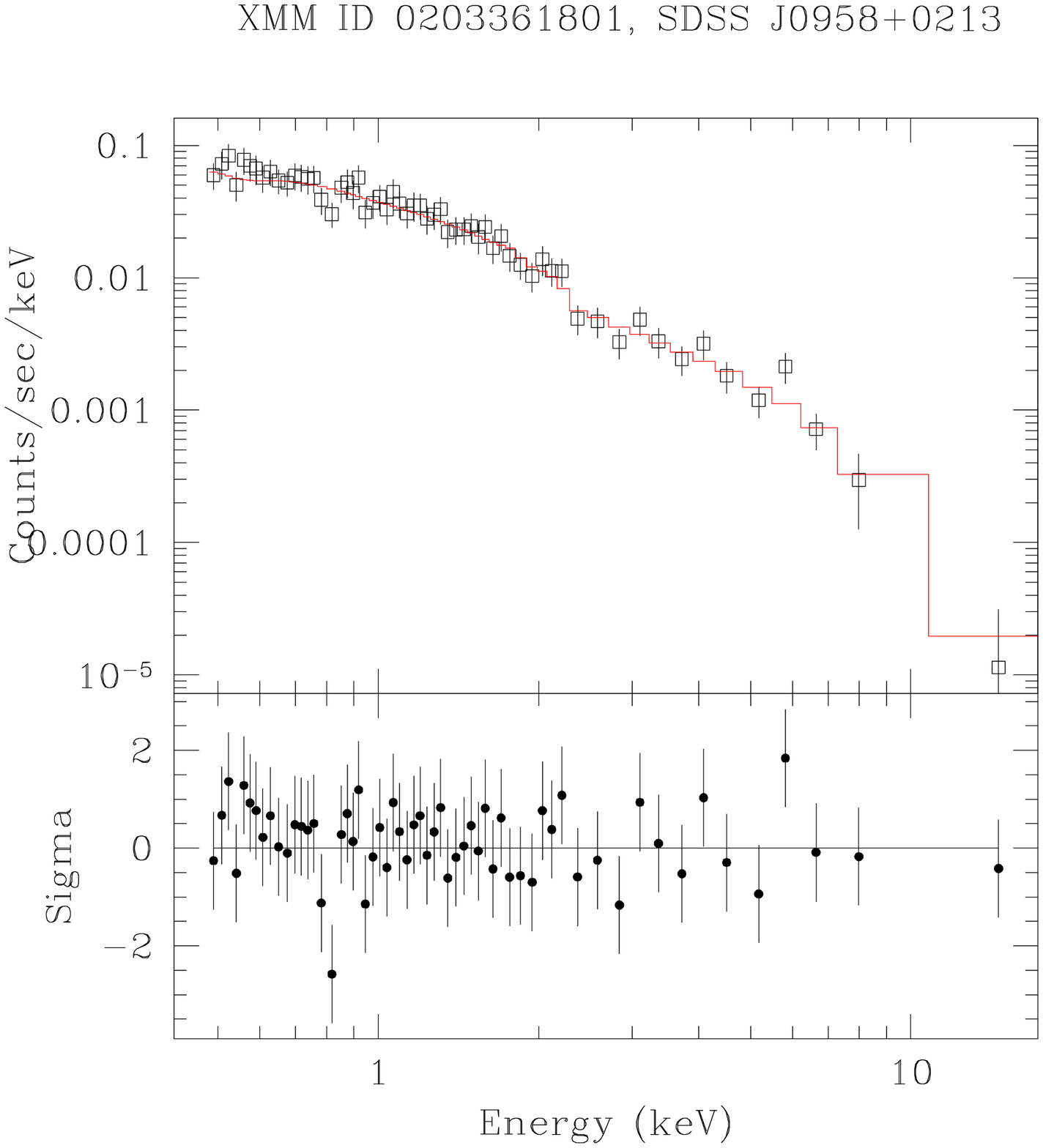}
\includegraphics[width=1.4in]{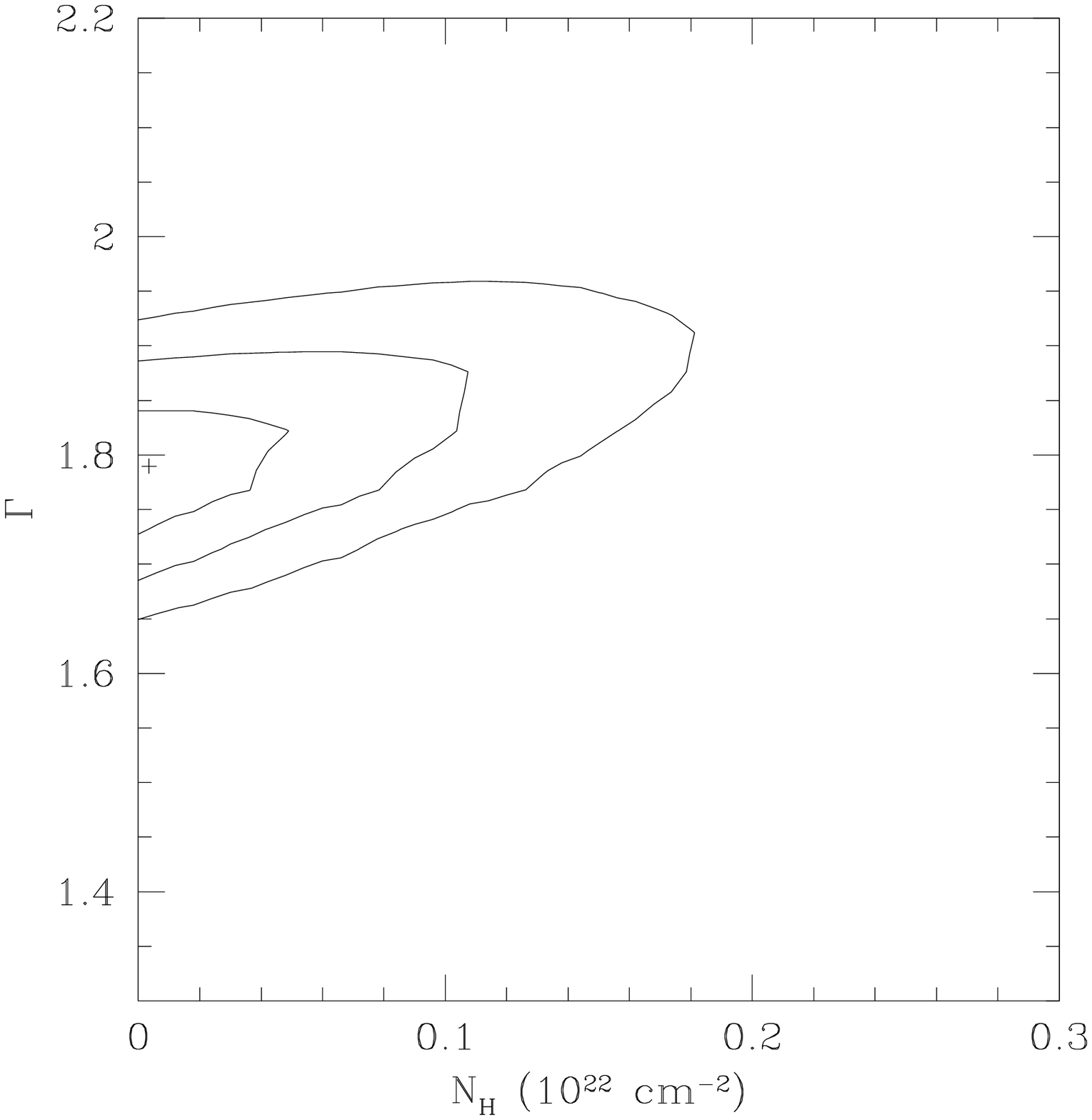}
\includegraphics[width=1.4in]{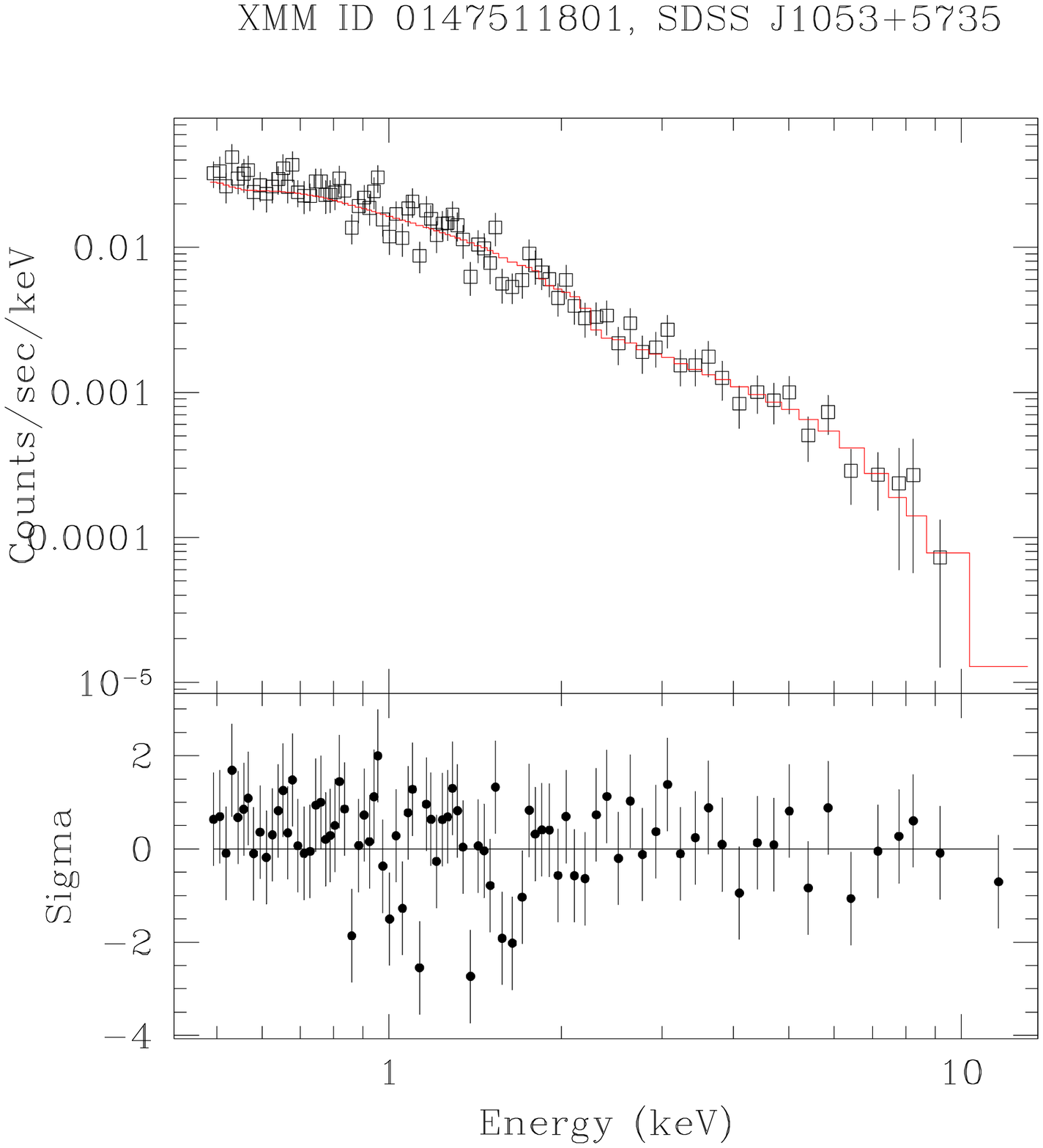}
\includegraphics[width=1.4in]{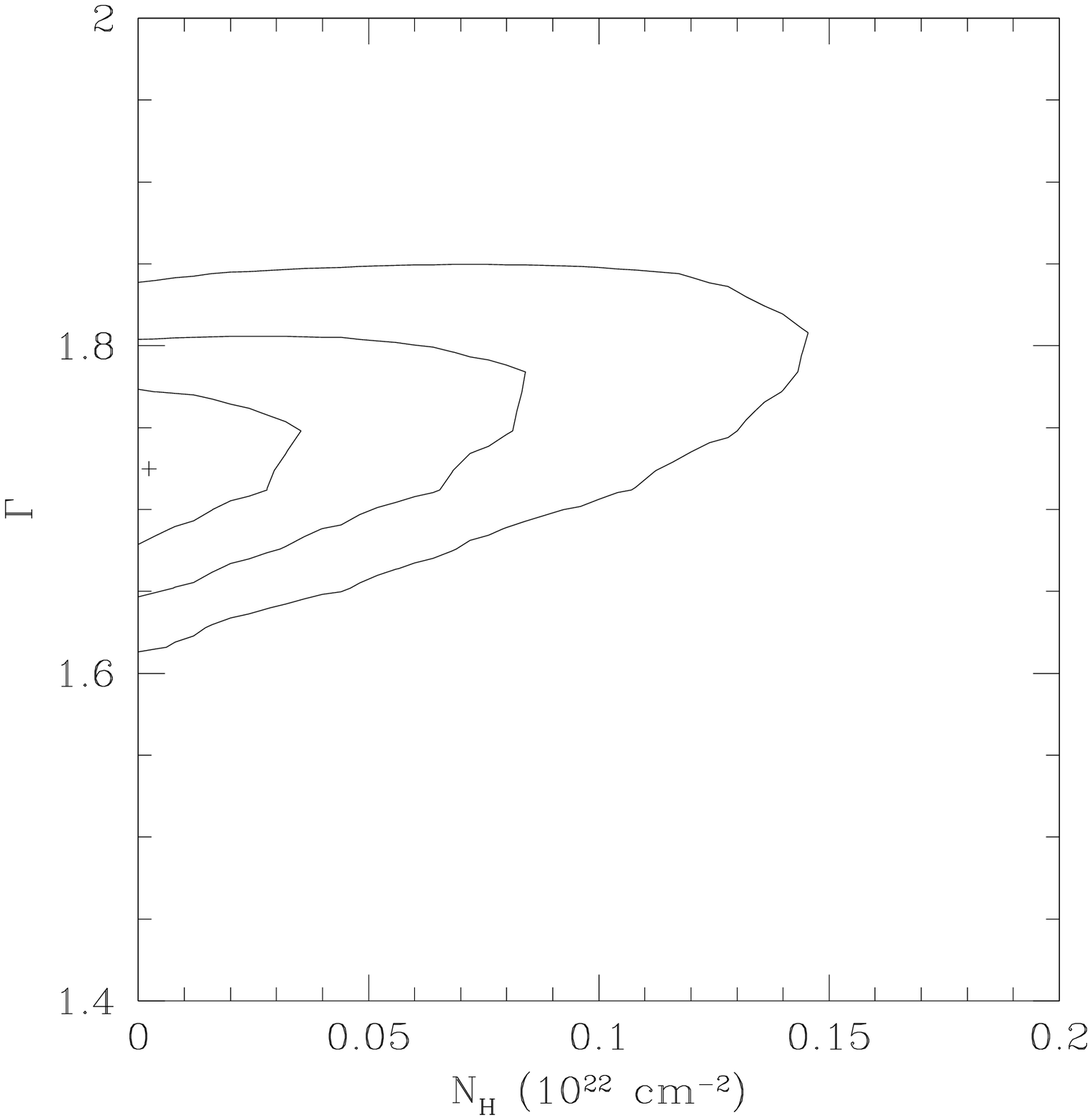}
\includegraphics[width=1.4in]{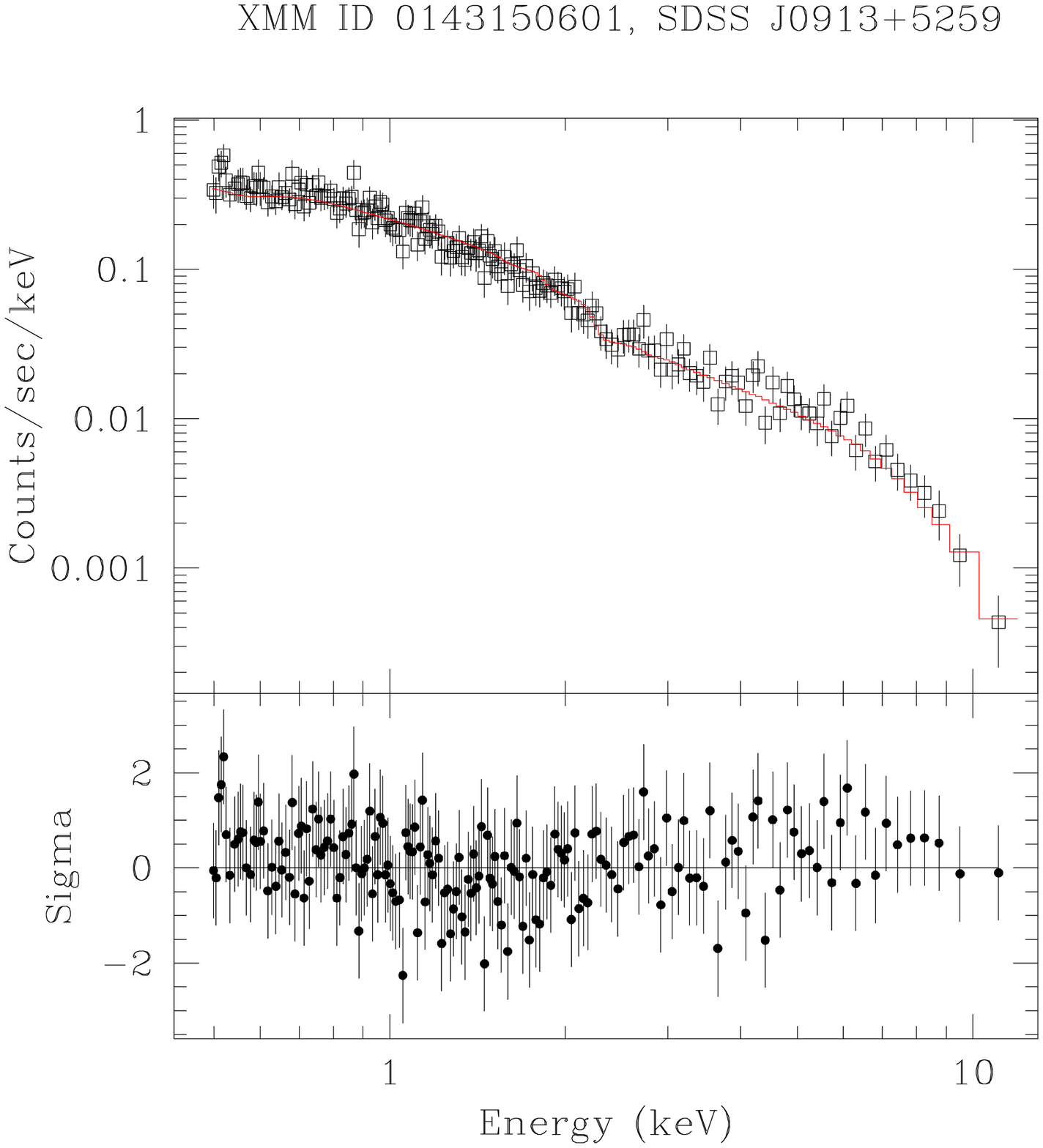}
\includegraphics[width=1.4in]{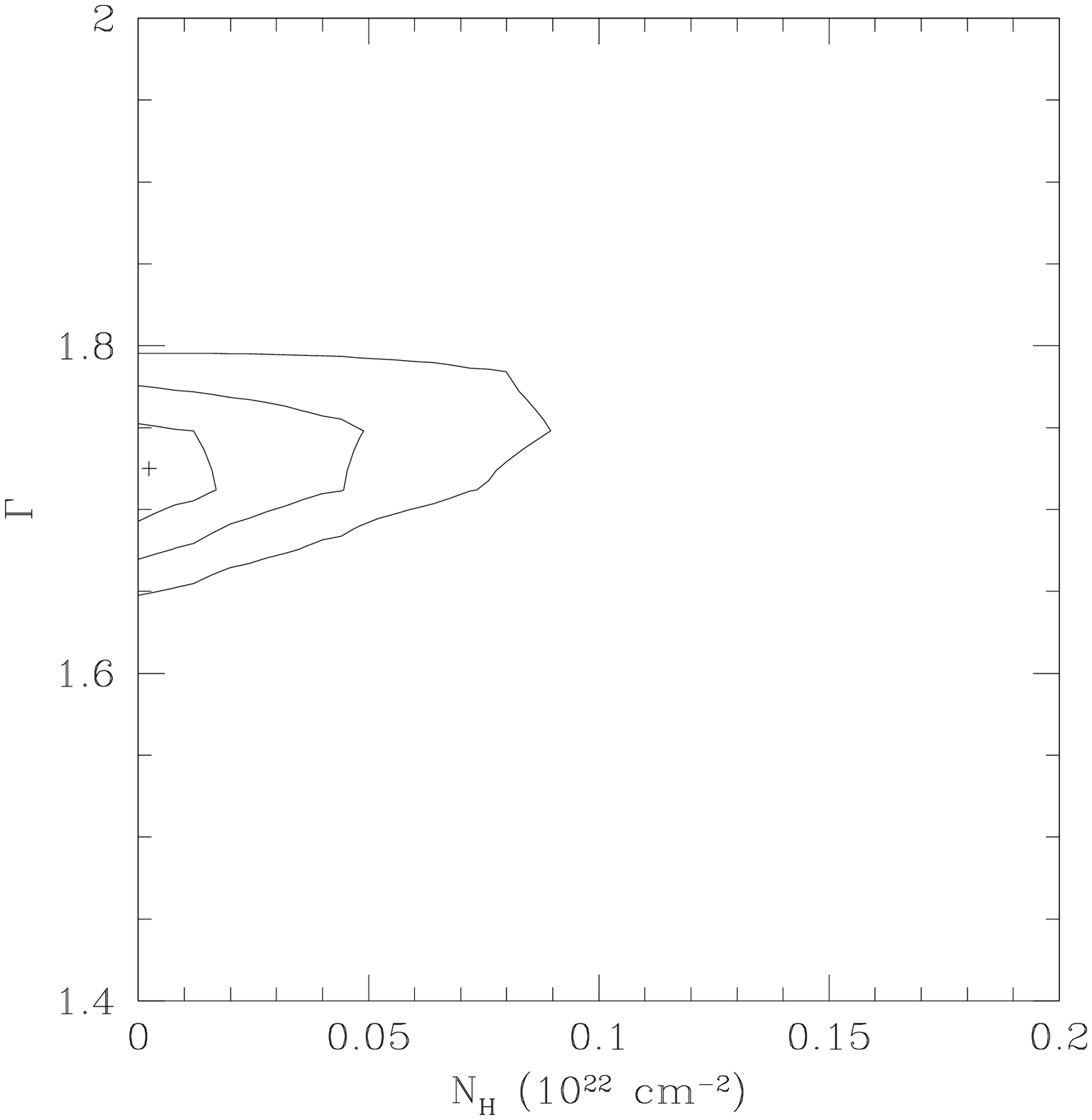}\\
\caption{For each pair of figures, the left column plots the data for sources with S/N $>$ 4 on a log 
scale, photons s$^{-1}$ keV$^{-1}$ 
versus energy (keV).  The data are fit with model D (power-law + intrinsic absorption, data binned to 
15 counts per bin).  The model, which has been folded through the detector response, is plotted as a 
histogram over the data.  Residuals are plotted underneath the fits in units of standard deviation.  
The right column contains $\Gamma$-N$_H$ contour plots, where the $1\sigma$, $2\sigma$ and $3\sigma$ 
contours are plotted as solid lines.  
The plots correspond to sources \#7-\#17, starting with \#7 in the top left panel.  Each pair of plots 
is labeled with the SDSS name and XMM observation ID.}
\label{fig:Xray fits}
\end{figure*}

Tables ~\ref{table:abs fits} and ~\ref{table:FeK fits} list the best-fit spectral slope and the upper-limits 
on other spectral parameters for rejected models B and D (Table~\ref{table:abs fits}) and E 
(Table~\ref{table:FeK fits}).  Figure~\ref{fig:Xray fits} shows the data, fits, residuals and 1$\sigma$, 
2$\sigma$ and 3$\sigma$ $\Gamma$-N$_H$ contours for the sources with S/N $>$ 4 fitted with model D 
(power-law + intrinsic absorption).  


\subsection{\it X-ray Absorption}

Of the 11 sources (and 14 observations) with S/N $>$ 4, none show significant intrinsic absorption, 
and the F-test demonstrates that all sources prefer a simple power-law model.  N$_H$ upper limits 
(Model D) for the five highest S/N sources range from 3 x 10$^{20}$ cm$^{-2}$ to 2 x 10$^{21}$ cm$^{-2}$ 
at the 90\% confidence level.  

Absorption can also be indicated when the best-fitting power-law is unusually hard.  
\citet{Mateos05} finds an average X-ray photon index $<\Gamma>$ $\sim$ 1.96 and intrinsic dispersion 
$\sigma_\Gamma = 0.4$ for a sample of 1137 AGN found in XMM-Newton fields, with fluxes ranging from 
10$^{-15}$ - 10$^{-12}$ ergs cm$^{-2}$ s$^{-1}$.  \citet{Mateos05} find a systematic hardening of 
the average X-ray spectra towards 
fainter fluxes, which they interpret as a higher degree of photoelectric absorption among fainter AGNs.  
Therefore, we define an unusually hard spectrum to have $\Gamma < 1.5$ at 90\% confidence.  Two of 
14 sources with S/N $>$ 4 have unusually hard spectral slopes by this measure: 
SDSS J1226+0130 (\#8) and SDSS J1435+4841 (\#9).  All of the quasars with high enough S/N to fit $\Gamma$ 
have best-fit spectral slopes flatter than the \citet{Mateos05} average.  

For the five sources with 2 $<$ S/N $<$ 4 (fit with model A), the S/N is not high enough to fit 
a power-law or intrinsic absorption.  One indicator of X-ray absorption for these sources is 
X-ray weakness relative to the optical luminosity, defined as $\alpha_{ox} < -1.8$, where $\alpha_{ox}$ 
is the spectral index from 2500 $\mbox{\AA}$ to 2 keV.  Such small values of $\alpha_{ox}$ are a common 
characteristic of BAL quasars, which are known to be heavily absorbed objects in the X-rays 
\citep{Mathur95, GM96}.

To calculate $\alpha_{ox}$ for the red sample, we use the 2 keV flux (F$_{2keV}$) given by the X-ray 
fits, and we derive the 2500 $\mbox{\AA}$ flux (F$_{uv}$) by interpolating a power-law between the 
nearest two of the \emph{ugriz} magnitudes.  We then calculate $\alpha_{ox,corr}$ by correcting 
$\alpha_{ox}$ for absorption.  We deredden F$_{uv}$ using $E(B-V)_{spec}$ and the 
\citet{Prevot84} SMC extinction curve, where $A_\lambda$ = E(B-V) 1.39 $\lambda^{-1.2}$.  For 
objects with X-ray S/N $>$ 4, we do not find significant intrinsic absorption, so we use the 
90\% upper-limit on N$_H$ to correct F$_{2keV}$.  For objects with X-ray S/N $<$ 4, we cannot fit 
intrinsic absorption, so we make no correction to F$_{2keV}$.  Table~\ref{table:aox} gives the 
measured and corrected F$_{uv}$ and F$_{2keV}$, $\alpha_{ox}$ (uncorrected), and 
$\alpha_{ox,corr}$ (corrected for absorption).  
\begin{table}
\begin{center}
\caption{X-ray Strength Relative to the Optical}
\label{table:aox}
\begin{tabular}{lcccccc}
\tableline
\tableline
ID & F$_{uv}$\tablenotemark{a} & F$_{uv,corr}$\tablenotemark{b} & F$_{2keV}$\tablenotemark{c} & F$_{2keV,corr}$\tablenotemark{d} & $\alpha_{ox}$ & $\alpha_{ox,corr}$\tablenotemark{e}\\
\tableline
1 & 3.87 & 6.04    & $\leq$2.45  & $\leq$2.45 & $\leq$-1.61 & $\leq$-1.69\\
2 & 3.19 & 3.99    & 0.33 & 0.33 & -1.91 & -1.95\\
3a & 4.39 & 7.54   & $\leq$0.17  & $\leq$0.17 & $\leq$-2.07 & $\leq$-2.16\\
3b & 4.39 & 7.54   & 3.08 & 3.08 & -1.59 & -1.68\\
4 & 2.85 & 15.05   & 2.81 & 2.81 & -1.54 & -1.82\\
5a & 2.69 & 3.83   & 0.38 & 0.38 & -1.86 & -1.92\\
5b & 2.69 & 3.83   & 0.67 & 0.67 & -1.77 & -1.83\\
6 & 5.00 & 15.66   & 0.15 & 0.15 & -2.12 & -2.31\\
7 & 2.53 & 5.82    & 2.75 & 29.0 & -1.52 & -1.27\\
8 & 5.65 & 15.34   & 0.77 & 7.24 & -1.87 & -1.66\\
9a & 2.10 & 5.06   & 1.87 & 45.1 & -1.55 & -1.17\\
9b & 2.10 & 5.06   & 1.38 & 3.67 & -1.61 & -1.59\\
10 & 2.64 & 5.67   & 4.32 & 10.0 & -1.45 & -1.44\\
11 & 5.34 & 17.66  & 4.57 & 8.44 & -1.56 & -1.66\\
12 & 7.39 & 9.88   & 7.36 & 10.6 & -1.54 & -1.52\\
13 & 1.33 & 2.64   & 9.72 & 10.9 & -1.20 & -1.30\\
14 & 2.20 & 5.09   & 7.79 & 8.02 & -1.32 & -1.46\\
15 & 2.21 & 6.47   & 21.9 & 22.6 & -1.15 & -1.33\\
16a & 4.42 & 9.48  & 14.7 & 15.1 & -1.33 & -1.45\\
16b & 4.42 & 9.48  & 15.2 & 15.6 & -1.33 & -1.45\\
17a & 53.67 & 95.96 & 63.5 & 64.5 & -1.51 & -1.60\\
17b & 53.67 & 95.96 & 63.8 & 64.4 & -1.51 & -1.60\\
\tableline
\end{tabular}
\end{center}
\tablenotetext{a}{Rest-frame 2500 $\mbox{\AA}$ fluxes are in units of 10$^{-28}$ ergs cm$^{-2}$ s$^{-1}$ Hz$^{-1}$.}

\tablenotetext{b}{The 2500 $\mbox{\AA}$ flux is corrected using E(B-V)$_{spec}$.}

\tablenotetext{c}{Rest-frame 2 keV fluxes are in units of 10$^{-32}$ ergs cm$^{-2}$ s$^{-1}$ Hz$^{-1}$.}

\tablenotetext{d}{For sources with S/N $>$ 4, the 2 keV flux is corrected for the N$_H$ 90\% confidence upper-limit.
(No 2 keV flux correction is made for sources with IDs \#1-7.)}

\tablenotetext{e}{$\alpha_{ox,corr}$ is calculated with F$_{uv,corr}$ and F$_{2keV,corr}$ so that it is corrected for 
dust-reddening and, where possible, an upper-limit on X-ray absorption.}  
\end{table}

\begin{figure}
\centering
\plotone{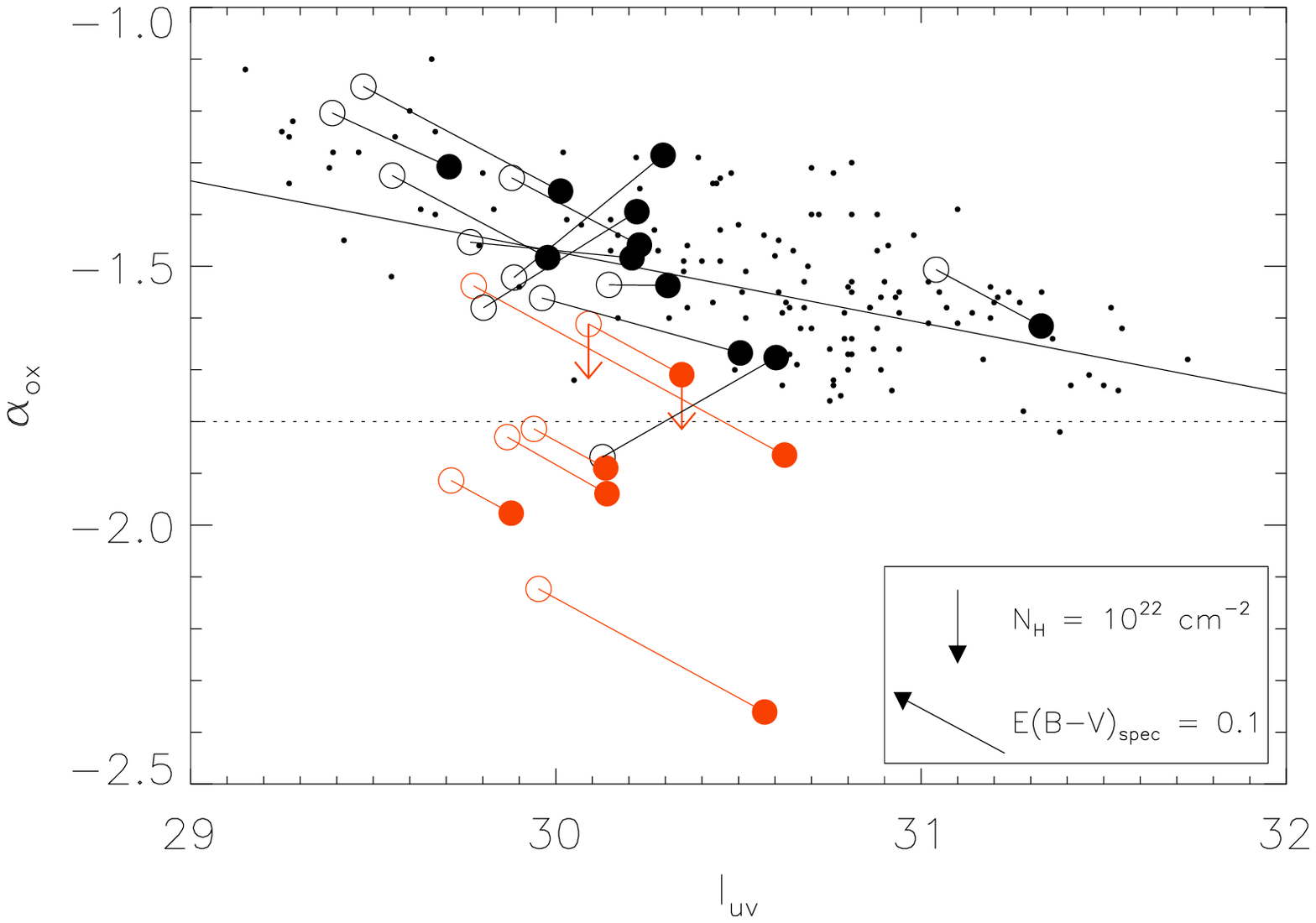}
\caption{$\alpha_{ox}$ vs. {\it l$_{uv}$} for the red quasar sample, where {\it l$_{uv}$} is the log of the
2500 $\mbox{\AA}$ luminosity in units of ergs cm$^{-2}$ s$^{-1}$ Hz$^{-1}$.  The unfilled circles give the measured 
{\it l$_{uv}$} and $\alpha_{ox}$ values for the red quasar sample.  The filled circles give {\it l$_{uv,corr}$} 
and $\alpha_{ox,corr}$, where the 2500 $\mbox{\AA}$ flux has been corrected for E(B-V)$_{spec}$ and the 2 keV flux 
has been corrected for the upper-limit on $N_H$.  The six sources with X-ray S/N $<$ 4 are corrected for 
E(B-V)$_{spec}$ only and not N$_H$; these sources are marked in red.  The S06 $\alpha_{ox} - l_{uv}$ 
correlation is plotted as a solid line and the S05 sample is overplotted in black points.  
The dotted line marks $\alpha_{ox}$ = -1.8.  Sources with $\alpha_{ox} < -1.8$ are defined to be X-ray weak.  
The legend shows the correction to $\alpha_{ox}$ and {\it l$_{uv}$} for an E(B-V)$_{spec}$ = 0.1 and the 
correction to $\alpha_{ox}$ for an N$_H$ upper-limit of 10$^{22}$ cm$^{-2}$.  Both of the corrections shown 
in the legend were calculated for the sample’s median redshift, z=1.3.}
\label{fig:aox-luv}
\end{figure}

Figure~\ref{fig:aox-luv} plots $\alpha_{ox}$ against the log of the 2500 $\mbox{\AA}$ luminosity 
({\it $l_{uv}$}) for the red sample.  The unfilled circles give the measured $\alpha_{ox}$ and 
{\it $l_{uv}$} values for the red sample.  The filled circles give $\alpha_{ox,corr}$ and 
{\it $l_{uv,corr}$}, the 2500 $\mbox{\AA}$ luminosity corrected for absorption.  Solid lines 
connect the uncorrected values to 
the corrected values for each source.  The five sources with X-ray S/N $<$ 4 are marked in red.  
The \citet{Strat05} sample and $\alpha_{ox} - l_{uv}$ correlation are plotted as black points 
and a solid line, respectively, for comparison.  The dotted line marks $\alpha_{ox}$ 
= -1.8; those sources with $\alpha_{ox} < -1.8$ are X-ray weak relative to the optical luminosity.  
All five sources with 2 $<$ S/N $<$ 4 are X-ray weak, possibly because we were unable to correct the 
X-ray flux for absorption.  All eleven sources with S/N $>$ 4 are X-ray normal once corrected for 
absorption ($\alpha_{ox,corr} > -1.8$). 

Correction for absorption can have a large effect on F$_{2keV}$, which can in turn have a large effect on 
$\alpha_{ox}$.  The largest correction in our sample, for an N$_H$ upper-limit of 10$^{23}$ cm$^{-2}$, led 
to a factor of 9.5 change in F$_{2keV}$.  An additional correction for dust-reddening changed F$_{uv}$ 
by a factor of 3, leading to $\Delta$($\alpha_{ox}$) = 0.2.  Sources with low S/N and/or high N$_H$ 
upper-limits will have a large but undetermined uncertainty in $\alpha_{ox}$, which makes it 
difficult to tell if the source is intrinsically weak relative to the optical, or if the X-ray 
weakness is due only to absorption.

There is a weak correlation between $\alpha_{ox}$ and $\Gamma$ (Spearman Rank = 0.13) such that 
X-ray weak objects are slightly more likely to have flatter spectral slopes.  

\section{DISCUSSION}

The redshift selection minimizes the chance that host galaxy emission or Ly$\alpha$ forest absorption 
produce the observed red colors of these quasars.  This 
leaves two possible causes of red colors: dust-reddening or intrinsically red optical continua.  
Table~\ref{table:classify} summarizes the source characteristics of the SDSS/XMM red quasar sample, 
which guide the classification of a source as `absorbed' or `intrinsically red.'
We discuss dust-reddened quasars in $\S$4.1 and intrinsically red quasars in $\S$4.2.  $\S$4.3 discusses 
two quasars that we were not able to classify.  We comment on the general properties 
of the red sample in terms of the $\alpha_{ox}$-{\it $l_{uv}$} relation ($\S$4.5) and 
the X-ray spectral slope ($\S$4.4).  Finally, we discuss properties of the intrinsically red quasars 
in (\S4.5).
\begin{table*}{\footnotesize
\tablewidth{0pt} 
\begin{center}
\caption{Summary of Source Characteristics \label{table:classify}}
\begin{tabular}{llccccccc}
\tableline
\tableline
ID                                             &
SDSS name                                      &
$\Gamma$                                       &
$\alpha_{ox,corr}$\,\tablenotemark{a}          &
$N_{H,x}$                                      &
E(B-V)$_{spec}$                                &
$N_{H,x}$/                                     &
Optical                                        &
Notes                                          \\
                                               &               
                                               &            
                                               & 
                                               & 
                                               & 
10$^{22}$ cm$^{-2}$                            & 
$E(B-V)_{spec}$\,\tablenotemark{b}             &               
Continuum\,\tablenotemark{c}             &    \\        
\tableline
\tableline
\multicolumn{9}{c}{Absorbed Quasars}\\
\tableline
1 & SDSS J0944+0410  & 1.9                       & $\leq$-1.69 & -                & -                   & -               & {\bf Dust} & {\bf BAL}\\
3 & SDSS J1652+3947  & 1.9                       & {\bf -1.92} & -                & 0.08$\pm$0.02       & -               & {\bf Dust} & \\
4 & SDSS J0728+3708  & 1.9                       & {\bf -1.82} & -                & {\bf 0.25$\pm$0.02} & -               & {\bf Dust} & \\
6 & SDSS J2217-0821  & 1.9                       & {\bf -2.31} & -                & {\bf 0.17$\pm$0.02} & -               & {\bf Dust} & \\
7 & SDSS J1533+3243  & 1.8$^{+0.5}_{-0.4}$       & -1.27       & {\bf $\leq$7.0}  & {\bf 0.12$\pm$0.01} & {\bf $\leq$100} & {\bf Dust} & \\
8 & SDSS J1226+0130  & {\bf 1.0$^{+0.4}_{-0.4}$} & -1.66       & {\bf $\leq$13.4} & {\bf 0.15$\pm$0.01} & {\bf $\leq$143} & {\bf Dust} & NAL\\
9 & SDSS J1435+4841  & {\bf 1.2$^{+0.3}_{-0.3}$} & -1.38       & {\bf $\leq$6.8}  & {\bf 0.13$\pm$0.01} & {\bf $\leq$100} & {\bf Dust} & NAL\\
11 & SDSS J1114+5315 & 1.5$^{+0.3}_{-0.3}$       & -1.66       & $\leq$0.9        & {\bf 0.18$\pm$0.03} & {\bf $\leq$8}   & Undefined  & \\
\tableline
\multicolumn{9}{c}{Intrinsically Red Quasars}\\
\tableline
5  & SDSS J1002+0203 & 1.9                    & -1.87 & -                & {\bf 0.052$\pm$0.005}           & -               & {\bf Red power-law} & \\
12 & SDSS J1133+4900 & 1.4$^{+0.5}_{-0.4}$    & -1.52 & {\bf $\leq$0.8}  & {\bf 0.04$\pm$0.01}             & $\leq$20        & {\bf Red power-law} & \\
13 & SDSS J1254+5649 & 1.6$^{+0.1}_{-0.1}$    & -1.30 & {\bf $\leq$0.2}  & {\bf 0.10$^{+0.11}_{-0.08}$}    & $\leq$3         & {\bf Red power-law} & \\
14 & SDSS J0959+0209 & 1.75$^{+0.09}_{-0.09}$ & -1.46 & {\bf $\leq$0.05} & {\bf 0.12$\pm$0.03}             & {\bf $\leq$0.6} & Undefined & \\
15 & SDSS J0958+0213 & 1.78$^{+0.07}_{-0.06}$ & -1.33 & {\bf $\leq$0.05} & {\bf 0.16$^{+0.09}_{-0.07}$}    & {\bf $\leq$0.5} & {\bf Red power-law} & \\
16 & SDSS J1053+5735 & 1.75$^{+0.04}_{-0.04}$ & -1.45 & {\bf $\leq$0.06} & {\bf 0.11$\pm$0.02}             & {\bf $\leq$0.9} & {\bf Red power-law} & \\
17 & SDSS J0913+5259 & 1.70$^{+0.03}_{-0.03}$ & -1.60 & {\bf $\leq$0.03} & {\bf 0.086$^{+0.004}_{-0.003}$} & {\bf $\leq$0.6} & {\bf Red power-law} & \\
\tableline
\multicolumn{9}{c}{Unclassified Quasars}\\
\tableline
2  & SDSS J0232-0731 & 1.9                 & -1.95 & -         & 0.03$\pm$0.02          & -        & Dust & \\
10 & SDSS J2217-0812 & 1.9$^{+0.4}_{-0.3}$ & -1.44 & $\leq$1.8 & 0.11$^{+0.03}_{-0.02}$ & $\leq$20 & Undefined & \\
\tableline
\end{tabular}

\tablenotetext{a}{$\alpha_{ox,corr}$ is calculated with F$_{uv,corr}$ and F$_{2keV,corr}$ so that it is corrected for 
dust-reddening and, where possible, for X-ray absorption upper-limits.}

\tablenotetext{b}{The measured gas/dust ratio relative to the Galactic value, 5.8~x~10$^{21}$~cm$^{-2}$ 
\citep{Bohlin78, Kent91}.}

\tablenotetext{c}{The optical continuum shape is defined by the relationship between the relative colors 
(\S3.4, Table~\ref{table:relcolors}).}
\end{center}

\tablecomments{For absorbed quasars, a value is bold-faced if the source matches one of the criteria for dust-
reddening in the optical/UV or absorption in the X-rays.  Criteria for dust-reddening are: a) strong 
spectroscopic reddening (E(B-V)$_{spec}$ $>$ 0.1) detected to $>3\sigma$ confidence and b) an optical 
continuum shape that indicates dust-reddening.  The criteria for X-ray absorption are: a) weak 
X-ray flux relative to the optical ($\alpha_{ox} < -1.8$), b) a flat X-ray spectral slope ($\Gamma < 1.5$ 
to greater than 90\% confidence), and c) a large gas column upper-limit (N$_H$ upper-limit $>$ 10$^{22}$ cm$^{-2}$ 
if intrinsic absorption fits are applied.  
For intrinsically red quasars, a value is bold-faced if the source matches one of the criteria of an 
intrinsically red power-law: a) reddening is mild (E(B-V)$_{spec} <$ 0.1) or detected at $< 3\sigma$, b) 
an optical continuum shape that contra-indicates dust-reddening, c) a low X-ray gas column upper-limit 
(N$_H$ upper-limits $<$ 10$^{22}$ cm$^{-2}$), and d)  gas/dust ratio smaller than the Galactic value.}
}
\end{table*}

\subsection{\it Dust-reddened Quasars with X-ray Absorption}

By the criteria shown in Table~\ref{table:classify}, 8 of 17 quasars in the sample are dust-reddened.  
E(B-V)$_{spec}$ is significant at $\geq$3$\sigma$ for all eight sources.  Seven of 8 sources have 
optical continuum shapes consistent with dust-reddened curvature.  (The exception, 
SDSS J1114+5315 (\#11), has larger errors on the relative colors, resulting in an ambiguous 
continuum shape.  However, strong E(B-V)$_{spec}$ and X-ray weakness argue in favor of 
dust-reddening.) 

SDSS J0944+0410 (\#1) is a low-ionization BAL quasar \citep{Voit93}, with Si IV, C IV, C III, 
C II] and Mg II absorption troughs of widths $\sim$2,000-5,000~km~s$^{-1}$ (Fig.~\ref{fig:SDSS spectra}).  
BALs are known to be heavily absorbed in the optical and the X-rays \citep{Green01}.  This source 
is undetected in the X-rays (upper-limit in Fig.~\ref{fig:aox-luv}).  Due to the numerous absorption 
lines and a strong FeII bump, we were not able to fit a reddening to the spectral continuum, 
so the flatness of the spectrum could be due either to line absorption, continuum absorption, 
or an intrinsically different continuum.  However, some research suggests that BALs tend to have 
redder continua than non-BALs, and dust-reddened templates appear to be a good fit to BAL 
continua \citep{YV99, Bro01, Tolea02, Reich03}.  We therefore conclude that this quasar is likely to have 
dust-reddening and X-ray absorption.  

All 8 sources match some of the criteria for X-ray absorption.  For the three sources with 
2 $<$ S/N $<$ 4 (where X-rays are detected but no X-ray spectral fits can be made), all are X-ray 
weak when the optical luminosity is corrected for dust-reddening ($\alpha_{ox,corr} <-1.8$).  These sources 
are discussed further in \S4.4.

Four of the 8 dust-reddened quasars have X-ray S/N $>$ 4, which allows us to fit a power-law (model C) 
and a power-law plus intrinsic absorption (model D).  All four sources 
prefer model C over model D, but two quasars have abnormally flat ($\Gamma$$<$1.5) 
photon indices to 90\% confidence: SDSS J1226+0130 (\#8) and SDSS J1435+4841 (\#9).  
This may be a sign of large N$_H$ columns.  In addition, the upper-limits on N$_H$ for all 
four sources are rather high (9 x 10$^{21}$ - 1 x 10$^{23}$ cm$^{-2}$, Table~\ref{table:abs fits}), and are consistent 
with the gas column predicted from E(B-V)$_{spec}$ ($N_{H,opt}$ = 7.0 x 10$^{20}$ - 1 x 10$^{21}$ cm$^{-2}$) 
using the Galactic gas-to-dust ratio \citep{Bohlin78, Kent91}.  Even the 
ten times higher N$_H$ expected from an SMC gas-to-dust ratio \citep{Issa90}, 
which is more typical for quasars \citep{Macc81, Mai01, Wilkes02}, 
is still below the N$_H$ upper-limit found in the X-rays.  

Note that the quasars classified as dust-reddened occupy the lower half of the red quasar 
sample with respect to S/N.  The low S/N for these quasars is likely due to the photoelectric absorption 
of the X-rays, as described above.  The effect of N$_H$­­ on the X-ray spectrum is 
further discussed in $\S$4.5.

\subsection{\it Intrinsically Red Quasars}

Seven sources display mild or insignificant dust-reddening: E(B-V)$_{spec}$ $\leq$ 0.1 or 
E(B-V)$_{spec}$ detected at $<3\sigma$.  Since the reddening in these sources is indistinguishable 
from a red power-law (\S3.2), we classify them as intrinsically red using two additional indicators: 
(1) the continuum shape as determined by the relative photometry (\S3.2) and (2) the lack of 
X-ray absorption (\S3.4).  

The optical continuum shape is inconsistent with  dust-reddening for all but one object.  
The relative colors for SDSS J0959+0209 (\#14) cannot distinguish between dust curvature 
and a red power-law; however, this source meets all of the other `intrinsically red' criteria.  
Six of seven sources also display no significant X-ray absorption, with low N$_H$ upper-limits of 
$\sim$ 10$^{20}$ - 10$^{21}$ cm$^{-2}$.  These same six sources are X-ray normal 
($\alpha_{ox,corr} > -1.8$), and are not abnormally X-ray flat ($\Gamma$$>$1.5 to 90\% confidence).  
It is not surprising then that these six sources have the highest X-ray S/N of the red quasar sample, 
due to the absence of X-ray absorption.  The exception is SDSS J1002+0203 (\#5), 
whose low S/N (S/N = 2.85) prohibits fitting a power-law or intrinsic 
absorption.  This source is X-ray weak ($\alpha_{ox}$ = -1.82), but the low amount of 
dust-reddening (E(B-V)$_{spec}$ = 0.052 $\pm$0.005) suggests that the X-ray weakness is 
intrinsic.  If we assume an SMC gas/dust ratio, then a reddening of 0.05 leads to an 
X-ray absorption column of N$_H$ = 3 x 10$^{21}$ cm$^{-2}$.  However, a column of less 
than 10$^{22}$ cm$^{-2}$ will not significantly affect the rest-frame 2 keV flux \citep{Tucker75}.  
  
\begin{figure}
\centering
\plotone{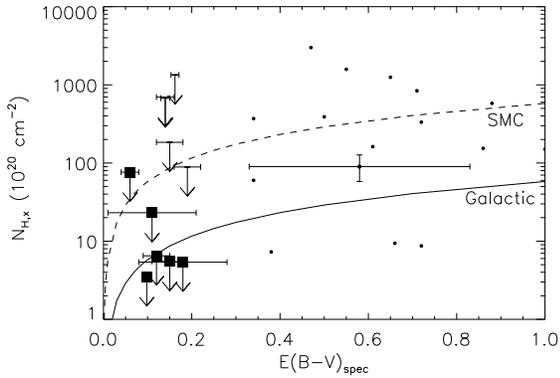}
\caption{$N_H$ (10$^{20}$ cm$^{-2}$) vs. $E(B-V)_{spec}$ for the eleven quasars with X-ray S/N $>$ 4.  
Because the $N_H$ column is a 90\% upper-limit for all of the quasars in the red sample, this results 
in an upper-limit on the gas/dust ratio.  Sources classified as ‘intrinsically red’ ($\S$4.2) are marked 
with a solid circle.  The Galactic (solid line) and SMC (dashed line) gas-to-dust ratios are plotted 
for reference.  Points from \citet{Mai01} are plotted as black points, with median error bars 
shown for one of these points.}
\label{fig:NH-E(B-V)}
\end{figure}

For the four sources with the highest X-ray S/N, SDSS J0959+0209 (\#14), SDSS J0958+0213 (\#15), SDSS 
J1053+5735 (\#16) and SDSS J0913+5259 (\#17), the low N$_H$ upper-limits imply gas-to-dust ratio 
upper-limits (Table~\ref{table:abs fits}) up to a factor of $\sim$2 smaller than the Galactic value 
\citep{Bohlin78, Kent91} and a factor of $\sim$20 below the SMC value \citep{Issa90} 
(Figure~\ref{fig:NH-E(B-V)}). The small dust-to-gas ratio found for these four objects contrasts 
with previous studies, which have 
shown that an SMC-type extinction curve and high gas-to-dust ratio is most appropriate for 
quasar absorption \citep{Macc81, Mai01, Wilkes02}.  For 
example, \citet{Mai01} collected X-ray and optical information for 19 objects from the literature.  
They find gas-to-dust ratios 3-100 times higher than Galactic for sixteen AGNs.  The three AGNS 
with gas-to-dust ratios a factor $\sim$1.5 below the Galactic value were all low luminosity AGNs 
($L_x < 10^{42}$ ergs s$^{-1}$).  For comparison, all of the red quasars have $L_x > 10^{42}$ ergs 
s$^{-1}$, with the exception of the BAL quasar (\#1).  The smaller filled circles in Fig.~\ref{fig:NH-E(B-V)} mark points 
from \citet{Mai01}.  Median error bars are displayed on one of the \citet{Mai01} points.  

The unusually low apparent gas-to-dust ratios for the four high X-ray S/N objects (\#14-17) 
suggest that dust-reddening and X-ray absorption are not applicable in these cases.  We therefore 
take the unusually low gas-to- dust ratios of these four sources to be another indicator 
of an intrinsically red continuum.  
\begin{figure}
\centering
\plotone{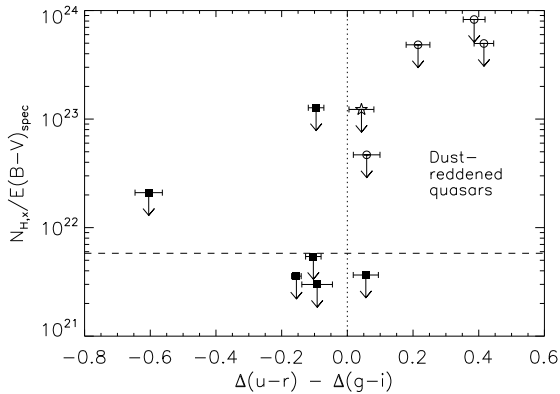}
\caption{The gas/dust ratio, N$_{H,x}$-E(B-V)$_{spec}$,  plotted against the 
$\Delta(u-r) - \Delta(g-i)$ color.  The vertical dotted line at $\Delta(u-r) - \Delta(g-i)$ 
= 0 divides the quasars with dust-reddened continuum shapes from the 
quasars with red power-law continua, as described in \S3.4.  The horizontal dashed line shows 
where the gas/dust ratio equals the Galactic value; sources below this line have unusually low 
gas/dust ratios and therefore are considered intrinsically red (\S4.2).  Sources ultimately 
classified as dust-reddened are plotted as open circles, while sources classified as 
intrinsically red are plotted as filled squares.  The unclassified source (\#10, $\S$4.3) 
is plotted as a star.}
\label{fig:NH/E(B-V)-color}
\end{figure}

The 'intrinsically red' classification is shown in Figure~\ref{fig:NH/E(B-V)-color}, where we plot the 
gas/dust ratio, N$_{H,x}$/E(B-V)$_{spec}$, against the $\Delta(u-r) - \Delta(g-i)$ color.  
The vertical dotted line at $\Delta(u-r) - \Delta(g-i)$ = 0 divides the quasars with 
dust-reddened continuum shapes ($\Delta(u-r) - \Delta(g-i) >$ 0) from the quasars with red 
power-law continua ($\Delta(u-r) - \Delta(g-i) <$ 0), as discussed in \S3.4.  The horizontal 
dashed line shows where the gas/dust ratio equals the Galactic value.  The sources in all 
but the top right quadrant are inconsistent with dust-reddening due to their low gas/dust 
ratios relative to the Galactic value and/or the lack of dust curvature in their optical 
continua.

Table~\ref{table:relcolors} shows that the dust-reddened quasars are more likely to have the 
largest color excesses, while the intrinsically red quasars are more likely to have the smallest.  
This difference is significant in a K-S test at a probability of 1.7\% that the two populations 
come from the same parent color distribution.  This suggests that, while dust-reddening can occur 
to arbitrarily high values of E(B-V), the intrinsic mechanism cannot steepen the continuum beyond 
a critical limit.

\begin{figure}
\centering
\plotone{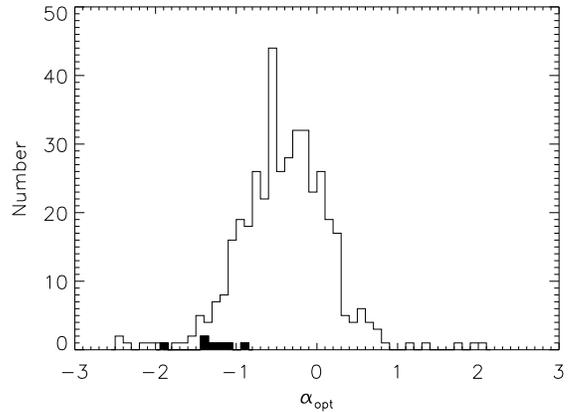}
\caption{The optical spectral indices for the Large Bright Quasar Survey \citep{Francis91}
are shown in an open histogram.  Overplotted is a solid histogram showing the optical spectral indices
for the intrinsically red quasars.}
\label{fig:aopt hist}
\end{figure}

The mean optical slope of the intrinsically red quasar candidates is $<\alpha_{opt}>$ = -1.30, 
where F$_{\nu}$ $\propto$ $\nu^{\alpha}$, with a dispersion $\sigma_{\alpha_{opt}}$ = 0.34.  
Figure~\ref{fig:aopt hist} shows a solid histogram of the optical slopes for the intrinsically red quasars, 
compared with an open histogram of the optical slopes for the Large Bright Quasar Survey 
\citep{Francis91}.  The intrinsically red quasars clearly lie in the red tail of the 
LBQS distribution.  For comparison, the power-law obtained for the Vanden Berk composite 
SDSS quasar spectrum (2001)\nocite{VdB01} between 1300 and 5000 $\mbox{\AA}$ is $\alpha_{opt}$ = -0.44.  

\subsection{\it Unclassified Quasars}

The classification of two quasars is undetermined.  SDSS J0232-0731 (\#2) is X-ray weak 
($\alpha_{ox}$ = -1.95), but the signal is too weak to fit $\Gamma$ or intrinsic absorption 
(S/N = 3.23).  Dust-reddening is mild, and is significant only at 1.5$\sigma$ (E(B-V)$_{spec}$ = 
0.03 $\pm$0.02).  However, the optical continuum shape is consistent with dust-type curvature 
(Table~\ref{table:relcolors}). 

SDSS J2217-0812 (\#10) is also ambiguous because, while it has normal $\Gamma$ and 
$\alpha_{ox,corr}$ values ($\Gamma$ = 1.9$^{+0.4}_{-0.3}$, $\alpha_{ox,corr}$ = -1.44), the upper-limit 
on N$_H$ is fairly high at 1.8 x 10$^{22}$ cm$^{-2}$, and the gas-to-dust ratio upper-limit is between 
the Galactic and SMC values.  The optical continuum shape is consistent with both dust-reddened 
curvature and a red power-law within the photometric error bars on the relative colors.  
While absorption is not specifically indicated, it also cannot be ruled out.

\subsection{\it Red Quasars and the $\alpha_{ox}$-{\it $l_{uv}$} relation}

Several studies \citep{AT82, Vign03, Strat05, Steff06}, have investigated an anticorrelation between 
$\alpha_{ox}$ and {\it $l_{uv}$} 
(the log of the 2500 $\mbox{\AA}$ luminosity) in radio-quiet quasars.  Figure~\ref{fig:aox-luv} 
(described in $\S$3.2) shows that the majority of the sample is consistent with 
the Strateva et al. (2005, hereafter S05)\nocite{Strat05} $\alpha_{ox}$ - {\it $l_{uv}$} sample.  
Five sources, though, are X-ray weak, even after a correction for E(B-V)$_{spec}$ is applied, 
and appear inconsistent with the S05 $\alpha_{ox}$ - {\it $l_{uv}$}­ sample.  
All five have S/N$<$4, so no correction can be made for N$_H$.

We have made a statistical comparison between the red quasar sample and the S05 sample.  The 
S05 sample was selected from the SDSS using Data Release 2 photometry, so aside from updates 
to the SDSS quasar selection process between DR2 and DR3, the S05 quasar selection is the same 
as that of the red quasar sample, except for the red color cut.  We simplified the comparison 
of the two samples by noting that the red sample luminosity is spread over a single decade, 
and for S05, $\alpha_{ox}$ changes by only 0.16 in this luminosity range.  We thus restricted 
the luminosity range of both samples so that we could apply the one-dimensional Kolmogorov-Smirnov 
(K-S) test.  The observed luminosity range of both the red quasar and the S05 samples are restricted 
to {\it $l_{uv}$} = 29.4-30.3 when comparing the uncorrected red quasar sample (Fig.~\ref{fig:aox-luv}, 
open circles) to the S05 sample.  Since correcting $\alpha_{ox}$ for dust-reddening shifts the UV 
luminosity to a brighter interval, we restrict the luminosity range to {\it $l_{uv}$} = 29.7-30.8 when 
comparing the corrected red quasar sample (Fig.~\ref{fig:aox-luv}, filled circles) to the S05 sample.  
This excludes only one member of the red quasar sample (\#17).

First we compare the complete red quasar sample (excluding the high $l_{uv}$ source \#17) 
with the S05 sample.  We find a K-S probability of 0.02\% that the two samples come from 
the same parent population.  Next, we consider only the 11 sources for the K-S test 
where we were able to correct for X-ray absorption (i.e. S/N $>$ 4, black circles in Fig.~\ref{fig:aox-luv}), 
so that we can compare the uncorrected and 
corrected $\alpha_{ox}$- {\it $l_{uv}$} distributions to the S05 sample.  Comparing 
first the uncorrected red quasar sample to the S05 sample, we find a probability of 10\% 
that the two samples come from the same parent population.  Then we compare the corrected 
red quasar sample to the S05 sample, and we find a a K-S probability of 76\%.  

Six of 11 quasars with S/N$>$4 are intrinsically red, so this suggests that intrinsically red 
quasars are not intrinsically X-ray weak compared to typical quasars, with the exception of 
SDSS J1002+0203 (\#5).  However, their peculiar SEDs make the physical interpretation unclear.

For sources with S/N$<$4, it is plausible that correcting for absorption 
would significantly increase the K-S probability if the low S/N is due to 
large N$_H$ columns.  We find that if we apply a uniform correction to all 
of the low S/N quasars, an absorption column of N$_{­­H}$ $>$ 10$^{23}$ 
cm$^{-2}$ is required to achieve a K-S probability of at least 1\% that the 
red quasar sample is drawn from the same parent population as the S05 sample.

\subsection{\it X-ray Power-law Index}

\begin{figure}
\centering
\plotone{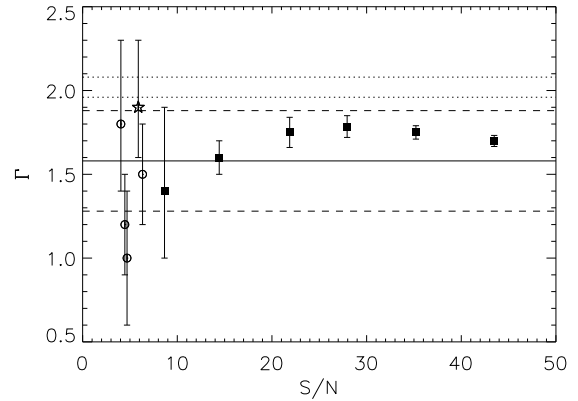}
\caption{Best-fit X-ray spectral slope ($\Gamma$) vs. signal-to-noise.  Objects classified as ‘intrinsically 
red’ are marked as filled squares, while the dust-reddened objects are marked as open circles.  
The unclassified source (\#10, $\S$4.3) is plotted as a star.
The average $\Gamma$ and the intrinsic dispersion for the red sample are plotted as solid and dashed 
lines, respectively.  The average $\Gamma$ for the \citet{Mateos05} sample ($<\Gamma>$ = 1.86) 
and the \citet{Main07} sample ($<\Gamma>$ = 2.08) are shown as dotted lines.}
\label{fig:Gamma-SNR}
\end{figure}

The mean X-ray photon index for sources with S/N $>$ 4 is $<\Gamma>$ = 1.58$\pm$0.09, with an 
intrinsic dispersion $\sigma_\Gamma$ = 0.3.  Figure~\ref{fig:Gamma-SNR} plots $\Gamma$ vs. S/N for each source 
with S/N $>$ 4, with the mean $\Gamma$ plotted as a solid line for reference, and the intrinsic 
dispersion plotted as dashed lines.  For sources with multiple observations, the photon 
indices are averaged and the error plotted is the rms of the measured errors.  

The mean $\Gamma$ in this sample is flatter than that of other samples of quasar X-ray spectra 
including XMM-Newton samples.  
For example, \citet{Mateos05} found a weighted mean $<\Gamma>$ = 1.96$\pm$0.01 with 
an intrinsic dispersion $\sigma_\Gamma$ = 0.4 for 1137 serendipitously-selected XMM-Newton AGN.  
An investigation of 86 bright AGN 
detected in the XMM-Newton COSMOS field \citep{Main07} gives $<\Gamma>$ = 2.08$\pm$0.08, with 
an intrinsic dispersion $\sigma_\Gamma$ = 0.24.  The two surveys are consistent within errors.
These mean values are plotted as dotted lines for reference in Figure~\ref{fig:Gamma-SNR}.  

The measurement of N$_H$­ can affect the measurement of $\Gamma$.  The contour plots in 
Figure~\ref{fig:Xray fits} show that the two variables are correlated so that the same spectrum may be fit 
with a large N$_H$ column and steep $\Gamma$, or with a smaller N$_H$ column and flatter $\Gamma$.  
Therefore, the flatter mean $\Gamma$ may be due to the effect of N$_H$ on the X-ray spectra with 
lower S/N.  Because the sources with higher X-ray S/N have tighter upper-limits, the $\Gamma$-N$_H$ 
correlation is less noticeable for these sources (e.g. Fig.~\ref{fig:Xray fits}, \#14, 15, 16, 17).  

Considering only the the intrinsically red group (squares in Fig.~\ref{fig:Gamma-SNR}), excluding \#5 due its 
low X-ray S/N, we find a mean $\Gamma$~=~1.66$\pm$0.08, with an intrinsic dispersion of only 
$\sigma_\Gamma$~=~0.06.  This tight intrinsic dispersion shows that most of the variance in the 
intrinsically red group comes from measurement errors.  Since the contour plots in Figure~\ref{fig:Xray fits} 
show that $\Gamma$ and N$_H$ are not strongly correlated for any of the intrinsically red quasars, 
the flatter X-ray spectra cannot be explained by absorption.

A significantly flatter $\Gamma$ may be explained via relation to other quantities.  
The X-ray spectral slope of normal quasars does not depend on optical luminosity or redshift 
\citep{Shem05, Vign05}, but does depend slightly on X-ray luminosity 
\citep{Dai04, Saez08}.  However, the $\Gamma$-L$_x$ correlation contains 
too much intrinsic scatter (e.g. Figure 8 of Saez et al. 2008)\nocite{Saez08} to make conclusive statements 
about the intrinsically red quasars.  The $\Gamma$-L$_x$ corrrelation is not significant 
in the red quasar sample.

\begin{figure}
\centering
\plotone{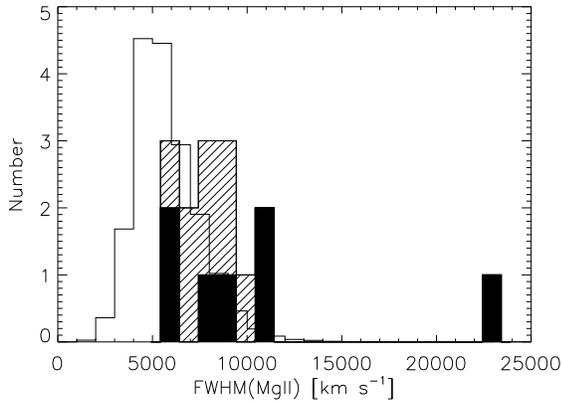}
\caption{The histogram shows the MgII line widths (FWHM) of the full sample in \citet{Shen07} 
(open histogram), the dust-reddened red quasars (line histogram), and the intrinsically red quasars 
(filled histogram).}
\label{fig:MgII hist}
\end{figure}

\begin{figure}
\centering
\plotone{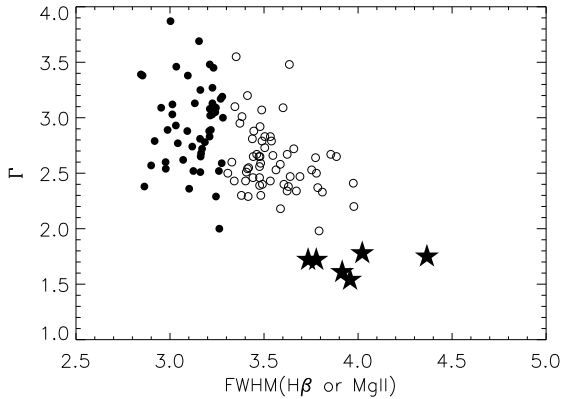}
\caption{The well-known FWHM(H$\beta$-$\Gamma$ relation is shown, using the \citet{Grupe04} 
sample of Seyferts for reference.  Broad-line Seyfert 1's are plotted as open circles and 
narrow-line Seyfert 1's are plotted as filled circles.  For the red quasar sample, the width of 
the MgII line is plotted instead of the H$\beta$ due to its availability and similarity to H$\beta$ 
(\S4.4).  The intrinsically red quasars are plotted as filled stars.  $\Gamma$ is not well-determined 
for the absorbed quasars due to the high upper-limits on N$_H$, so we do not plot them on this plot.}
\label{fig:FWHM-Gamma}
\end{figure}

The strongest correlation reported between X-ray parameters and other variables is between $\Gamma$ 
and the H$\beta$ emission-line FWHM, where 
flatter X-ray spectral slopes correlate with broader emission lines \citep{Laor97, Shem06}.  
The H$\beta$ emission line width is in turn believed to be  
anti-correlated with the accretion rate \citep{BG92,BB98}.
Narrow Line Seyfert 1's (NLS1) lie on one end of the $\Gamma$-H$\beta$ correlation, with steep X-ray
spectral slopes, narrow H$\beta$ lines \citep{BB98,Grupe04}, and accretion rates close to the 
Eddington rate \citep{Kom07}.  While the H$\beta$ line is not visible in the optical spectra for 
the red quasar sample, the broad MgII line is visible in all of the intrinsically red spectra and 
has an ionization potential similar to H$\beta$, and a FWHM that correlates well with FWHM(H$\beta$) 
\citep{MJ02}.  The MgII line widths of all the red quasars are broader than average 
(Figure~\ref{fig:MgII hist}).  This may be due to our color and redshift selection criteria, 
which biases the red quasar sample towards objects with strong 
MgII lines.  However, the unabsorbed, flat X-ray spectral slopes combined with 
the broad MgII lines put the intrinsically red quasars on the correlation at the opposite end 
from NLS1's (Figure~\ref{fig:FWHM-Gamma})\footnote{We do not plot the absorbed quasars on this plot because, 
due to the high N$_H$ upper-limits, $\Gamma$ is not well-determined for these objects.}.  
NLS1's are believed to have accretion rates close 
to the Eddington rate because of their narrow H$\beta$ lines and steep X-ray spectra \citep{Kom07}.  
Steep red optical/UV continua seem to be an effective means of selecting this extreme population.  

\subsection{\it Properties of Intrinsically Red Quasars}
In this section, we investigate the basic properties of the intrinsically red quasars: 
their black hole masses (M$_{BH}$) and accretion rates ($\dot{M}$).  These values, and the 
values from which they are derived, are listed in Table~\ref{table:mdot}. 
The black hole masses are calculated by assuming virial 
motion of the broad-line clouds around the central black hole, an assumption justified 
from reverbration mapping \citep{PW00,OP02}.  The FWHM of the broad MgII line can then 
be used to determine the black hole mass using the following equation from \citet{MJ02}: 
\begin{displaymath} \frac{M_{BH}}{M_{\odot}} = 3.37 \left(\frac{\lambda L_{3000}}{10^{37} 
W}\right)^{0.47} \left[\frac{FWHM(MgII)} {km~s^{-1}}\right]^2 \end{displaymath}  
\begin{table*}
\begin{center}
\caption{Black hole masses and accretion rates of intrinsically red quasars}
\label{table:mdot}
\begin{tabular}{llccccc}
\tableline
\tableline
ID & SDSS name & FWHM(MgII)                 & M$_{BH}$                 & L$_{bol}$                              & $\dot{M}$\tablenotemark{a}                & $\dot{M}/\dot{M_{Edd}}$ \\
   &           & \footnotesize{km s$^{-1}$} & \footnotesize{10$^9$ M$_\odot$} & \footnotesize{10$^{45}$ ergs s$^{-1}$} & \footnotesize{M$_\odot$ yr$^{-1}$} &                        \\

\tableline
5  & SDSS J1002+0203 & 10,800 & 9.1  & 2.5 & 0.5 & 0.002 \\
12 & SDSS J1133+4900 & 9,100  & 3.3  & 4.2 & 0.7 & 0.010 \\
13 & SDSS J1254+5649 & 8,200  & 0.68 & 1.6 & 0.3 & 0.019 \\
14 & SDSS J0959+0209 & 23,200 & 9.9  & 1.6 & 0.3 & 0.001 \\
15 & SDSS J0958+0213 & 10,500 & 1.3  & 2.5 & 0.5 & 0.015 \\
16 & SDSS J1053+5735 & 5,400  & 0.76 & 3.4 & 0.6 & 0.035 \\
17 & SDSS J0913+5259 & 6,000  & 3.9  & 2.6 & 0.5 & 0.005 \\
\tableline
\end{tabular}
\end{center}
\tablenotetext{a}{The accretion rate calculated from L$_{bol} = \eta \dot{M}$ c$^2$, 
where $\eta$ = 0.1.}


\end{table*}

The luminosity at 3000 $\mbox{\AA}$ (L$_{3000}$) gives the radius of the BLR as determined by the R$_{BLR}$-L 
relation obtained from reverberation studies \citep[e.g., ][]{Kaspi00,Bentz06}.  This relation 
was derived for a normal BBB-dominated SED.  Since the BBB appears to be missing from the 
intrinsically red quasars, the ionizing continuum may be lower than normal; this would result in 
an overestimate of the black hole mass.  \citet{Shang05} show that the BBB peaks at 
$\sim$1100 $\mbox{\AA}$, so $\alpha_{ox}$ is a good estimate of the ionizing continuum.  
Since the intrinsically red quasars have $\alpha_{ox}$ values within the normal range 
(cf. Steffen et al. 2006)\nocite{Steff06}, we can assume that the R$_{BLR}$-L relation is 
still valid for the intrinsically red quasars.

The apparent black hole masses for the intrinsically red quasars are relatively high, ranging 
from $\sim$10$^9$ to $\sim$10$^{10}$~$M_\odot$ (cf. Shen et al. 2007)\nocite{Shen07}.  

We obtain the accretion rate from the bolometric luminosity: L$_{bol}$ = $\eta \dot{M}$c$^2$.
We assume an efficiency $\eta$ = 10\%.  Standard bolometric correction factors 
\citep[e.g., ][]{Elvis94, Kaspi00} 
are not reliable for the intrinsically red quasars with their atypical SEDs.  
Instead, we integrated the optical power-law ($\alpha_{opt}$) from 2,500 to 10,000~$\mbox{\AA}$, 
the nominal UV-to-X-ray power-law ($\alpha_{ox}$) from 2,500~$\mbox{\AA}$ to 2~keV, 
and the X-ray power-law ($\Gamma$) from 2-10 keV to give a rough approximation of 
the bolometric luminosity.  Changing the break point between the two power-laws from 
2500 $\mbox{\AA}$ to, e.g. 1150 $\mbox{\AA}$ \citep{Shang05}, or even to 1000 $\mbox{\AA}$, 
has a small effect on L$_{bol}$.  Since object \#17 is a lensed quasar, we reduce its 
luminosity by the modeled magnification factor, M=15 \citep{Chartas00}. 
Although we have no mid-IR data for these quasars, IR emission accounts for $\sim$40\% 
of the bolometric luminosity for a typical quasar SED \citep{Rich06}.  Therefore, if 
the intrinsically red quasars are similar, the IR emission will not change the total 
luminosity, and so the accretion rate, by more than a factor of a few.  To constrain 
this estimate, we searched the IRAS Point Source Catalog and the Faint Source Catalog 
(FSC) for the red quasars, but found no matches.  Sources in the 
FSC\footnote{http://irsa.ipac.caltech.edu/IRASdocs/surveys/fsc.html} 
have flux densities greater than 200 mJy at 60 $\mu$m, which gives an upper-limit to 
$\lambda$L$_\lambda$(60 $\mu$m)/L$_{bol} \lesssim$ 7.0.  This constrains the IR 
emission to be less than an order of magnitude of the bolometric luminosity.  
The estimated bolometric luminosities give a lower bolometric correction compared to previous work.  
L$_{bol}$/$\lambda$L$_{3000}$ = 2.6 $\pm$ 1.6 for the intrinsically red sample, 
a factor 2.4 lower than that obtained by \citet{Elvis94} (6.3 $\pm$ 3.1).
\begin{figure}
\centering
\plotone{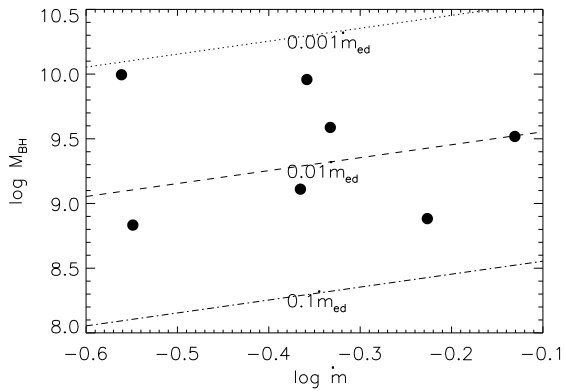}
\caption{Black hole mass (M$_\odot$) is plotted against the accretion rate (M$_\odot$ yr$^{-1}$).  
The Eddington accretion rate is plotted for a reference as a solid line, while correspondingly smaller
fractions of the Eddington luminosity are plotted as dash-dotted, dashed, and dotted lines respectively, 
as labeled in the plot.}
\label{fig:MBH-mdot}
\end{figure}

Using L$_{bol}$, we find that the intrinsically red quasars have relatively low 
accretion rates (Figure~\ref{fig:MBH-mdot}).  The intrinsically red quasars lie 
clustered between $\sim$0.001-0.03~$\dot{M_{edd}}$.  In comparison, typical 
accretion rates for optically or UV-selected quasars range from $\dot{M}/\dot{M_{edd}} 
\sim$ 0.03 - 10.0, with a median value of $\sim$0.6 \citep{WHD04,Bonning07}.  
The low accretion rates of the intrinsically red quasars are in agreement with their 
position on the $\Gamma$-FWHM(MgII) correlation (Fig.~\ref{fig:FWHM-Gamma}, \S4.5)

\section{INTRINSICALLY RED QUASARS: PHYSICAL MODELS}

We have found a substantial subset of red quasars whose red colors are probably due 
to intrinsically red optical emission in the optical/UV rather than dust-reddening.  
This subset comprises a large fraction (7/17 $\sim$ 40\%) of SDSS quasars with ($g - r$) 
$\geq$ 0.5.  Note that this is based on color selection; selection via relative colors 
(e.g. $\Delta$($g - i$) $>$ 0.35) would likely change the frequency of intrinsically red 
quasars because dust-reddened quasars have a redder color distribution than 
intrinsically red quasars (\S4.2).  For redshifts z $\sim$ 1 - 1.5, the Small Blue Bump reddens 
the observed $g - r$ colors resulting in similar samples, whether selected by color or color excess.  
However, for redshifts z $\sim$ 1.5 - 2, the $g - r$ color decreases; for this range, selection via color 
excess would increase the number of quasars with smaller color excesses, thereby possibly increasing 
the fraction of intrinsically red quasars.  

Quasars with intrinsically red SEDs have been discussed before (e.g., Risaliti et al. 2003; 
Hall et al. 2006, see \S1)\nocite{Risaliti03, Hall06}.  For example, the intrinsically red 
quasars found by \citet{Hall06} 
do not show absorption in the optical or X-ray continua, yet they also have steep 
optical slopes ($<\alpha_{opt}>$ = -1.0, $\sigma_{\alpha_{opt}}$ = 0.2), determined via 
fits to the SDSS spectra.    

Intrinsically steep optical power-laws suggest that these quasars have a different continuum 
emission mechanism from ordinary quasars.  We consider 3 possibilities to explain the red 
optical power-laws: (1) strong synchrotron emission visible in the optical, (2) a high-temperature BBB, 
which exposes the red power-law that may underlie typical optical emission, or (3) a low-temperature BBB resulting 
from a low accretion rate, such that the steep, high-energy tail of the BBB is visible in the optical.  

\subsection{\it Synchrotron Emission}

Synchrotron radiation could result in a red optical power-law.  However, typical
synchrotron emission can be discounted for two reasons.  
First, synchrotron radiation is strongest in the radio, with a turnover frequency at 
$\nu_m$ = 10$^{11}$ Hz \citep{Marscher95}.  Therefore, we would expect strong radio emission
in all of the intrinsically red quasars, when in fact only one can be classified as radio-loud 
(Table~\ref{table:sample}).  Second, even if a higher than normal turnover frequency allowed the synchrotron 
emission to dominate in the optical, the superposition of synchrotron emission on a BBB would 
result in a `U'-shaped spectrum (Fig. 2 of Francis et al. 2000)\nocite{Francis00}.  
This is not observed in any of the optical spectra.  

\subsection{\it Exposed Underlying Power-law}

A red power-law is sometimes 
proposed to underlie the BBB in other quasars \citep{MS82, Ward87, Law05}.  
If the BBB were removed or diminished, this underlying power-law would be exposed.  One way of 
``removing'' the BBB is to shift it to higher temperatures.  
Possible examples of high-temperature BBB's exposing 
a red, underlying optical power-law are known: the narrow-line Seyfert 1 galaxies RE J1034+396 
\citep{Puch95a} and RE J2248-511 \citep{Puch95b}.  In these objects, 
the optical power-law slopes are -1.3 and -0.9, respectively.  RE J1034+396 is an 
EUV-selected source with an ultrasoft X-ray excess that peaks near 0.4 keV.  The effective 
maximum temperature of the accretion disk is T $\sim$ 10$^6$ K (log $\nu_{max}\sim$16.8), 
more than an order of magnitude hotter than typical quasar accretion disk temperatures 
\citep[T $\sim$ 10$^{4.5}$ K, log $\nu_{max}\sim$15.3, ][]{MS82}.  
The optical spectrum of this source shows no signs of a BBB.  
RE J2248-511 also has a high-energy turnover at around 0.25 keV, but a blackbody is a 
poor fit to the X-ray data.  

The presence of an underlying power-law may be explained by weak synchrotron emission from an 
associated jet emitting at a large angle to the line of sight.  This synchrotron emission would 
necessarily be weak for a radio-quiet object, which is why the BBB must be removed or otherwise 
modified in order for the red power-law to be visible in the optical. Alternatively, unsaturated 
Comptonization of a seed spectrum can produce a power-law over seven decades in frequency 
\citep[e.g., ][]{Mar82}.  

Assuming a simple model of the accretion disk as a sum of blackbodies, \citet{Law05} showed 
that quasar SEDs scale homologously with temperature, such that hotter AGN have thermal 
peaks at higher frequencies.  If the local temperature of the disk is due to the release 
of the gravitational binding energy, then the temperature of the accretion disk scales 
homologously with the accretion rate.  Since the M$_{BH}$ determines the inner radius of the 
accretion disk, scaling with black hole mass is model-dependent and non-homologous; however, 
simple models show that a less massive black hole will have a hotter maximum effective disk 
temperature \citep{Law05}.  Therefore, high accretion rates and/or low black hole masses could 
shift the BBB to higher temperatures/energies.  

Neither high accretion rates nor low black hole masses are observed in the intrinsically red 
quasars (\S4.6),  but since the SDSS/XMM red quasars have relatively high redshifts (z $\sim$ 1-2), 
a soft X-ray turnover similar to that found in the \citet{Puch95a, Puch95b} objects would be shifted to 
$\sim$0.4 keV/(1+z) = 0.1-0.2 keV.  This turnover would lie below the observed XMM-EPIC 
energy band (0.5-10 keV).  A hidden BBB in the EUV or soft X-rays would increase the 
bolometric luminosity, thereby increasing the accretion rates.  However, an increase in the ionizing 
continuum would mean that the black hole mass calculations are underestimates.  Since the effective 
maximum disk temperature goes roughly as T $\propto$ ($\dot{M}^{1/4}$M$_{BH}^{-1/2}$), a NLS1 
($<$M$_{BH}>$ $\sim$ 10$^6-10^7$ M$_{\odot}$, Grupe \& Mathur 2004)\nocite{GM04} can 
achieve an effective maximum disk temperature of 10$^6$ K at normal accretion rates, but 
highly super-Eddington conditions (L/L$_{Edd} \sim 10^3$) are required for a quasar 
with M$_{BH} \sim 10^9 $M$_{\odot}$ to reach the same temperature.  If the black hole masses 
are underestimates, then even higher accretion rates are required.  
Therefore, a high-temperature BBB is unlikely to explain the intrinsically red quasars.  


\subsection{\it The Tail of a Low-Temperature Big Blue Bump}

Low disk temperatures due to low accretion rates and/or large black hole masses should shift 
the peak of the Big Blue Bump to longer, near-infrared wavelengths (Frank, King \& Raine 2002, p. 90) 
\nocite{FKR}.  Our bolometric 
luminosity estimates for the intrinsically red quasars suggest accretion rates 
$\sim$10-100 times lower than typical quasars (Fig.~\ref{fig:MBH-mdot}, \S4.6).  
The low accretion rate hypothesis is supported by the uniformly flat X-ray spectral slopes 
and broad MgII lines of the intrinsically red quasars (Fig.~\ref{fig:FWHM-Gamma}, \S4.5).   
We can determine the temperature of the thermal peak expected for the calculated accretion 
rate using the scaling relation developed by \citet{Law05}.  A typical quasar has an 
effective maximum temperature $\sim$10$^{4.8}$ K \citep{MS82}; this corresponds to a BBB peak at 
log $\nu_{max}~\sim$~15.6.  
Assuming a conservative accretion rate of 0.1$\dot{M_{Edd}}$ for the \citet{Elvis94} SED, 
an intrinsically red quasar accreting at a rate lower by a factor 10 will be a factor of 
(10)$^{1/4}\sim$1.8 lower in temperature, so the BBB will peak at log $\nu_{max}$~$\sim$~15.3.  
The most extreme accretion rate (0.001$\dot{M_{Edd}}$, \#14) gives a BBB peak at log $\nu_{max}$~$\sim$~15.1 



\begin{figure*}
\centering
\includegraphics[width=3.1in]{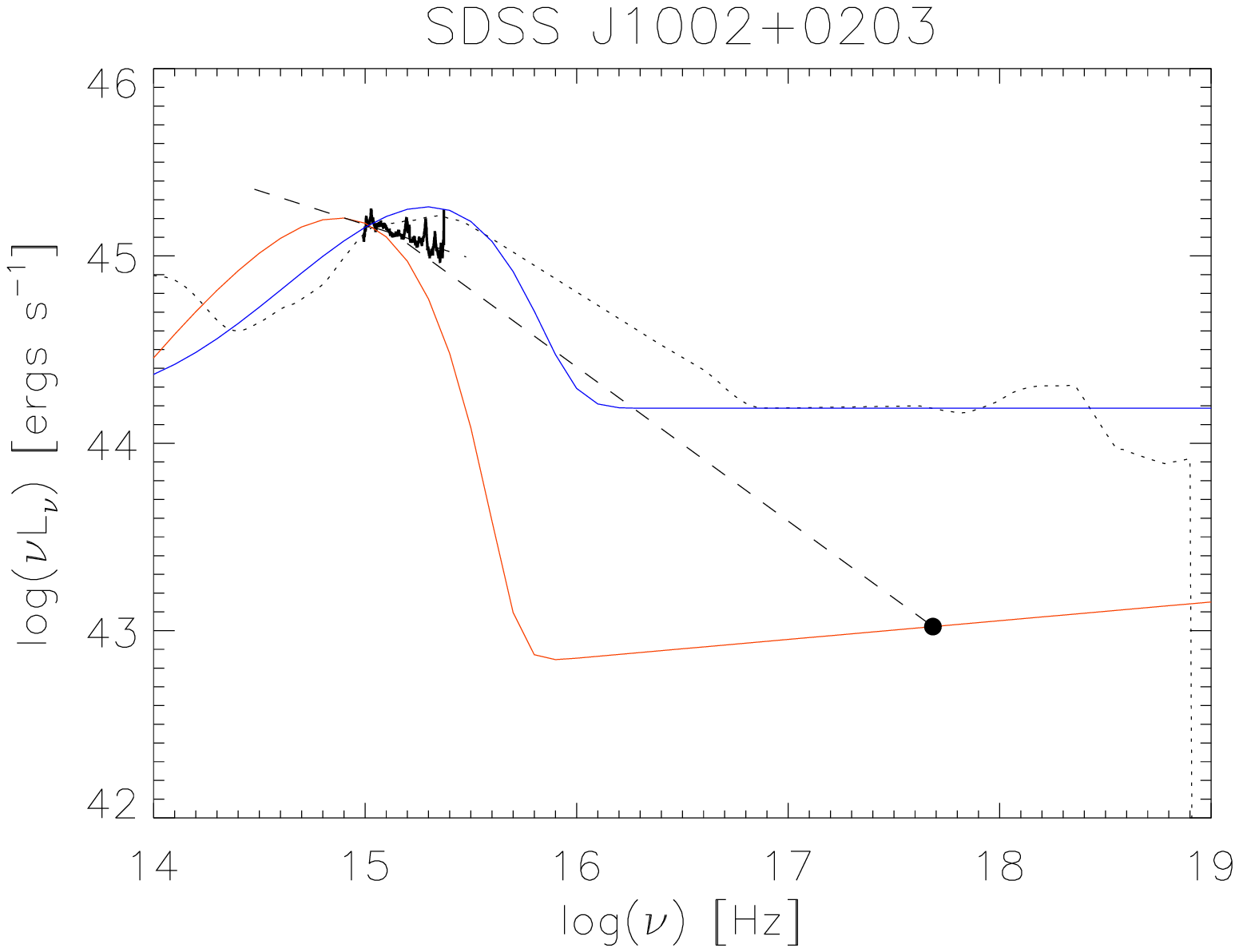}
\includegraphics[width=3.1in]{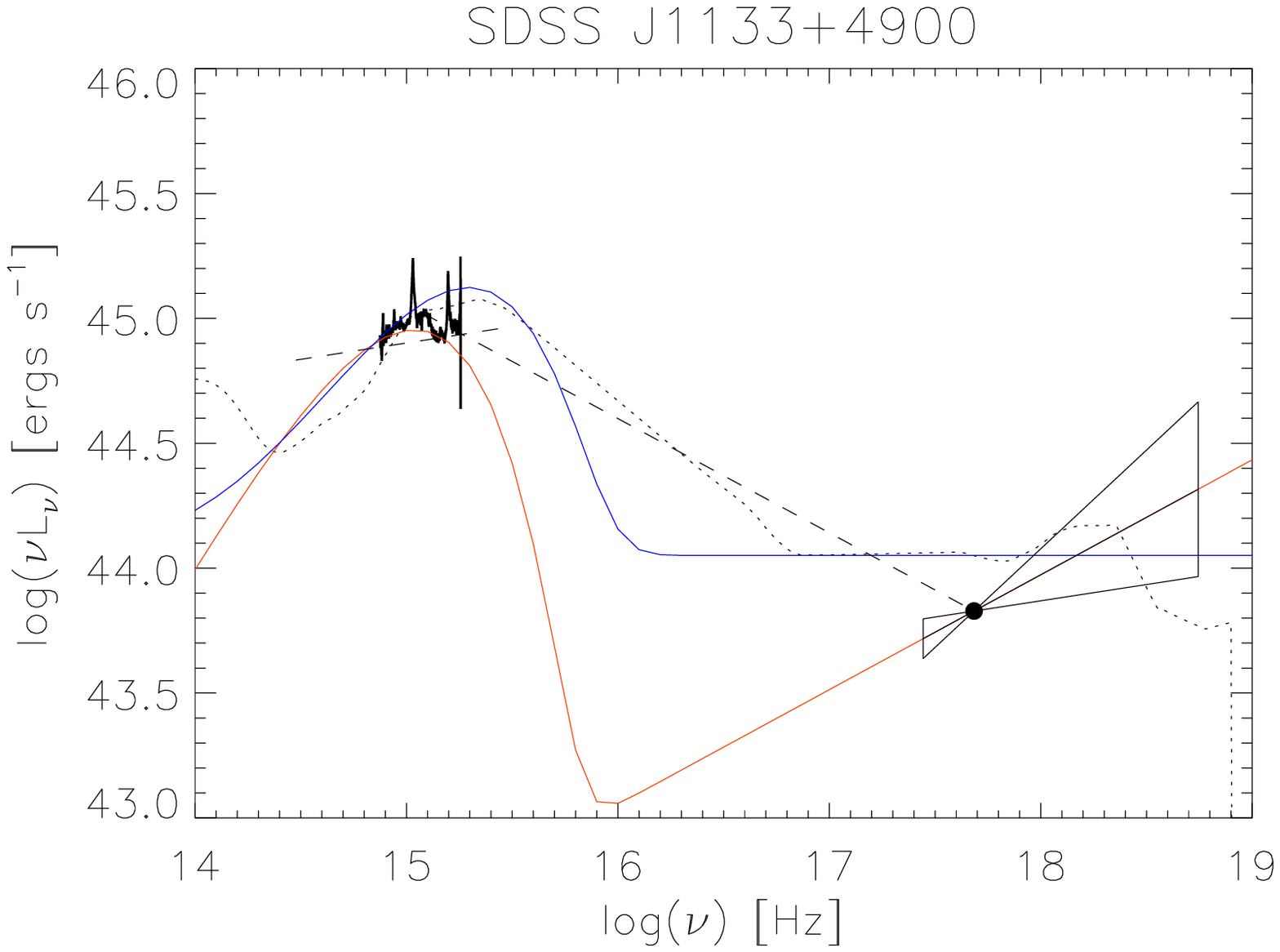}
\includegraphics[width=3.1in]{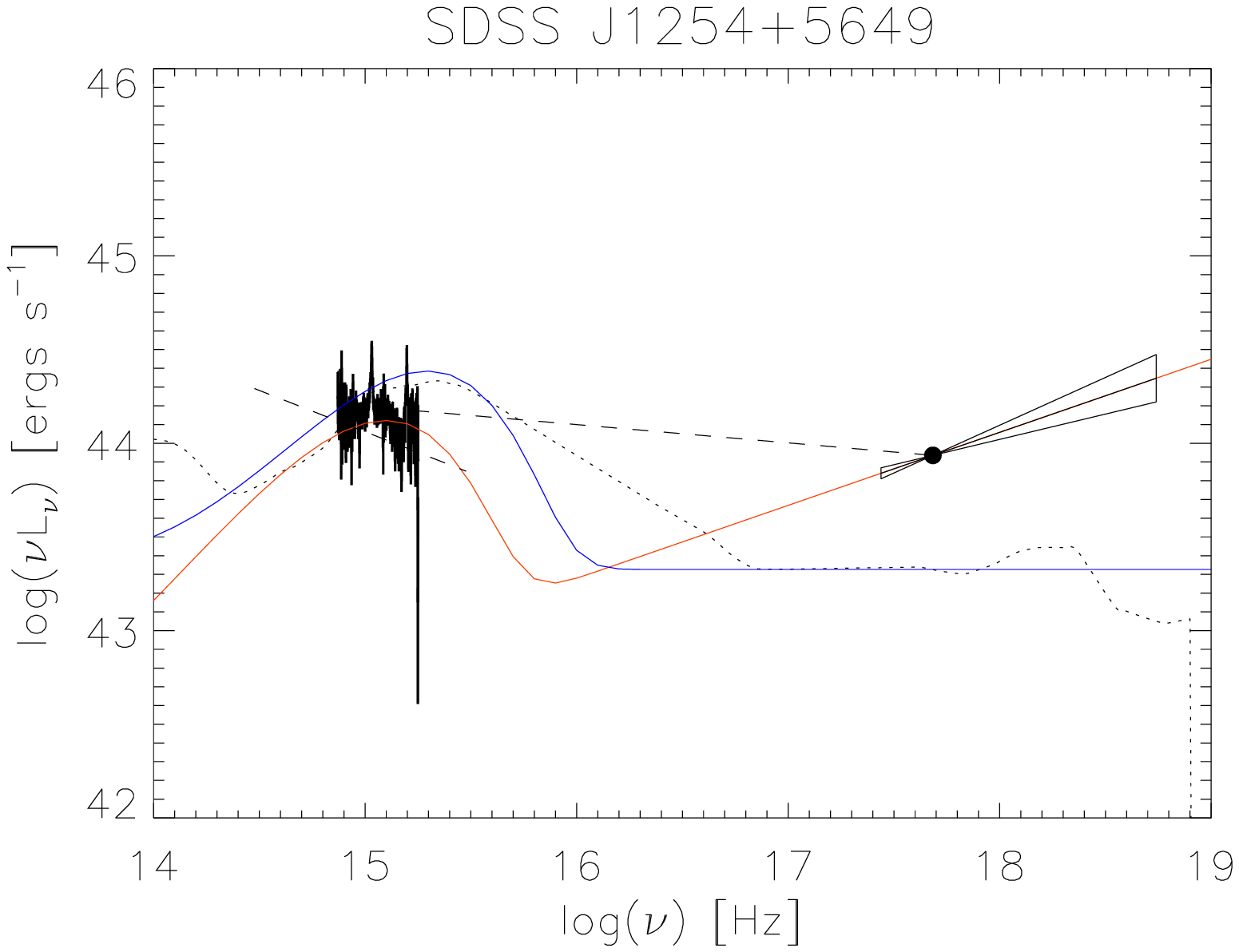}
\includegraphics[width=3.1in]{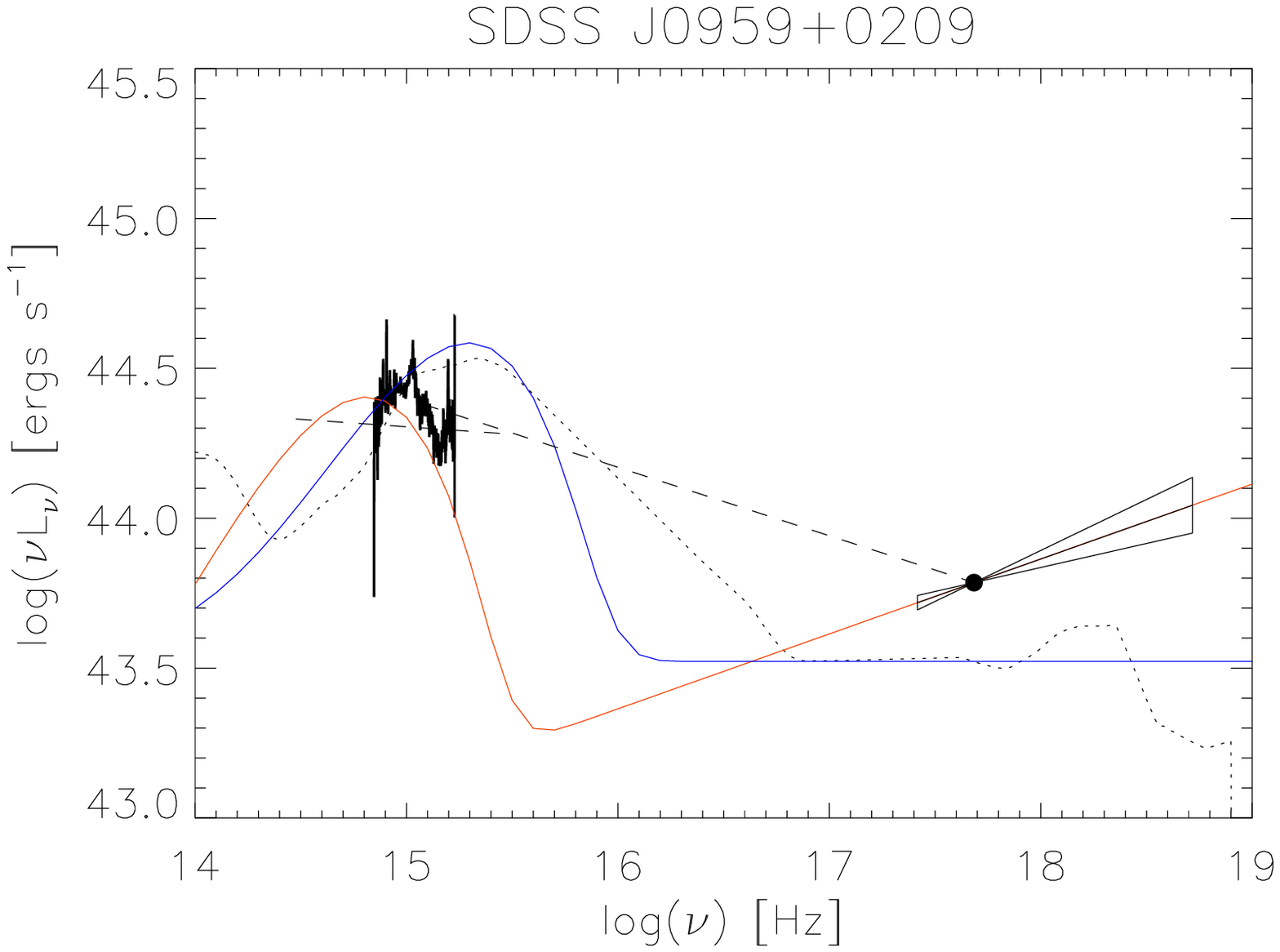}
\includegraphics[width=3.1in]{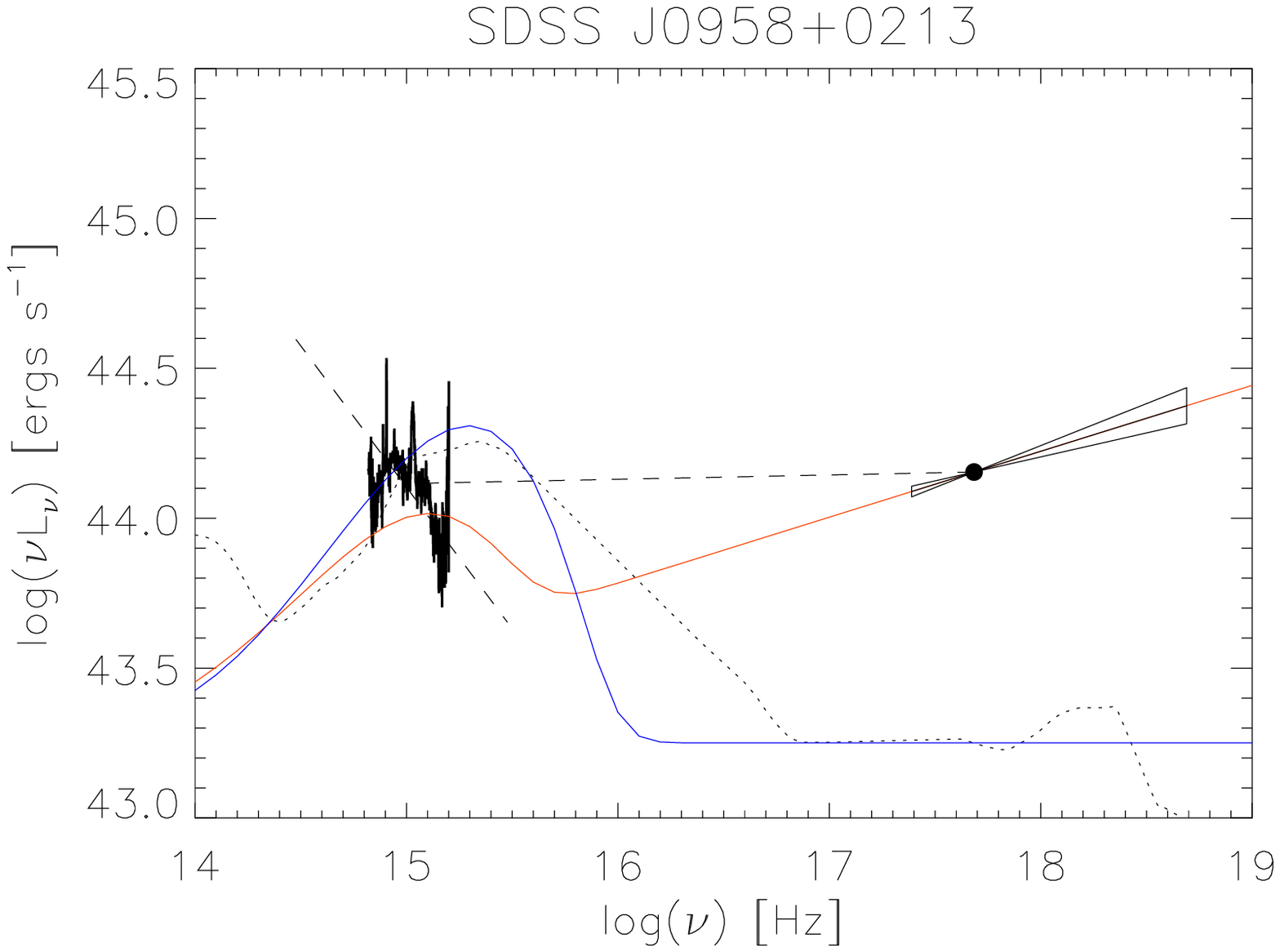}
\includegraphics[width=3.1in]{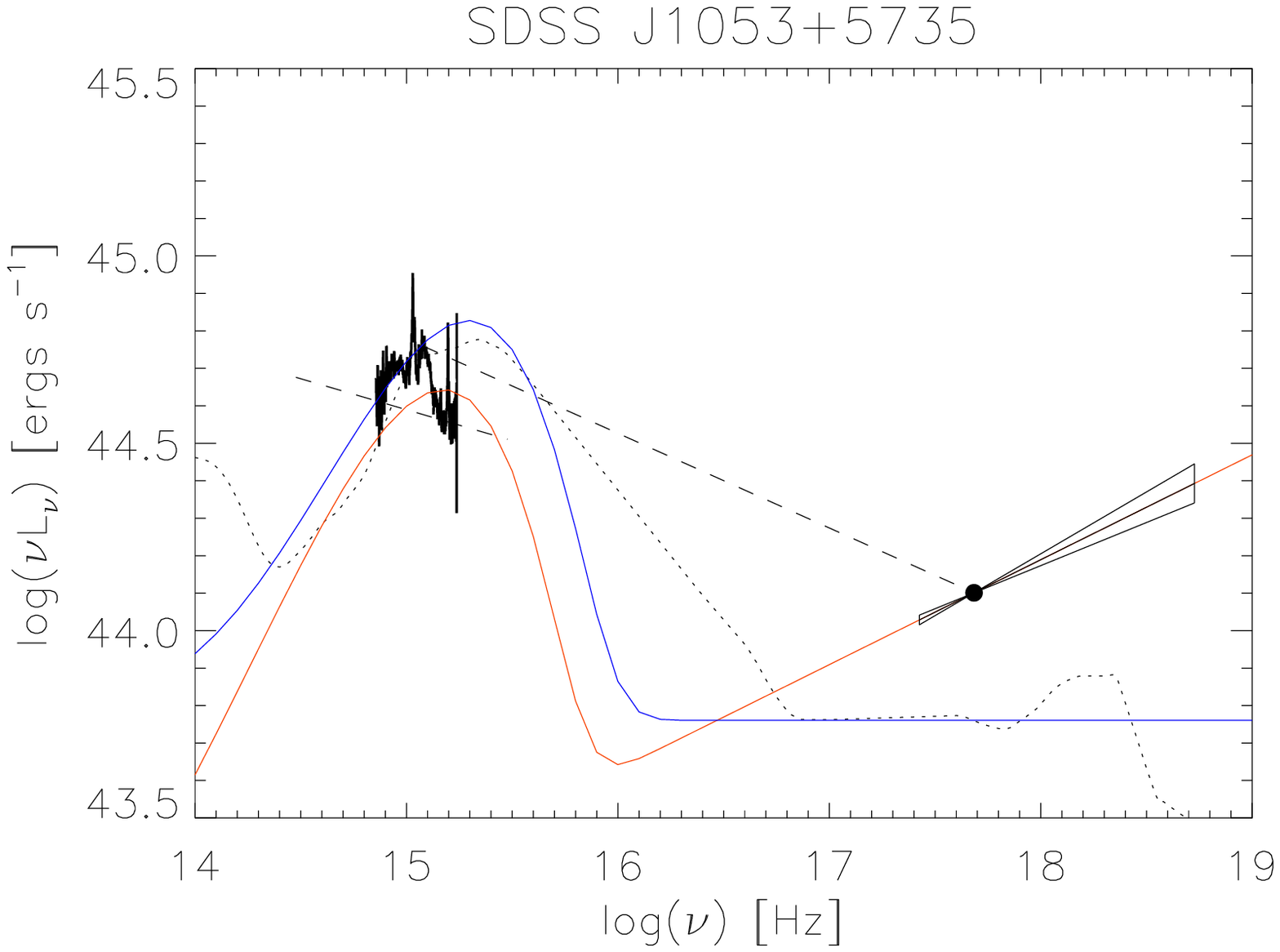}
\includegraphics[width=3.1in]{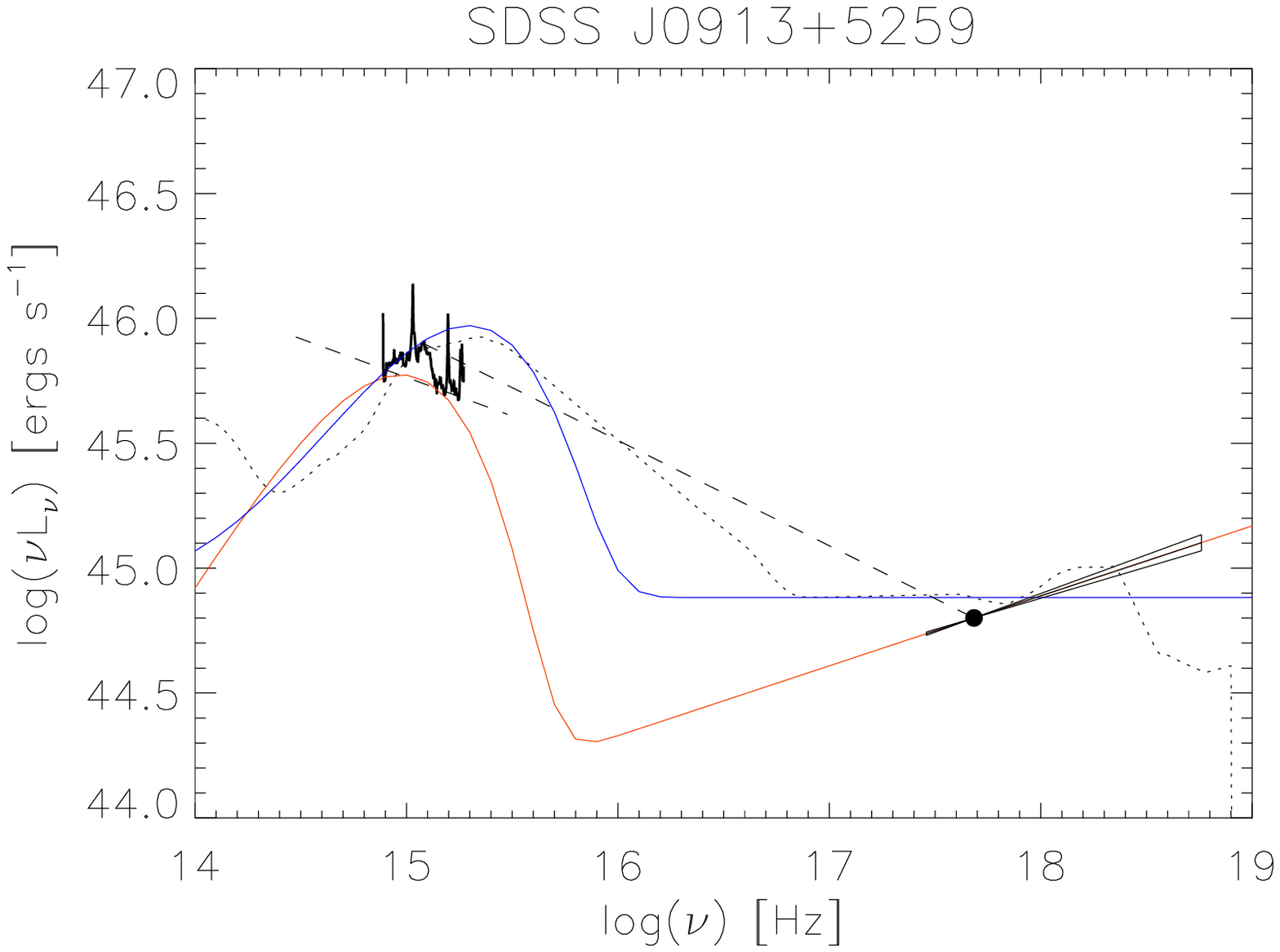}\\
\caption{The SED is plotted as a solid black line for the intrinsically red sources 
\#5,\#12-\#17, starting with \#5 in the top left panel,.  
Units are log($\nu$L$_\nu$)~[ergs s$^{-1}$] vs. log($\nu$)~[Hz].  
The \citet{Elvis94} SED is overplotted for reference as a 
dotted black line.  A toy SED model employing a simple `disk' shape and parameters as described in 
\S5.3 is overplotted as a solid, blue line to match the \citet{Elvis94} SED.  
A second toy model is calculated, where all the parameters remain the same except that the 
cut-off temperature is reduced by the amount expected from the calculated accretion rate (\S5.3).  
This second model is overplotted as a solid, red line.  
The optical power-law ($\alpha_{opt}$) and the optical-to-X-ray power-law ($\alpha_{ox}$) are 
overplotted as dashed lines.  90\% errors on the X-ray power-law are plotted as a bow tie.
Each plot is labeled with the SDSS name.}
\label{fig:SED}
\end{figure*}

We can model the low disk temperature expected from a low accretion rate using a simple template SED.  
The CLOUDY package approximates the AGN SED with a toy model \citep{Ferland01}:
\begin{displaymath}F_\nu = A\nu^{-\alpha_{uv}} e^{-h\nu/kT_{cut}} + B\nu^{\alpha_x}\end{displaymath}
The A and B constants are normalizations for the optical/UV and X-ray terms, respectively.  
The UV and X-ray slopes are power-laws ($\alpha_{uv}$ and $\alpha_{x}$), where F$_\nu \propto 
\nu^\alpha$.  The high-energy cut-off, kT$_{cut}$, corresponds to the maximum accretion disk temperature.  
We performed a qualitative fit, first to the \citet{Elvis94} SED, and then to the intrinsically 
red quasar SED.  For the \citet{Elvis94} SED, we use $\alpha_{uv}$ = -0.5, $\alpha_x$ = -1, and T$_{cut}$ = 
10$^{4.8}$ K.  For the intrinsically red SEDs, we change $\alpha_x$ to the value obtained in the 
X-ray fit, and we change T$_{cut}$ by the amount expected for the lower accretion rate, calculated 
according to the \citet{Law05} scaling relation.  Since the UV power-law defines the rise of the 
BBB, we do not change $\alpha_{uv}$.
Figure~\ref{fig:SED} shows the SED data and model fit for all seven intrinsically red quasars.  
The figure demonstrates that a lower disk temperature qualitatively reproduces the 
observed red power-law for four objects.  Two objects (\#15 and \#16) are better fit with a 
an accretion rate an order of magnitude smaller than that shown.  A third object (\#5) is not 
fit well by any model.  

\section{CONCLUSIONS AND FUTURE WORK}

We have cross-correlated the SDSS DR3 Quasar Catalog \citep{Schnei05} with the XMM-Newton archive, 
selecting the reddest ($g - r$ $\geq$ 0.5), moderate-redshift (0.9 $<$ z $<$ 2.1) quasars.  We 
obtain a sample of 17 quasars, 16 of which are detected in the X-rays.  Using both optical 
and X-ray data to constrain dust-reddening and absorption allowed us to distinguish between 
obscured and intrinsically red quasars, although two cases remain ambiguous.

Eight quasars are dust-reddened in the optical and, while the X-ray data prefer an unabsorbed 
power-law, the upper-limits are high enough that X-ray absorption at the level expected for an 
SMC dust-to-gas ratio is allowed.  For the three quasars with high enough X-ray S/N to fit a 
power-law + intrinsic absorption model, we obtain upper-limits of 3 - 13 x 10$^{22}$ cm$^{-2}$.  
For five sources, the X-ray S/N is too low to fit spectral models.  However we can obtain a 
lower-limit to N$_H$ by comparing the $\alpha_{ox}-l_{uv}$ distribution of the red quasar 
sample to that of \citet{Strat05}, a recent study that uses a similar selection method, aside 
from the red color cut.  If the two samples come from the same parent distribution, an 
absorbing column of at least 10$^{23}$ cm$^{-2}$ is required for the five red quasars.  

Seven quasars display no evidence of X-ray absorption, and dust-reddening is contraindicated 
by the continuum shape as determined by the optical photometry.  These quasars seem to form 
a group with intrinsically red power-laws in the optical/UV.  It seems that these intrinsically 
red power-laws may be caused by lower accretion rates: low accretion rates are derived from 
M$_{BH}$ and L$_{bol}$ estimates, although the M$_{BH}$ values may be biased high by the red 
SED's.  Moreover the unusually broad MgII emission lines place these quasars on the $\Gamma$-
FWHM(MgII) relation, also suggesting lower accretion rates.  

The intrinsically red quasars are a substantial population with extreme characteristics 
that are well-suited to X-ray 
follow-up.  Since 7 intrinsically red quasar candidates have been found in the 1\% of the 
SDSS DR3 quasar catalog that overlaps with archival XMM-Newton observations, as many as 700 intrinsically 
red candidates may exist in the complete catalog.  Since the DR6 has more than 
doubled the number of SDSS quasars, we can expect 1600 intrinsically red quasar candidates, 
16 of which have been observed serendipitously with the XMM-Newton.  The intrinsically red 
quasar candidates in this paper also merit further observations in order to confirm the lack of
dust-reddening and pinpoint the physical cause of the red optical power-laws. For example, 
NIR (e.g. JHK) spectra would include H$\alpha$ and H$\beta$, and so could give an independent 
measure of dust-reddening via the Balmer decrement.  NIR measurements will also constrain a 
low-temperature disk model, as well as L$_{bol}$ and so $\dot{M}$.  With optical and X-ray 
spectra already available, the intrinsically red quasar candidates described in this paper 
are well-suited to more complex SED modeling.  

\acknowledgements

The authors thank Gordon Richards for his help in navigating the SDSS database, 
Joanna Kuraszkiewicz for her help with understanding IRAF, and Bo$\dot{z}$ena Czerny 
for helpful conversations regarding SED models.  We also thank the anonymous referee for 
helpful comments which improved the presentation of this work.  This paper is based on 
observations obtained with XMM-Newton, an ESA science mission with 
instruments and contributions directly funded by ESA Member States and NASA, and the Sloan 
Digital Sky Survey (SDSS).  Funding for the SDSS and SDSS-II has been provided by the 
Alfred P.  Sloan Foundation, the Participating Institutions, the National Science Foundation, 
the U.S.  Department of Energy, the National Aeronautics and Space Administration, the 
Japanese Monbukagakusho, the Max Planck Society, and the Higher Education Funding Council 
for England.  This research also made use of the NASA/ IPAC Infrared Science Archive, 
which is operated by the Jet Propulsion Laboratory, California Institute of Technology, 
under contract with the National Aeronautics and Space Administration.  This work has 
been partially funded by NASA Grants NASA NNX07AI22G and NASA GO6-7102X G06-7102X.



\end{document}